\documentclass{aa_gaia}  

\usepackage[draft]{hyperref}

\usepackage{xcolor}
\definecolor{dark-red}{rgb}{0.9,0.0,0.0}
\definecolor{dark-blue}{rgb}{0.15,0.15,0.9}
\definecolor{dark-green}{rgb}{0.15,0.8,0.15}
\definecolor{medium-blue}{rgb}{0,0,0.9}
\hypersetup{
    colorlinks, linkcolor=red,
    citecolor={dark-blue} , urlcolor={medium-blue}
}
\usepackage{changepage} 

\usepackage{graphicx}
\usepackage{txfonts}
\usepackage{soul}
\usepackage{relsize}
\begin{document}

   \title{The CARMENES search for exoplanets around M dwarfs}

   \subtitle{First visual-channel radial-velocity measurements and orbital
parameter updates of seven M-dwarf planetary systems\thanks{Based on observations collected at 
the European Organisation for Astronomical Research in the Southern Hemisphere under ESO 
programmes 072.C-0488, 072.C-0513, 074.C-0012, 074.C-0364, 075.D-0614, 076.C-0878,
077.C-0364, 077.C-0530, 078.C-0044, 078.C-0833, 079.C-0681, 183.C-0437, 60.A-9036, 
082.C-0718, 183.C-0972, 085.C-0019, 087.C-0831, 191.C-0873 }}

    \author{T.\,Trifonov\inst{1}
     \and M.\,K\"urster\inst{1}
     \and M.\,Zechmeister\inst{2}
     \and L.\,Tal-Or\inst{2}
     \and J.\,A.\,Caballero\inst{3,5}
     \and A.\,Quirrenbach\inst{5}
     \and P.J.\,Amado\inst{6}      
     \and I.\,Ribas\inst{7}
     \and A.\,Reiners\inst{2} 
     \and S.\,Reffert\inst{5}
% ...
}
   \institute{Max-Planck-Institut f\"ur Astronomie,
              K\"onigstuhl 17, D-69117 Heidelberg, Germany\\
              \email{trifonov@mpia.de}
         \and Institut f\"ur Astrophysik, Georg-August-Universit\"at, 
              Friedrich-Hund-Platz 1, 37077 G\"ottingen, Germany
         \and Centro de Astrobiolog\'ia (CSIC/INTA), Instituto Nacional de T\'ecnica Aeroespacial,
              Ctra de Torrej\'on a Ajalvir, km 4, 28850 Torrej\'on de Ardoz, Madrid Spain
         \and Th\"uringer Landessternwarte Tautenburg, Sternwarte 5, D-07778 Tautenburg, Germany
         \and Zentrum f\"ur Astronomie der Universt\"at Heidelberg, Landessternwarte,
              K\"onigstuhl 12, D-69117 Heidelberg, Germany
         \and Instituto de Astrof\'isica de Andaluc\'ia (IAA-CSIC), Glorieta de la Astronom\'ia s/n, 
              E-18008 Granada, Spain
         \and Institut de Ci\`encies de l’Espai (CSIC-IEEC), Campus UAB, c/ de Can Magrans s/n, 
              E-08193 Bellaterra, Barcelona, Spain
% ...
}

   \date{Received 25 June 2017 / Accepted 20 September 2017}
 

  \abstract
   {
  The main goal of the CARMENES survey is to find Earth-mass planets around nearby M-dwarf stars. 
  Seven M dwarfs included in the CARMENES sample had been observed before with HIRES and HARPS and
either were reported to have one short period planetary companion  
 (GJ\,15\,A, GJ\,176, GJ\,436, GJ\,536 and GJ\,1148) or are multiple planetary systems (GJ\,581 and GJ\,876).  
   }   
   {
 We aim to report new precise optical radial velocity measurements
for these planet hosts and test the overall capabilities of CARMENES.  
   }
   {We combined our CARMENES precise Doppler measurements with those available from HIRES and HARPS and 
   derived new orbital parameters for the systems. Bona-fide single planet systems were fitted with a Keplerian model.
   The multiple planet systems were analyzed using a self-consistent dynamical model 
   and their best fit orbits were tested for long-term stability.
   }
   {We confirm or provide supportive arguments for planets around all the investigated stars except for GJ\,15\,A, 
   for which we find that the post-discovery HIRES data and our CARMENES data do not show a signal at 11.4 days. 
   Although we cannot confirm the super-Earth planet GJ\,15\,Ab, we
   show evidence for a possible long-period ($P_{\rm c}$ = 7030$_{-630}^{+970}$ days) Saturn-mass 
   ($m_{\rm c} \sin i$ = 51.8$_{-5.8}^{+5.5}$~$M_\oplus$) planet around GJ\,15\,A.
   In addition, based on our CARMENES and HIRES data we discover a second planet around 
   GJ\,1148, for which we estimate a period $P_{\rm c}$ = 532.6$_{-2.5}^{+4.1}$ days, 
   eccentricity $e_{\rm c}$ = 0.342$_{-0.062}^{+0.050}$ 
   and minimum mass $m_{\rm c} \sin i$ = 68.1$_{-2.2}^{+4.9}$~$M_\oplus$.
   }
   {
   The CARMENES optical radial velocities have similar precision and overall scatter when 
   compared to the Doppler measurements conducted with HARPS and HIRES. 
   We conclude that CARMENES is an instrument that is up to the challenge of 
   discovering rocky planets around low-mass stars.}


   \authorrunning{Trifonov et al.}
   \titlerunning{First CARMENES visual-channel radial-velocity measurements}

   \maketitle


\section{Introduction}

The quest for extrasolar planets around M dwarfs via precise Doppler measurements
is almost two decades old \citep{Marcy1998,Delfosse1998,Marcy2001,Endl2003,Kurster2003, Bonfils2005,Butler2006,Johnson2010}.
To date we are aware of at least 20 planet candidates orbiting nearby M-dwarf stars detected by the radial velocity
(RV) method \citep[][]{Bonfils2013, Hosey2015}, but the real number is likely to be much larger given the 
fact that the vast majority (70--80\%) of the stars in the solar neighborhood are yet poorly explored M dwarfs.
Indeed, the recent discoveries of planets in the habitable zone around 
Proxima Centauri \citep{Escude2016} and LHS 1140 \citep{Dittmann2017}, and the multiple planet system 
around the ultra-cool M-dwarf star TRAPPIST-1 \citep{Gillon2017}
provide strong evidence for an enormous population of potentially habitable planets around red dwarfs.

\begin{table*}{}      
\centering    

\caption{List of CARMENES known exoplanet host stars studied in this paper with their physical characteristics.}   
\label{table:1} 

\resizebox{0.7\textheight}{!}{\begin{minipage}{\textwidth}
\centering

 \begin{tabular}{llccccccc @{ }lccc}
\hline
\noalign{\vskip 0.5mm}
\hline
\noalign{\vskip 0.9mm}

$Karmn$    & GJ  & SpT$^a$ & $M^b$ & $d^a$    & $K_{\rm s}^a$ &  $P_{\rm rot}^a$    &  SA$^c$    \\  
         &     &         & $[M_{\odot}]$ &  [pc] &  [mag]   &  [d]               &   [m\,s$^{-1}$\,yr$^{-1}$]   \\  
\hline
\noalign{\vskip 0.9mm}

 J00183+440    &   15\,A   & M1.0\,V & 0.414~$\pm$~0.012  & ~~3.562~$\pm$~0.039  & 4.018~$\pm$~0.020     &   ~~44.0~$\pm$~0.5          &   0.698   \\
 J04429+189    &   176     & M2.0\,V & 0.504~$\pm$~0.013  & ~~9.406~$\pm$~0.053  & 5.607~$\pm$~0.034     &   ~~40.6~$\pm$~0.4          &   0.363   \\
 J11417+427    &   1148    & M4.0\,V & 0.357~$\pm$~0.013  & 10.996~$\pm$~0.051   & 6.822~$\pm$~0.016     &   ~~73.5~$\pm$~0.4          &   0.086   \\ 
 J11421+267    &   436     & M2.5\,V & 0.436~$\pm$~0.012  & ~~9.748~$\pm$~0.029  & 6.073~$\pm$~0.016     &   ~~39.9~$\pm$~0.8          &   0.328   \\
 J14010--026   &   536     & M1.0\,V & 0.530~$\pm$~0.011  & 10.418~$\pm$~0.055   & 5.683~$\pm$~0.020     &   ~~43.3~$\pm$~0.1          &   0.245   \\
 J15194--077   &   581     & M3.0\,V & 0.323~$\pm$~0.013  & ~~6.304~$\pm$~0.014  & 5.837~$\pm$~0.023     &   132.5~$\pm$~6.3           &   0.218   \\
 J22532--142   &   876     & M4.0\,V & 0.350~$\pm$~0.013  & ~~4.672~$\pm$~0.021  & 5.010~$\pm$~0.021     &   ~~81.0~$\pm$~0.8          &   0.147   \\ 

\hline
\noalign{\vskip 0.5mm}

\end{tabular}

\tablefoot{
a - $Carmencita$ Catalog and references therein,
b - Combined polynomial fit to the \citet{Benedict2016} and \citet{Delfosse2000} relations,
c - Positive RV drift due to secular acceleration.
}

\end{minipage}}
\end{table*}

M dwarfs are particularly suitable targets to detect temperate 
low-mass rocky planets primarily for two reasons:
(1) The lower masses of M dwarfs compared to those of solar-like
stars facilitate the detection of lower mass planets.
(2) Due to the low flux of M dwarfs, the habitable zone is located
closer-in than that of hotter and more massive stars. 
As a result, planets in the habitable zone of M dwarfs have shorter periods,
and thus higher Doppler signals than those orbiting in the habitable zones of more massive stars.
However, their active nature can also cause certain observational difficulties.
Starspots,  plages, or activity cycles can lead to line profile variations, 
which can be easily mistaken as an RV signal due to an orbiting planet. 
In addition, non-negligible stochastic stellar jitter can have velocity levels
of a few m\,s$^{-1}$, making the detection of low-mass planets challenging.
Therefore,  persistent observations with state-of-the-art RV 
precision instruments such as HARPS \citep[La Silla Observatory, Chile,][]{Mayor2003},
HARPS-N \citep[Roque de Los Muchachos Observatory, La Palma, Spain,][]{Cosentino2012}, 
or HIRES \citep[Keck Observatory, Hawaii, USA,][]{Vogt1994} 
are needed to disentangle the planet signal from stellar activity. 
Alternatively, precise RV measurements simultaneously obtained in the optical and in the
near-infrared (NIR) domains may provide more evidence in favor or against the planet hypothesis.

These issues and observational philosophy are addressed 
with the new CARMENES\footnote{Calar Alto high-Resolution search for
M dwarfs with Exo-earths with Near-infrared and optical
Echelle Spectrographs. \url{http://carmenes.caha.es}} instrument and survey 
\citep{Quirrenbach2014,Quirrenbach2016,Amado2013,Floriano2015}
using a high-resolution dual-channel (Visual: R = 94\,600, NIR: R = 80\,400) 
spectrograph installed at the 3.5\,m telescope of the Calar Alto Observatory (Spain). 
CARMENES is designed to provide precise RV measurements in the optical and
NIR wavelength regimes with a precision of 1--2 m\,s$^{-1}$.  
The science program with CARMENES started on Jan 1, 2016 
and its main goal is to probe $\sim$300 close M-dwarf stars
for the presence of  exoplanets, in particular Earth-mass planets in the habitable zone. 

In this paper we present results from observations of single and multiple planetary systems around seven well-known
M dwarfs based on precise Doppler measurements taken with the visual channel of CARMENES. While the performance of
the NIR channel will be the subject of a future study, the RV precision achievable with the visual channel is compared
with those achieved for the same stars with other state-of-the-art planet-hunting spectrographs working in the
visible such as HARPS and HIRES. We use the combined RVs to confirm or refute the existence of the announced planets, look
for new candidates, and refine the orbital parameters of the planets.

We organize this paper as follows: 
in Section \ref{Sec2}, we introduce the seven known M-dwarf planet hosts, for which we obtain Doppler data with CARMENES.
In Section \ref{Sec2a} we discuss the available RV data for these stars and we present our RV analysis strategy.
In Section \ref{Sec3}, we present our results and we discuss each single and multiple planet system individually.
In Section \ref{sect.4} we provide an overview of the CARMENES performance compared to HARPS and HIRES.
Finally, in Section \ref{Sec4}, we provide a summary of our results and our overall conclusions.

\begin{figure}[]
\begin{center}$
\begin{array}{cc} 
\includegraphics[width=9cm]{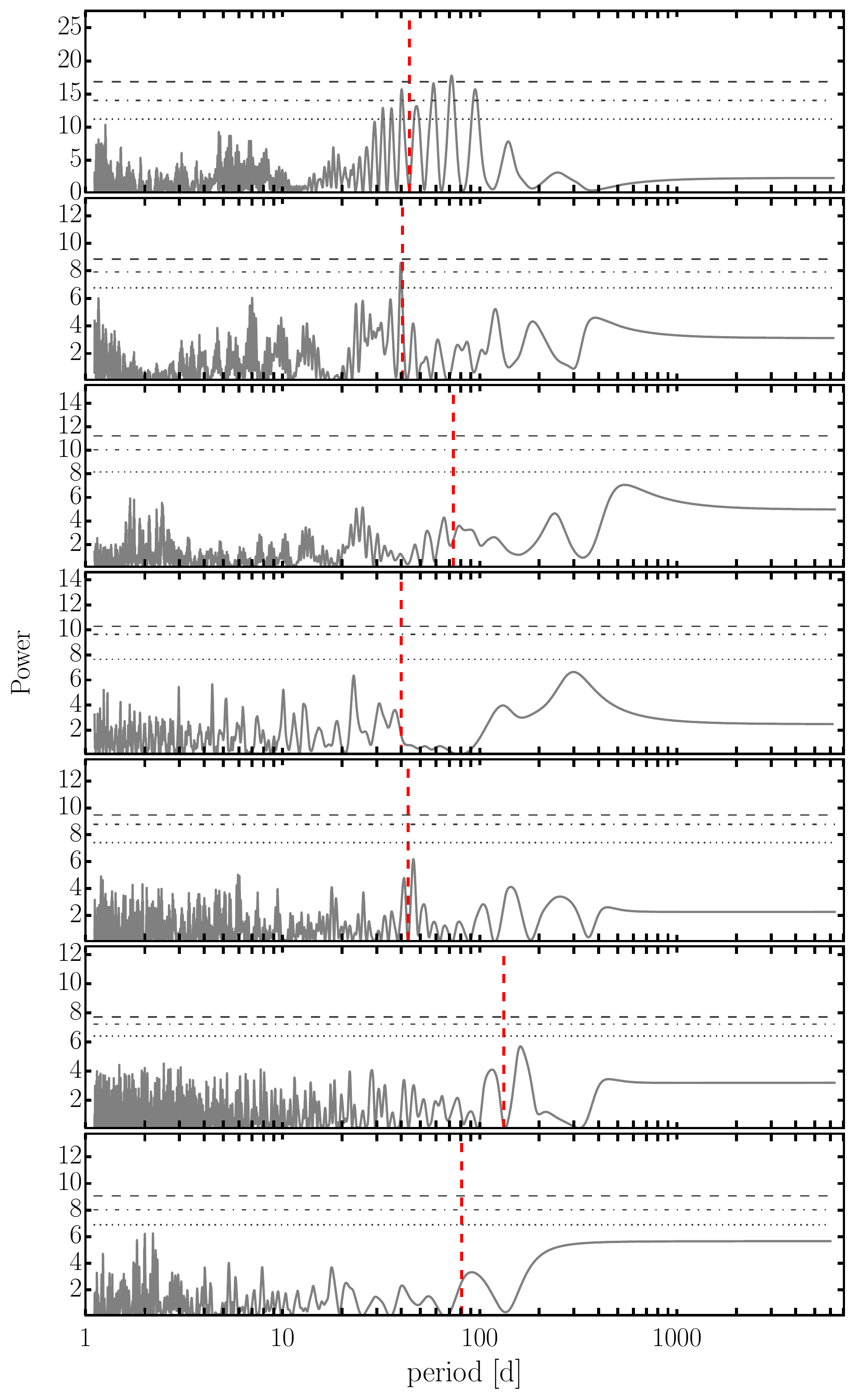}
\put(-220,400){GJ\,15\,A} \put(-220,345){GJ\,176} \put(-220,289){GJ\,1148} \put(-220,233){GJ\,436} \put(-220,177){GJ\,536} 
\put(-220,121){GJ\,581} \put(-220,65){GJ\,876} 
\\
\end{array} $
\end{center}

\caption{
GLS periodograms of the H$\alpha$ index obtained from CARMENES spectra
for the seven known planetary hosts.
Horizontal lines show  the bootstrapped FAP levels of 10\% (dotted line), 1\% (dot-dashed line) and 0.1\% (dashed line),
while red vertical lines indicate the stellar rotational periods listed in Table~\ref{table:1}.
The H$\alpha$ index analysis do not yield 
significant peaks at the known planet periods for these stars.
For GJ\,15\,A we identify several formally significant peaks between 40 and 100 days. 
For GJ\,176 and GJ\,536 the H$\alpha$ index peaks near the rotational periods of these two stars.
}   

\label{fig0} 
\end{figure}
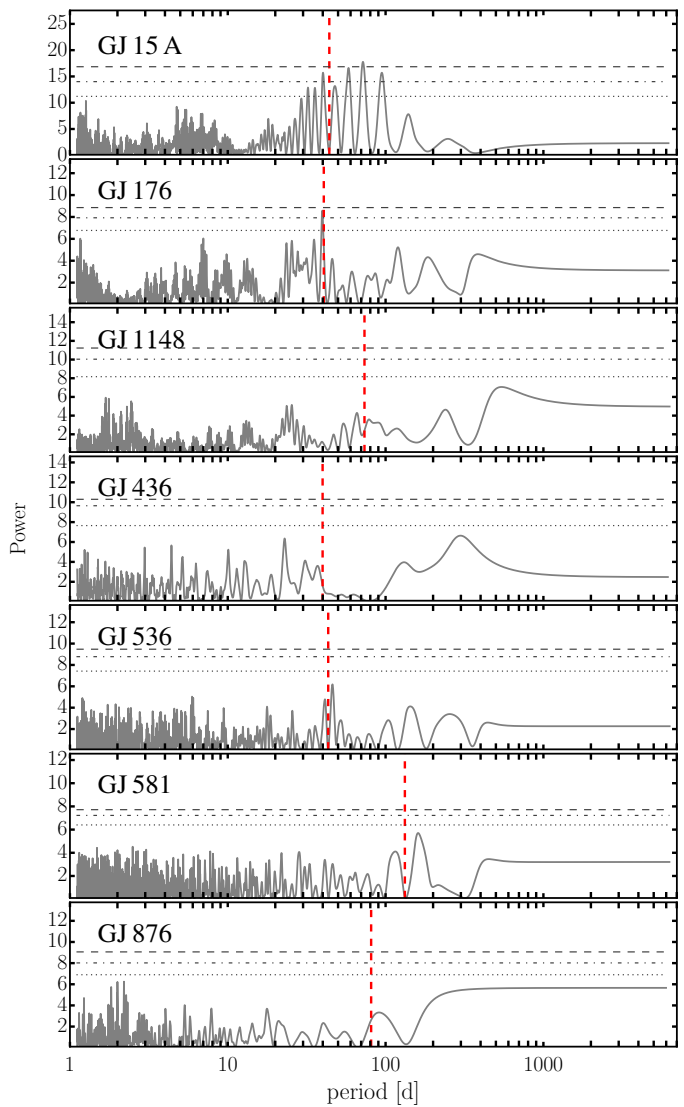

\section{The planetary systems}
\label{Sec2}

\subsection{Target selection}
\label{Sec2_targ}

The CARMENES GTO targets were selected from the $Carmencita$ catalog
\citep{Caballero2016} based on their observability from Calar Alto ($\delta > -23^\circ$), 
spectral types (M0.0--9.5\,V), $J$ magnitude for spectral sub-type (mean magnitude $\overline J$ = 7.7 mag), and 
status as bona-fide single stars with no evidence for a stellar companion within 5\,\arcsec. 
Given these selection criteria, several already known M-dwarf planetary systems were 
naturally included in the CARMENES sample.
These systems are the (presumably) single planet systems: 
\object{GJ\,15\,A} \citep{Howard2014},
\object{GJ\,436} \citep{Butler2004,Maness2007,Lanotte2014},
\object{GJ\,176} \citep{Endl2008,Forveille2009}, 
\object{GJ\,536} \citep{Mascareno2017} 
and 
\object{GJ\,1148} \citep{Haghighipour2010},
and the multiple planet systems: 
\object{GJ\,581} \citep{Bonfils2005,Udry2007,Mayor2009,Robertson2014}, 
and 
\object{GJ\,876} \citep{Marcy2001,Rivera2005,Rivera2010}.

There are actually several more M dwarfs with known planets in our sample, for example, 
GJ\,179 \citep{Howard2010},
GJ\,625 \citep{Mascareno2017c},
GJ\,628 \citep{Wright2016,Defru2017},
GJ\,649 \citep{Johnson2010} and  
GJ\,849 \citep{Butler2006}.
However, these stars either have planetary companions with very long orbital periods
exceeding the current temporal baseline of the survey,
or we have not yet collected a sufficient number of CARMENES data 
to adequately constrain their planetary architectures.
Therefore, we have chosen not to include these stars in this paper.

The seven selected stars are listed in Table~\ref{table:1}, sorted by their $Carmencita$ 
identifier $Karmn$, followed by their Gliese-Jahrei\ss{} 
\citep[GJ,][]{Gliese1991} catalog number, as well as by observational parameters,
such as spectral type, distance, $K_s$ magnitude and the estimated rotational period $P_{\rm rot}$. 
The M-dwarf mass estimates were derived using a combined polynomial 
fit of the $K_s$-band mass-luminosity relationships of \citet{Delfosse2000} and \citet{Benedict2016},
and thus represent an update with respect to the literature mass estimates.
These seven stars are typical red dwarfs with spectral types M1.0--4.0\,V and 
with masses in the range 0.32--0.53 $M_\odot$. 
Their relatively long stellar rotation periods, $P_{\rm rot}$,  and their the H$\alpha$ index 
activity indicator as defined in \citet[][]{Kuerster2003} suggest 
that these particular stars should not be strongly affected by stellar magnetic activity.
In Fig.~\ref{fig0} we show Generalized Lomb-Scargle \citep[GLS;][]{Zechmeister2009} 
power spectrum\footnote{In this work the GLS power spectrum is normalized 
following the \citet{Horne1986} normalization scheme, while
the false-alarm probability (FAP) levels of 10\%, 1\% and 0.1\% 
were calculated by bootstrap randomization creating 1000 randomly 
reordered copies of the data time-series \citep{Bieber1990, Kuerster1997}} 
periodograms of the  H$\alpha$ index activity indicator for the seven stars, obtained from the CARMENES spectra.
Our preliminary results of the H$\alpha$ index measurements 
show that the periodograms of GJ\,176 and GJ\,536 have strong peaks
near their stellar rotational periods, while for
GJ\,15\,A we find a forest of strong peaks between 40 and 100 days, likely induced by activity.
For the remaining four targets, we do not find
significant periodic signals at the known stellar rotational periods or the known planetary periods.
These stars have been already extensively studied for more activity indicators to ensure robust 
planet detection \citep[e.g.,][]{Queloz2001,Boisse2011,Robertson2014,Mascareno2015,Hatzes2016,Mascareno2017b}.

The relative proximity of these M dwarfs \citep[$\overline{d}$ = 7.9$\pm$3.0\,pc;][]{Caballero2016}
results in high proper motions and, hence, in a notable secular acceleration (SA) of the RV \citep[][]{Kurster2003}.
The SA is a positive, usually very small, RV drift, but it can  accumulate considerably 
as the baseline of the Doppler observations increases.
Therefore, in Table~\ref{table:1} we provide the SA estimates, 
which were calculated following \citet{Zechmeister2009b} from
the stars proper motions ($\mu_\alpha \cos{\delta}$, $\mu_\delta$)
and parallaxes ($\pi$) taken from the Tycho-Gaia Astrometric Solution (TGAS) catalogue of the
{\em Gaia} DR1 release \citep{Lindegren2016,Gaia2016,Prusti2016}.

The results of this work are derived from CARMENES observations of these planetary
hosts between January 2016 and April 2017.
Most of the planets in these systems were discovered or confirmed as part of high-precision
Doppler programs for M-dwarf planets either with HARPS \citep{Bonfils2005} or with HIRES \citep{Howard2009}. 
Therefore, these stars have excellent pre-existing RV data, which we 
use as a benchmark to study the overall precision of our visual CARMENES velocities.

\begin{table*}{}      
\centering    

\caption{Number of literature, archival and CARMENES Doppler measurements for the M-dwarf planet hosts 
and the number of planets assumed in this paper.}   
\label{table:2} 

\resizebox{0.75\textheight}{!}{\begin{minipage}{\textwidth}
\centering

 \begin{tabular}{lcccccccc @{ }lccc}
\hline
\noalign{\vskip 0.5mm}
\hline
\noalign{\vskip 0.9mm}

 GJ  &    HIRES$^a$ & HARPS$^b$ & HARPS-N & CARMENES & \# Planets  \\ 

\hline
\noalign{\vskip 0.9mm}
 
   15\,A&   358    & \dots          & \dots          & 92   &  1$^d$  \\
   176  &   111    & 70             & \dots          & 23   &  1  \\
   1148 &   125    & \dots          & \dots          & 52   &  2$^e$    \\
   436  &   356    & 169            & \dots          & 113  &  1  \\
   536  &    70    & 195            & 12$^c$         & 28    &  1  \\
   581  &   413    & 251            & \dots          & 20   &  3  \\
   876  &   338    & 256            & \dots          &  28  &  4  \\
 
\hline
\noalign{\vskip 0.5mm}
 
\end{tabular}

\tablefoot{
a - All HIRES data taken from \citet{Butler2017},
b - Publicly available ESO archive data re-processed with SERVAL,  
c - \citet{Mascareno2017}, 
d - Additional long-period planet candidate (see Section~\ref{33}),
e - We announce the discovery of GJ\,1148 c.  
}

\end{minipage}}
\end{table*}

\subsection{Literature overview}

\paragraph{GJ\,15\,A:} Using 117 HIRES RVs of GJ\,15\,A taken between 1997 and 2011, 
\citet{Howard2014} detected several distinct periodic signals in the Doppler time series.
The strongest periodogram peak was reported at 11.44\,d 
followed by a large number of significant peaks in the range 
of 30 to 120\,d, the strongest of which at $\sim$44.0\,d.
\citet{Howard2014} concluded that the 44.0 day Doppler signal 
and its neighboring peaks were artifacts of rotating spots
induced by stellar activity, since similar periodic variability 
was also detected in their optical photometry and in the Ca~{\sc ii} H\&K lines.
The strong $\sim$11.44 day period signal, however, could not 
be associated with activity and thus suggested a planetary interpretation.
The best Keplerian fit with 11.44 day periodicity was found 
to be consistent with a low-mass planet ($m \sin i$ = 5.35 $M_{\oplus}$) having a nearly circular orbit.

\paragraph{GJ\,176:} A Neptune mass ($m \sin i$ = 24.5 $M_{\oplus}$) 
planet with a period of 10.24 days around GJ\,176 was initially proposed by \citet{Endl2008} 
based on 28 RV measurements taken with the 
High-Resolution Spectrograph \citep[HRS;][]{Tull1998} at the Hobby-Eberly Telescope (HET).
However, soon after the discovery, the existence of the planet was questioned by \citet{Butler2009},
who failed to detect the planet in their 41 HIRES RVs taken between 1998 and 2008.
\citet{Butler2009} argued that the higher precision of HIRES when compared to HET-HRS
should have been advantageous in recovering the planetary signal,
but instead they found an RV scatter of about $\sim$4\,m\,s$^{-1}$,
mostly consistent with the estimated jitter for GJ\,176 combined with the instrumental noise.
\citet{Forveille2009} presented independent observations with HARPS, 
which confirmed a planet around GJ\,176, but in an 8.8-day orbit 
and with a lower RV semi-amplitude consistent with a super-Earth planet 
with a minimum mass of~$m \sin i$~=~8.4~$M_{\oplus}$.
 
\paragraph{GJ\,1148:} The moderately eccentric planet GJ\,1148 b ($e_{\rm b}$ = 0.31)    
 was discovered based on 37 velocities taken with HIRES \citep{Haghighipour2010}.
The RV signal is consistent with a planetary period of $\sim$41.4 days 
and a semi-amplitude $K$ = 34~m\,s$^{-1}$, corresponding to $m \sin i$ = 89 $M_{\oplus}$ (0.28 $M_{\rm Jup}$).
The RV data for GJ\,1148 are also compatible with a linear trend of $\sim$2.47\,m\,s$^{-1}$\,yr$^{-1}$,
suggesting a possible long-period companion to the system.
Additionally, \citet{Haghighipour2010} performed extensive photometric observations 
of GJ\,1148 and found a significant 98.1-day periodicity 
that most likely arises from spots on the rotating star.
\citet{Butler2017} have published an extended HIRES data set for 
GJ\,1148, which seems to show an additional signal in the 
one-planet fit residuals with a periodicity of $\sim$530 days. 
\citet{Butler2017} have classified this signal as a planetary ``candidate'',
but they neither provide an orbital solution for the possible second planet,
nor have they updated the orbital solution for GJ\,1148 b.

\paragraph{GJ\,436:} This star has a very well studied planet first 
discovered by \citet{Butler2004} using HIRES data.
GJ\,436\,b has a period of only $P_{\rm b}$ = 2.64 days, a minimum mass of 
$m_{\rm b} \sin i$ = 23 $M_{\oplus}$ and an eccentricity of~$e_{\rm b}$ = 0.15.
Later, \citet{Gillon2007} found that GJ\,436\,b is a transiting planet 
 with an estimated radius and mass comparable to that of Neptune and Uranus.
It was suggested that GJ\,436 has additional planets.
A long-period planet was suspected to gravitationally perturb GJ\,436\,b, thus leading to the planet's
surprising non-zero eccentricity \citep{Maness2007}, or a lower mass Super-Earth planet
suspected to be orbiting at a period of 5.2 days in a possible 2:1 Laplace mean-motion resonance 
\citep[MMR;][]{Ribas2008}, but such claims have not been confirmed.
Finally, by studying 171 precise HARPS velocities and {\em Spitzer} data, \citet{Lanotte2014}
concluded that present data support the presence of only a single planet around the host star.

\paragraph{GJ\,536:} \citet{Mascareno2017} reported the discovery of a super-Earth like planet
orbiting GJ\,536 by analyzing 158 HARPS and 12 HARPS-N RV measurements. 
According to them, GJ\,536\,b has an orbital 
period of 8.7076 $\pm$ 0.0025\,d and a minimum mass of $m \sin i$ = 5.36 $\pm$ 0.69 $M_{\oplus}$.
In addition to the planetary signal, a strong $\sim$43-d period is evident, but it was attributed 
to stellar rotation after analyzing the time series of the Ca~{\sc ii} H\&K and H$\alpha$ activity indicators.

\paragraph{GJ\,581:} This star has one of the most debated multiple 
planetary systems when it comes to the number of detected planets.
The first planet GJ\,581\,b, was discovered by \citet{Bonfils2005} followed by \citet{Udry2007}, 
who increased the planet count to three by discovering GJ\,581\,c and~d. 
The planetary system suggested by \citet{Udry2007} consists of three planets with orbital  
periods of $P_{\rm b,c,d} \approx$ 5.4, 12.9 and  83.6\,d 
and minimum masses of $m_{\rm b,c,d}  \sin i \approx$ 15.7, 5.0 and  7.7\,$M_{\oplus}$, respectively.
Later, \citet{Mayor2009}  revised the period of GJ\,581\,d to 66.8\,d and
discovered an additional 1.7-$M_{\oplus}$ mass planet at 3.15 days named GJ\,581\,e. 
A simultaneous analysis of the HIRES and HARPS data for GJ\,581 led \citet{Vogt2010} to increase 
the planet count to six by introducing GJ\,581~f and~g  with $P_{\rm f,g} \approx$ 433 and 37\,d,
suggesting a very compact system where all six planets must have near-circular orbits.
Since \citet{Vogt2010}, a number of independent studies have disputed some of these discoveries.
\citet{Forveille2011} and \citet{Tuomi2011}  strongly supported 
only four planetary companions, arguing against GJ\,581~f and~g.
\citet{Baluev2013} suggested that the impact of red noise on precise Doppler planet searches 
might lead to false positive detections and, therefore, even GJ\,581\,d  might not be real.
\citet{Robertson2014} corrected the available Doppler data for activity 
and also suggested that the signal of GJ\,581\,d  might be an artifact of stellar activity.
Finally, \citet{Hatzes2016} showed an anti-correlation of the 66.8\,d period 
with the H$\alpha$ equivalent width to confirm that the signal of GJ\,581\,d  is intrinsic to the star.
To our knowledge, the currently confirmed planets orbiting the GJ\,581 system are three (b, c, e), and
in our analysis we will assume this number.

\paragraph{GJ\,876:} This star has another well-studied planetary system,
currently known to host four planets, three of which are likely in 1:2:4 MMR.
The first planet GJ\,876\,b  was independently discovered by \citet{Marcy1998} and \citet{Delfosse1998}.
The planet was reported to have a period of $\sim$61 days and a minimum mass of 
$m \sin i$ $\approx$ 860\,$M_\oplus$,  which was the first discovery of a Jovian-mass planet around an M-dwarf star. 
However, after continued monitoring of this star using HIRES, \citet{Marcy2001} provided
strong evidence for a second planet with a minimum mass of 
$m \sin i$ $\approx$ 250\,$M_\oplus$ and a period of $\sim$30 days. 
\citet{Marcy2001} also showed that the planets interact so strongly that a double Keplerian fit is not a valid model. 
Instead, a three body Newtonian dynamical model is necessary 
to fit the data, showing  that GJ\,876\,b  and GJ\,876\,c  are in a strong 2:1 MMR.
After the discovery of GJ\,876\,c, a super-Earth planet with a short period of only 
1.94\,d was proposed by \citet{Rivera2005}. 
A dynamical model including a third planet yielded a significant improvement over
the two-planet model, suggesting that the innermost planet is real and designated as GJ\,876\,d.
A fourth $\sim$124 day planet named GJ\,876\,e was proposed by \citet{Rivera2010} because of an additional 
strong periodicity seen in the three-planet dynamical model.
We consider four confirmed planets orbiting GJ\,876.

\section{Observations and data}
\label{Sec2a}
 
\subsection{CARMENES data}
\label{Sec2.2}

The two CARMENES spectrographs are grism cross-dispersed, white pupil, 
\'e{}chelle spectrograph working in quasi-Littrow mode using a two-beam, two-slice image slicer. 
The visible spectrograph covers the wavelength range from 0.52\,$\mu$m to 1.05\,$\mu$m with 61 orders,
a resolving power of R = 94\,600, and a mean sampling of 2.8 pixels per resolution element.
However, in the standard configuration. Since the dichroic beam splitter in the front 
end splits the wavelength range around 0.97\,$\mu$m  and because of low sensitivity and flux levels at the blue end of the spectrum
effectively only 42 orders from 0.52\,$\mu$m to 0.97\,$\mu$m yield useful data in the visible  channel.
The spectrograph accepts light from two fibers; the first fiber carries the 
light from the target star, while the second fiber can either be used for 
simultaneous wavelength calibration or for monitoring the sky. 
The former configuration was used for all observations presented in this paper. 
The spectrograph is housed in a vacuum vessel and operated at room temperature. 
The detector is a back-side illuminated $4112 \times 4096$ pixel CCD. 
The CARMENES instrument is described in more detail in \citet{Quirrenbach2016} and in the references therein.

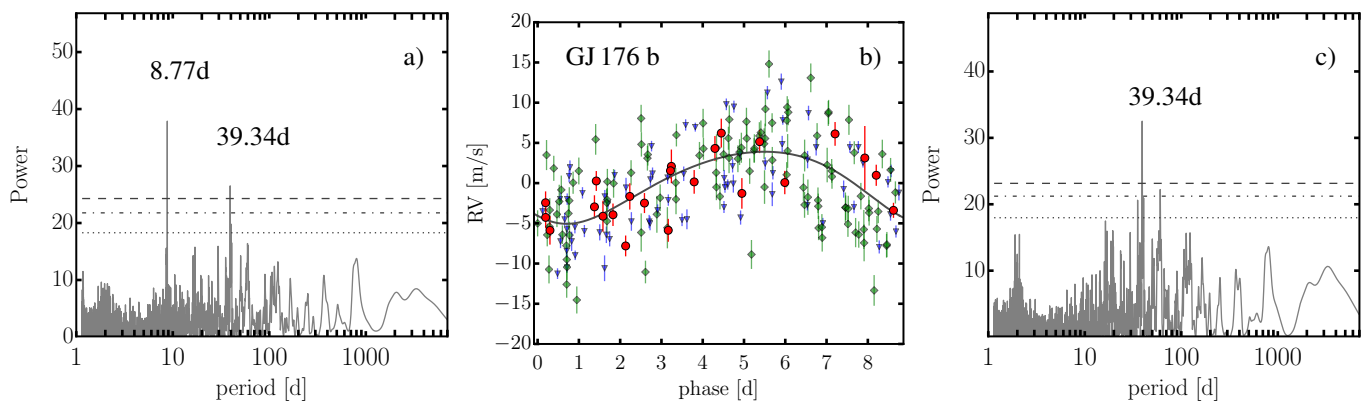
\begin{figure*}[tp]
\begin{center}$
\begin{array}{cc} 
\includegraphics[width=6cm]{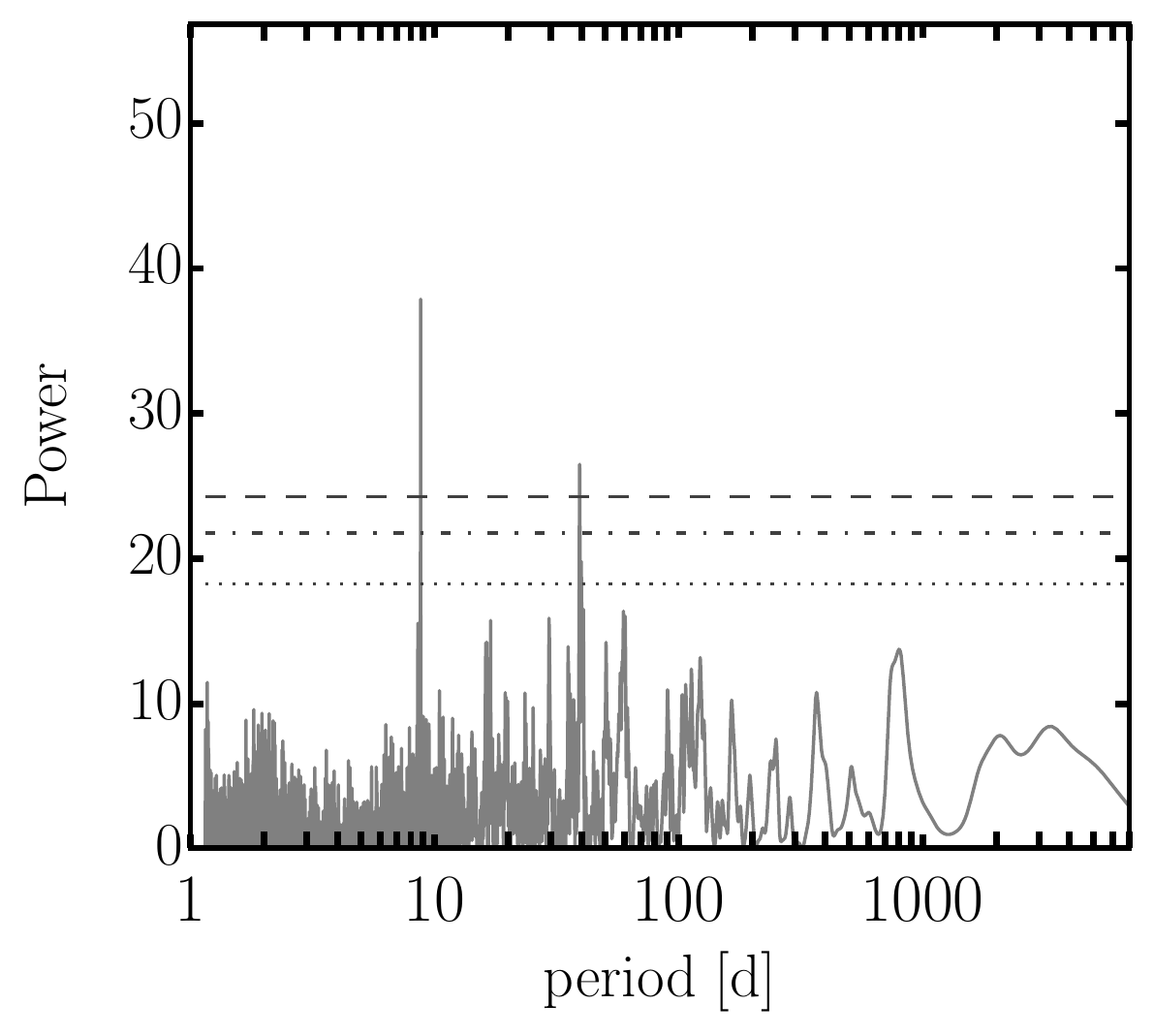}
\put(-115,125){8.77d} \put(-90,100){39.34d} \put(-20,130){a)} 
\includegraphics[width=6cm]{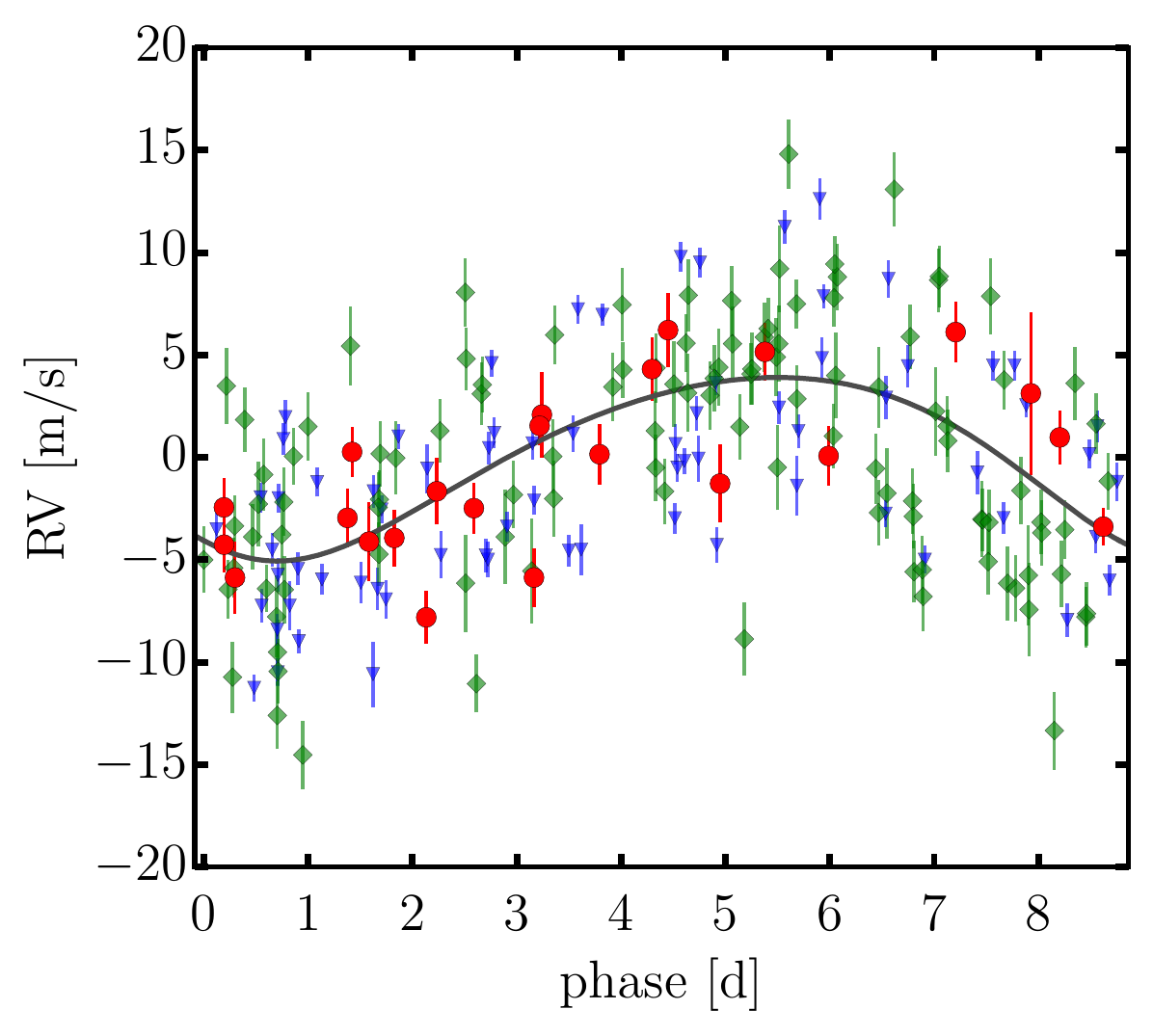} 
\put(-130,130){GJ\,176 b} \put(-20,130){b)} 
\includegraphics[width=6cm]{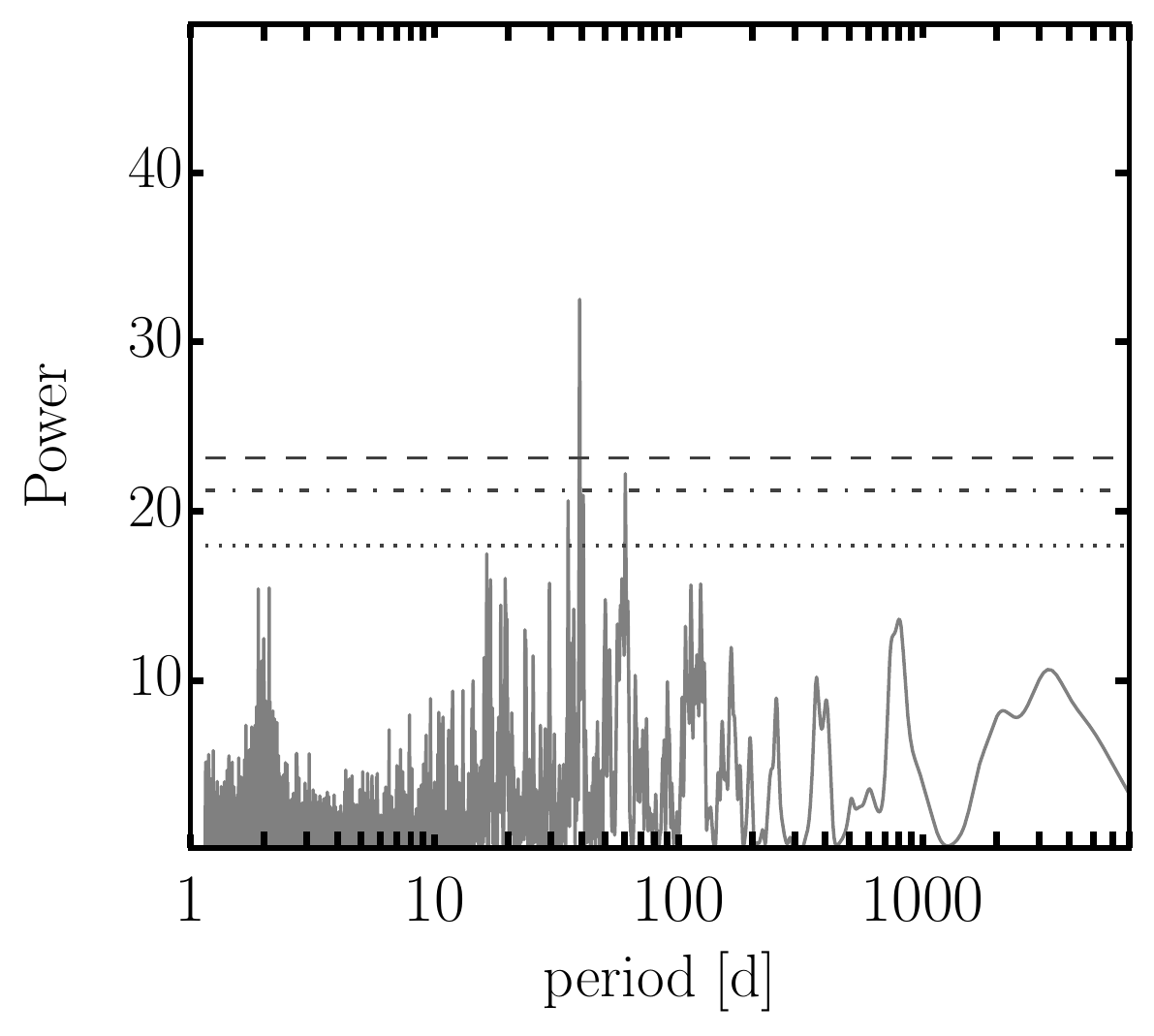}
\put(-90,115){39.34d} \put(-20,130){c)}  
\\
\end{array} $
\end{center}

\caption{
Panel a) GLS periodogram of the available Doppler data 
for GJ\,176, with horizontal lines showing the bootstrapped FAP levels of 10\% (dotted line), 
1\% (dot-dashed line) and 0.1\% (dashed line). 
Two distinct peaks above  the FAP = 0.1\% level can be seen at 
8.77\,d and 39.34\,d attributed to a planetary companion and stellar activity, respectively.
Panel b) Data from CARMENES (red circles), HARPS (blue triangles) and HIRES (green diamonds) 
phase-folded to our best Keplerian fit consistent with an 8.77\,d planet.
Panel c) GLS periodogram of the residuals after fitting the 8.77 d signal, revealing a 39.34\,d activity peak.
}   

\label{fig2} 
\end{figure*}

Standard processing of raw CARMENES spectra, such as bias, flat,
and cosmic ray correction are automatically performed using the CARACAL
\citep[CARMENES Reduction And Calibration,][]{Caballero2016b} pipeline.
The extraction of the spectra is based on flat-relative optimal extraction 
\citep[FOX; e.g.,][]{Zechmeister2014} and wavelengths are calibrated with algorithms described in 
\citet{Bauer2015}. The precise radial velocities are derived using our custom SERVAL 
\citep[SpEctrum Radial Velocity AnaLyser, ][]{Zechmeister2017} pipeline, 
which employs a $\chi^2$ fitting algorithm with one of the fit parameters being the RV. 
The observations are modeled with a template that is established from the observations 
following a suitable shifting and co-adding approach.
\citet{Escude2012} demonstrated that, in the case of  M dwarf stars, 
this method can provide higher RV precision than the method of cross-correlation 
with a weighted binary mask employed in the standard ESO HARPS pipeline.

The data presented in this paper were taken during the early 
phase of operation of the CARMENES visible-light spectrograph. 
During this time we identified a number of instrumental 
effects and calibration issues affecting the data on the m\,s$^{-1}$ level.
Therefore, taking advantage of the survey-mode observations, we calculated for each GTO 
night an instrumental nightly zero-point (NZP) of the RVs by using all the stars with 
small RV variability (RV-quiet stars) observed in that night. 
The sample of RV-quiet stars was defined as the sub-sample of CARMENES-GTO 
stars with RV standard deviation $<10$\,m\,s$^{-1}$. We then corrected 
each RV for its NZP and propagated the NZP error.
 
After 16 months of observations, the sample of RV-quiet stars includes $\sim200$ 
stars of which $10$--$20$ are observed in a typical night. 
Prior to the NZP calculation, we corrected each star's RVs for their own 
error-weighted average, replaced repeated exposures of a star in a given night 
by their median, and removed $4\sigma$ outliers. 
The NZP was then taken as the weighted-average RV of the observed RV-quiet stars. 
The NZP error was derived either from their RV uncertainties or from their RV standard 
deviation--whichever gave a larger value.
 
The median NZP uncertainty was found to be $\lesssim1$\,m\,s$^{-1}$, 
while their scatter is $\sim2.5$\,m\,s$^{-1}$. 
Only a few extreme NZPs were found to be as high as $\sim10$\,m\,s$^{-1}$. 
For the seven planetary systems investigated here, we found 
the NZP-corrected RVs to improve the $rms$ velocity 
dispersion around the best fit models  by $\sim25\%$ on average 
so we used them in our combined modeling with other-instrument's RVs.
We expect that a fuller understanding of the instrument 
will in the near future enable us to improve the 
calibration to the point where it is better than the present NZP correction scheme.
Examples for the improvement of the $rms$ of the time series of three stars 
due to the NZP correction are shown in Fig.~\ref{FigGam:A1}, while Fig.~\ref{FigGam:A2} 
provides a comparison of the pre-NZP and post-NZP correction for a larger sample of stars.
All CARMENES Doppler measurements and their individual formal uncertainties used
for our analysis in this paper are available in the Appendix (Tables A1-A9).

\subsection{Literature and archival data used in this paper}
\label{Sec2.3}

Table~\ref{table:2} provides the total number of available RVs for the seven M~dwarf planet hosts that we use for our analysis. 
RV data obtained with the HARPS and the HIRES spectrographs that have been in operation for much more than a decade dominate
over the RV data taken with the more recent instruments HARPS-N (only for GJ\,536) and our ongoing CARMENES survey.
GJ\,15\,A and GJ\,1148 have not been observed with HARPS since they are northern targets inaccessible from La Silla. 
For the rest, we used HARPS spectra from the ESO archive, which we re-processed with our SERVAL pipeline
for better precision and consistency.
All HIRES data for our selected targets were taken from \citet{Butler2017}, 
who released a large database of RV data collected over the past twenty years with HIRES.

Both HIRES and HARPS had a major optical upgrade since they were commissioned.
HIRES was upgraded with a new CCD in August 2004, while HARPS received new optical fibers in May 2015 \citep{LoCurto2015}. 
These upgrades aimed to improve the instrument's performance, 
but might also have introduced an RV offset between data taken before and after the upgrades. 
Further studies did not find a significant RV offset in HIRES \citep{Butler2017},
and in HARPS the offset is also close to zero in the case of M-stars  \citep{LoCurto2015}.
Therefore, we did not fit additional RV offsets between 
the pre- and post-upgrade HIRES and HARPS data in our analysis.

We modeled all available literature and archive RVs together with our CARMENES precise Doppler data.
In our analysis we used the individual RV data sets as they were, without
removing outliers or binning RVs into one measurement,
unless we find obviously wrong RV data (strong outliers over 10$\sigma$)
or heavy clustering of data with more than 5 RVs taken consecutively. 
We did not add stellar ``jitter'' quadratically to the RV error budget,
nor did we model the RV jitter variance of the data simultaneously 
with our orbital parameter optimization \citep[e.g.,][]{Baluev2009}.
All data sets were weighted by their nominal formal errors.
The main reason for analyzing the RVs in this way is simply because we 
know little about the stochastic stellar noise and active region evolution in M~dwarfs, 
their true orbital architecture (i.e., additional planets in the system 
and their mutual inclinations), or any instrumental low-amplitude systematics that 
might exist in CARMENES, HARPS and HIRES.
Thus, any unknown source of ``noise'' around our best-fitting model is 
accounted as a radial-velocity scatter (weighted $rms$) that we aim to study.

\subsection{RV modeling}

As a first step in our Doppler time series analysis we employed the GLS periodogram
to look for significant periodic signals that might be induced either by
known planetary companions, previously undiscovered planetary companions, or stellar activity.
The false-alarm probability (FAP) levels of 10\%, 1\% and 0.1\% were calculated by bootstrap randomization 
creating 1000 randomly reordered copies of the RV data and tested against the GLS algorithm.

To model the orbital parameters, we applied the Levenberg-Marquardt (L-M) based $\chi^2$ minimization technique 
coupled with two models. For {\em bona fide} single-planet systems we used
a Keplerian model, while the known multiple planet systems were fitted with a self-consistent N-body
 model based on the Gragg-Bulirsch-Stoer integration method \citep[][]{Press}. 
The N-body modeling scheme was fully described for the HD~82943 2:1 MMR system \citep{Tan2013}
and it was successfully applied to other multiple planet 
systems  such as HD~73526 \citep{Wittenmyer2014} and $\eta$~Ceti \citep{Trifonov2014}.

For both models the fitted parameters are the spectroscopic elements: radial velocity semi-amplitude~$K$, 
orbital period~$P$, eccentricity~$e$, longitude of periastron~$\varpi$, mean anomaly~$M$ and the velocity offset $\gamma$
for each data set included in the analysis; they are valid for the first observational epoch $T_0$.
For the N-body model we obtain the parameters in Jacobi coordinates \citep[e.g.,][]{LeeM2003}, 
which is a natural frame for analyzing an RV signal in multiple planet systems.
A final output from our models are also the best-fit reduced $\chi^2$ ($\chi_{\nu}^2$) 
and the individual data sets weighted $rms$ statistics, 
while the best-fit parameter uncertainties are determined by drawing 5000 
model-independent synthetic bootstrap samples from the available data \citep[e.g.,][]{Press}.
Each of the combined 5000 bootstrapped data sets is consecutively fitted with the corresponding 
Keplerian or N-body model, and from the resulting parameter distribution we obtain the 1-$\sigma$ asymmetric uncertainties.

\begin{figure*}[btp]
\begin{center}$
\begin{array}{cc} 
\includegraphics[width=6.0cm]{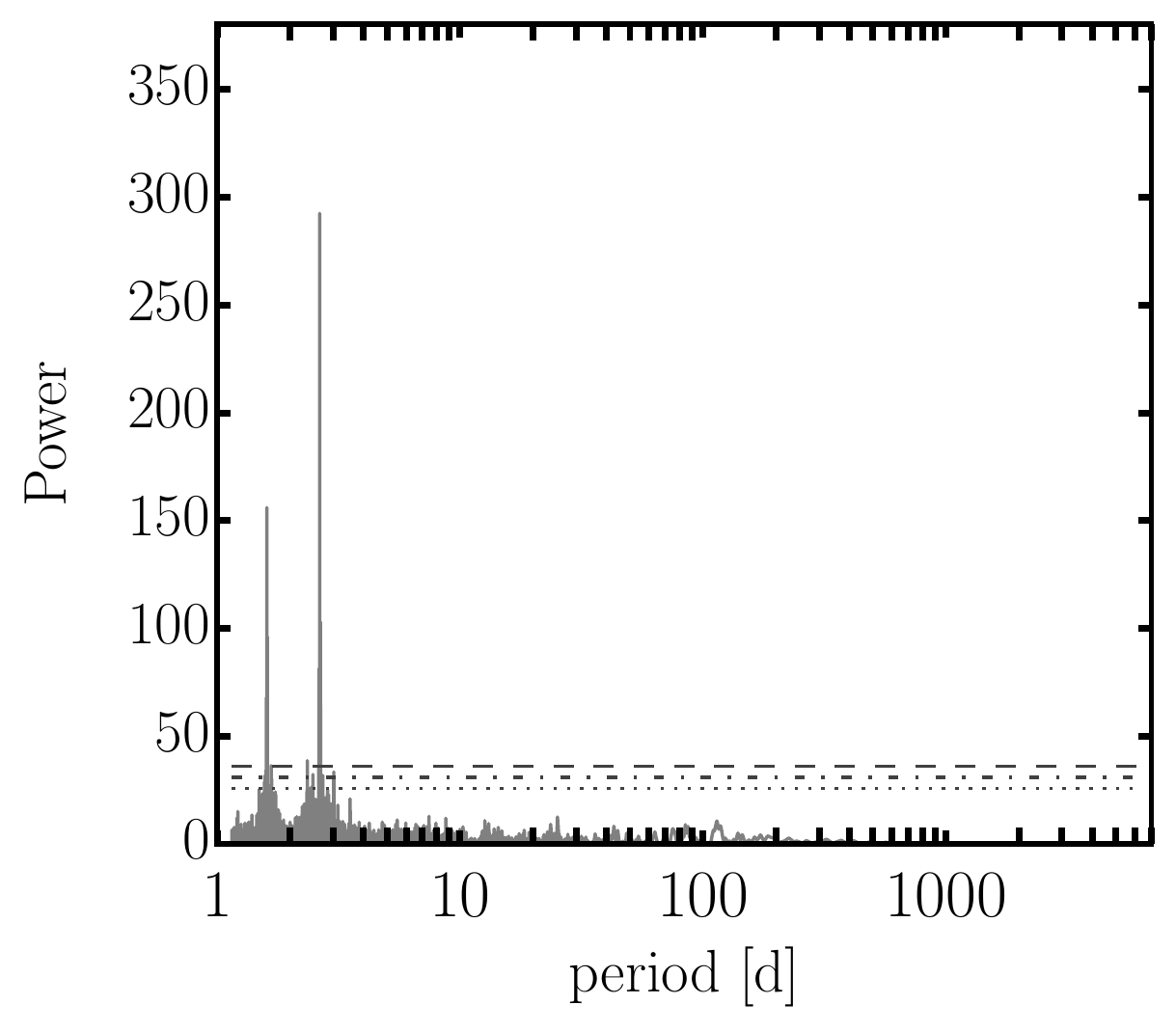} 
\put(-133,125){2.64d} \put(-138,85){1.60d} \put(-20,130){a)} 
\includegraphics[width=6.0cm]{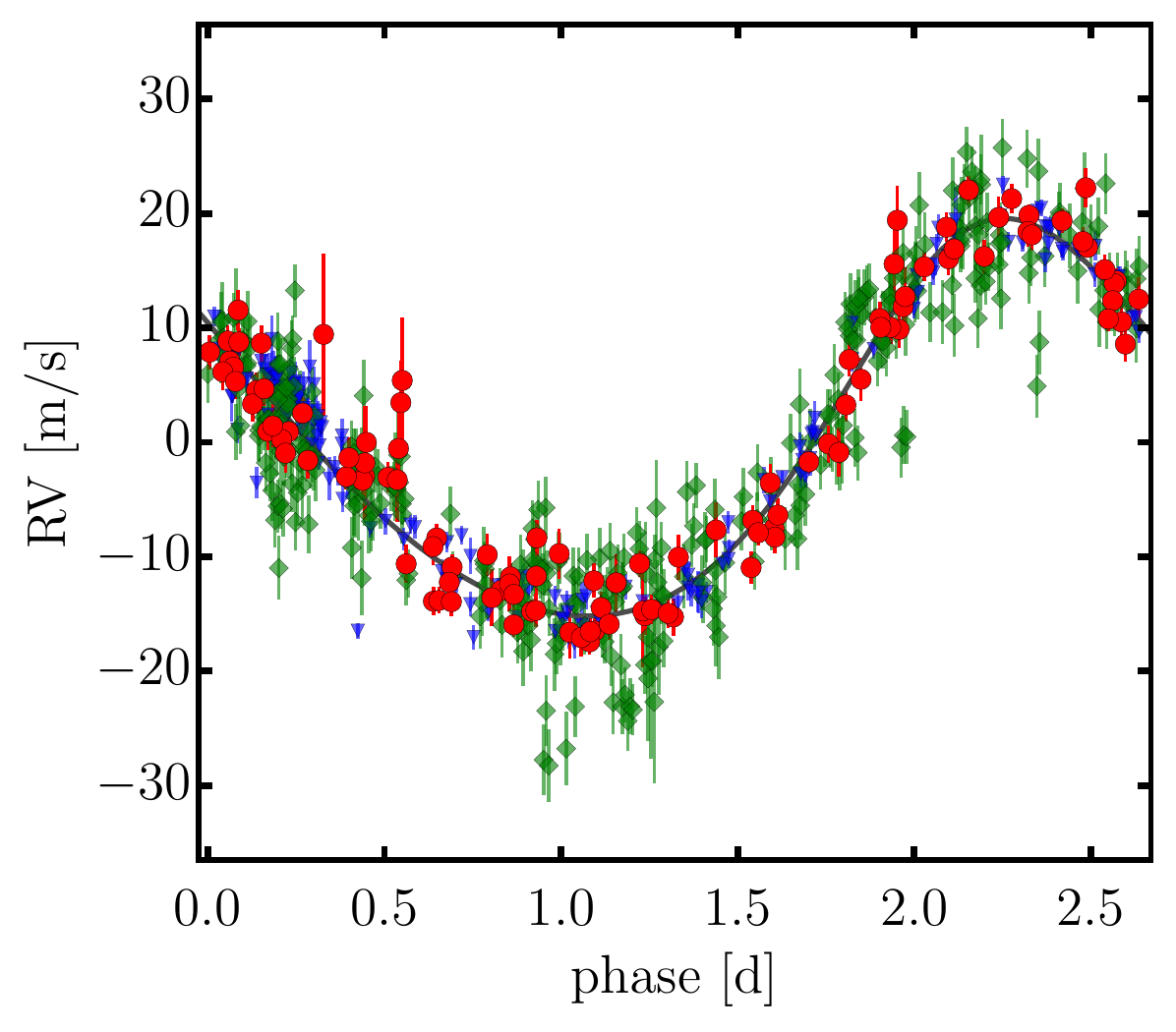} 
\put(-130,130){GJ\,436\,b} \put(-20,130){b)} 
\includegraphics[width=6.0cm]{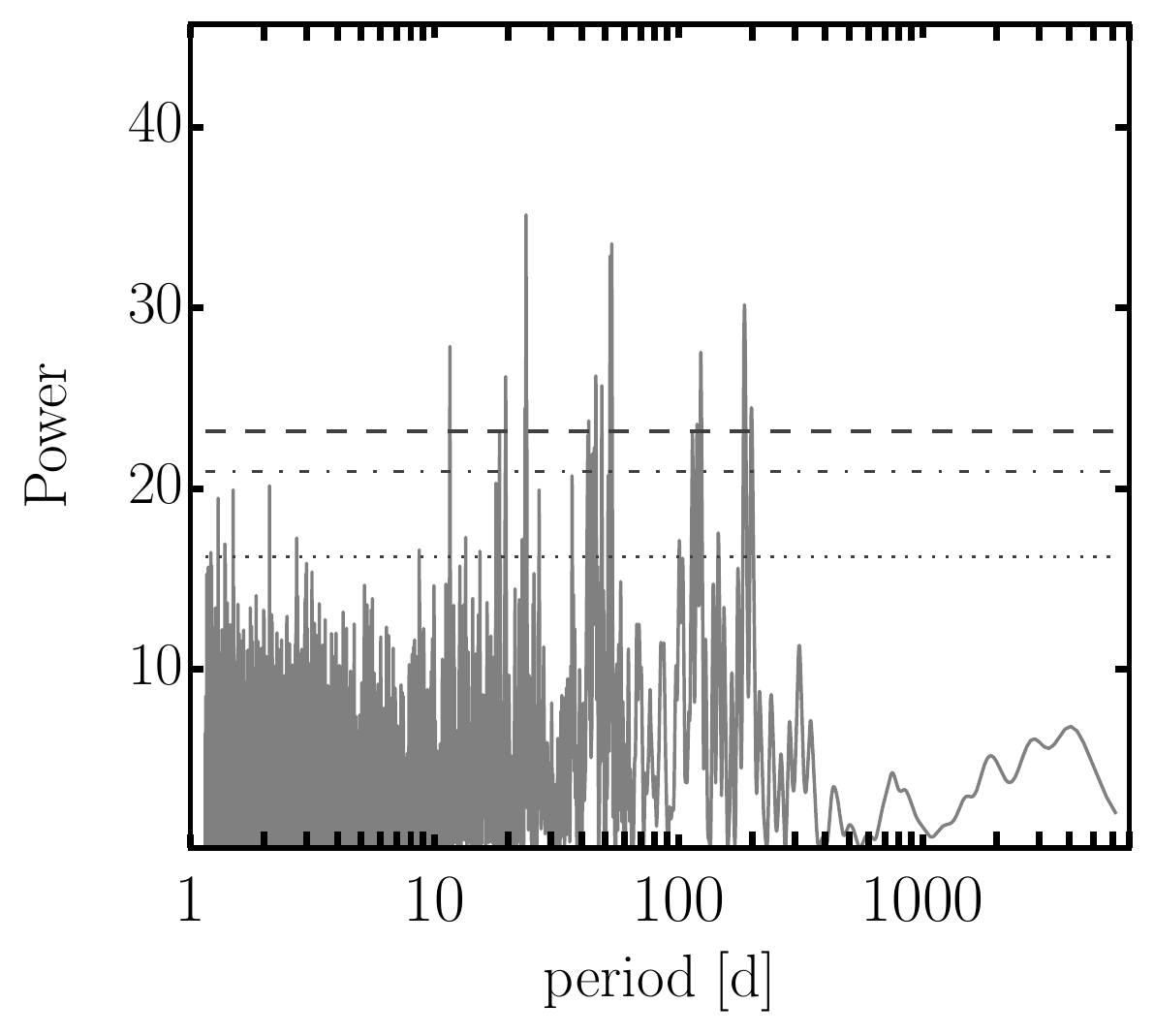} 
\put(-20,130){c)} \put(-100,130){23.7d} \put(-85,120){53.2d} \put(-70,110){186.5d} \put(-115,110){11.6d}
\\

\end{array} $
\end{center}

\caption{Same as Fig.~\ref{fig2}, but for GJ\,436.
Panels a) and b) show that the CARMENES, HARPS and HIRES
data sets used  for the construction of the best fit are fully
consistent with a planetary companion with a period of 2.64\,d.
In panel a) the 1.60\,d GLS peak is a 1-day alias of the 2.64\,d 
periodicity induced by the planetary companion.
Panel c) shows  the GLS periodogram of the best fit residuals, 
which yields several peaks above the FAP = 0.1\%, 
most likely due to  the observational window, stellar activity and their aliases. 
}   

\label{fig3} 
\end{figure*}
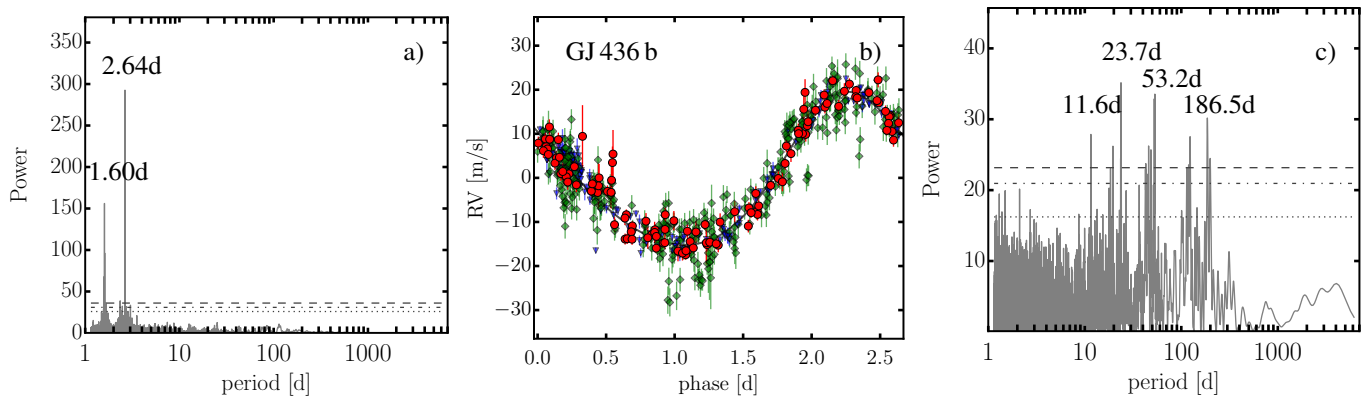

The best dynamical models are further tested for long-term dynamical stability 
using the {\it SyMBA} symplectic integrator \citep{Duncan1998}, modified to work with Jacobi input elements. 
We chose a maximum of 10 Myr of integration time, 
which we believe is adequate to test the long-term stability of our fits.
The time step we chose is 1\% of the period of the respective innermost planet,
thus allowing for precise orbital integrations.
We consider a best-fit orbit as unstable if at any given time
of the orbital evolution the planetary semi-major axes deviate by more than 
10\% from their initial values, or if eccentricities reach large values leading to crossing orbits.

\section{Results}
\label{Sec3}

\subsection{The single planet systems}

\subsubsection{GJ\,176}

For GJ\,176 we collected 23 precise CARMENES RVs between January 2016 and January 2017. 
Together with the 111 literature HIRES data and the 71 HARPS RV data points, a total data set of
 205  precise RV measurements is obtained that leads to
a Keplerian signal with the following orbital parameters: 
a planetary period $P_{\rm b}$ = 8.776\,d, orbital eccentricity $e_{\rm b}$ = 0.148,
and semi-amplitude $K_{\rm b}$ = 4.49 m\,s$^{-1}$, 
corresponding to a planet with a minimum mass of $m_{\rm b} \sin i$ $\sim$9.1 $M_{\oplus}$ 
and semi-major axis of $a_{\rm b}$ = 0.066\,au. 
The updated best-fit orbital parameters for GJ\,176\,b and their
bootstrap uncertainties and statistics can be found in Table~\ref{table:3}.

Figure~\ref{fig2}, panel a) shows  the GLS power spectrum of the combined data, which 
reveals the significant planetary signal at 8.77\,d.
In panel b) of Fig.~\ref{fig2} we show the combined data together with the Keplerian model 
phase-folded with the planetary period, while panel c) shows a GLS periodogram of the one-planet model residuals
revealing a significant signal at 39.34\,d (seen also in panel a)\,). 
\citet[][]{Forveille2009} attributed the $\sim$40\,d RV signal to stellar activity,
since it agrees well with the $\sim$40\,d periodicity found in their HARPS H+K and H$\alpha$ activity indices,
and the stellar rotation period of GJ\,176 \citep[see Table~\ref{table:1} and][]{Kiraga2007,Mascareno2017b}.
The activity nature of the 40\,d period was further confirmed by \citet{Robertson2015}, 
who noted that the HIRES and the HARPS data can reveal independently the 8.77\,d planetary signal, 
but the $\sim$ 40\,d RV signal is supported only by the HARPS RVs. 
We confirm these findings. By examining our one-planet fit residuals for each data set, 
we find that the $\sim$40\,d period is seen only in the HARPS RVs, 
but not in the HIRES nor in the CARMENES data.
\citet{Robertson2015} suggested that this peculiar absence of the 40\,d period in the HIRES data
is likely a result of the higher resolving power of HARPS (and as we think, also due to the bluer spectral region),
which makes it more sensitive to line profile variations induced by rotational modulation of stellar spots.
Furthermore, complementary to \citet[][]{Forveille2009}, our CARMENES H$\alpha$-index 
measurements also suggest a strong peak at a period of 39.70\,d (see Fig.~\ref{fig0}) 
leading to the conclusion that the most likely reason for the 39.34\,d signal in 
the HARPS residuals is indeed stellar activity.

From Fig.~\ref{fig2}, our CARMENES measurements follow well the best fit model GJ\,176\,b.  
Indeed, the CARMENES velocity scatter around the best fit is the lowest among the data sets 
included to construct the fit with a scatter of $rms_{\rm CARMENES}$ = 2.95 m\,s$^{-1}$, followed 
by HARPS with $rms_{\rm HARPS}$ = 4.16 m\,s$^{-1}$  and HIRES with $rms_{\rm HIRES}$ = 4.81 m\,s$^{-1}$.
The overall weighted $rms$ scatter around the best fit is $rms$ = 4.33 m\,s$^{-1}$, 
which is slightly smaller than the planetary signal. 
As discussed above, a possible reason for this somewhat large $rms$ seen in GJ\,176 
is the additional $\sim$40-d periodic stellar activity seen in the HARPS RVs, 
which we consider as part of the $rms$ scatter.

Although our CARMENES dataset is too small for an independent detection of GJ\,176\,b, 
the strongest GLS peak of the CARMENES data exceeds the 10\% FAP level at the expected planetary period.
A GLS test as a function of the number of data points 
shows that sequentially adding CARMENES data monotonously decreases the FAP of the planetary signal.
This is an indication that CARMENES RVs contain the planetary signal.
Additionally, a flat model with variable RV zero offset applied to the CARMENES data has 
$rms$ = 3.80 m\,s$^{-1}$, while a fit to the combined one-planet Keplerian fit 
leads to $rms$ = 2.95 m\,s$^{-1}$, showing an improvement 
(although insignificant according to an 
F-test\footnote{  We adopted an $F$-test approach for nested models \citep[see][]{Bevington2003},
where the F-ratio is defined as: $F = ({\Delta\chi^2 / \zeta}) / {\chi_{\nu2}^2}$,
where $\Delta\chi^2$ = $\chi_1^2 - \chi_2^2$ is the difference between 
two nested models with $p_1 < p_2$ fitting parameters, 
$\zeta$ = $p_2-p_1$ is the number of additional parameters being tested, 
and $\chi_{\nu2}^2$ is the reduced $\chi_2^2$ of the model with more parameters.}) 
when assuming a planet in an 8.77\,d orbit. 
We conclude that the CARMENES data acquired so far support the existence of the 8.77 day planet.

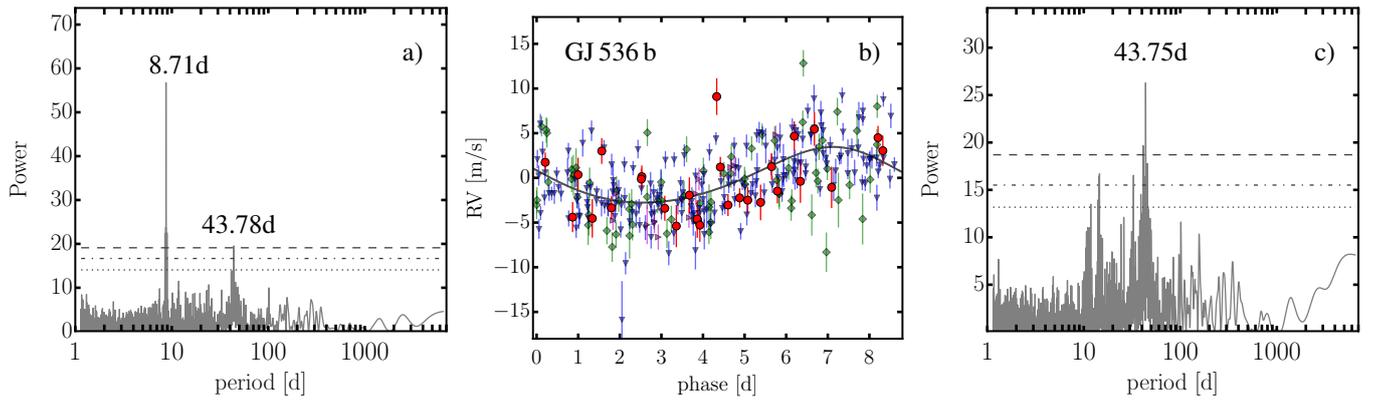
\begin{figure*}[tbp]
\begin{center}$
\begin{array}{cc} 
\includegraphics[width=6cm]{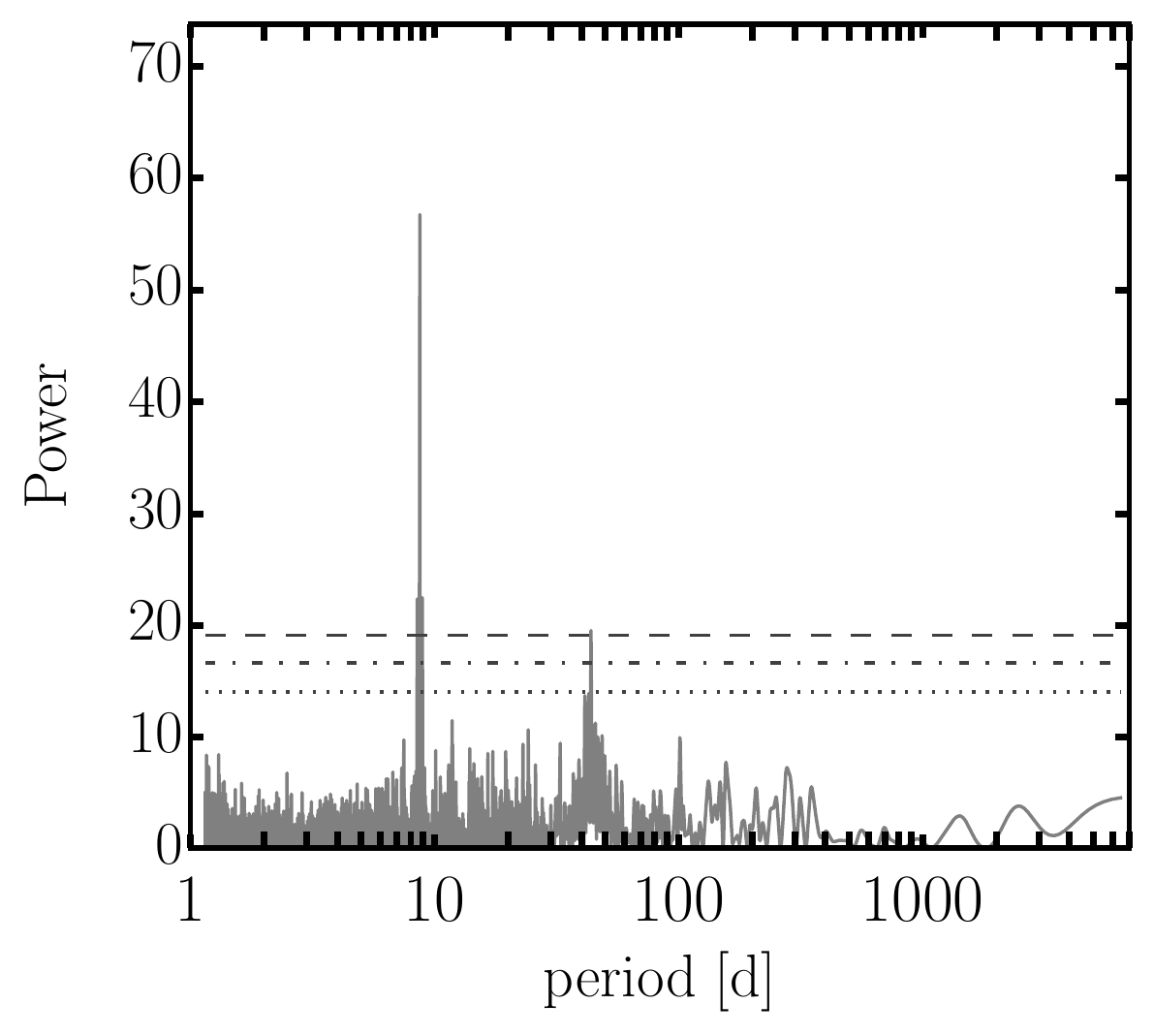} 
\put(-115,125){8.71d} \put(-95,65){43.78d} \put(-20,130){a)} 
\includegraphics[width=6cm]{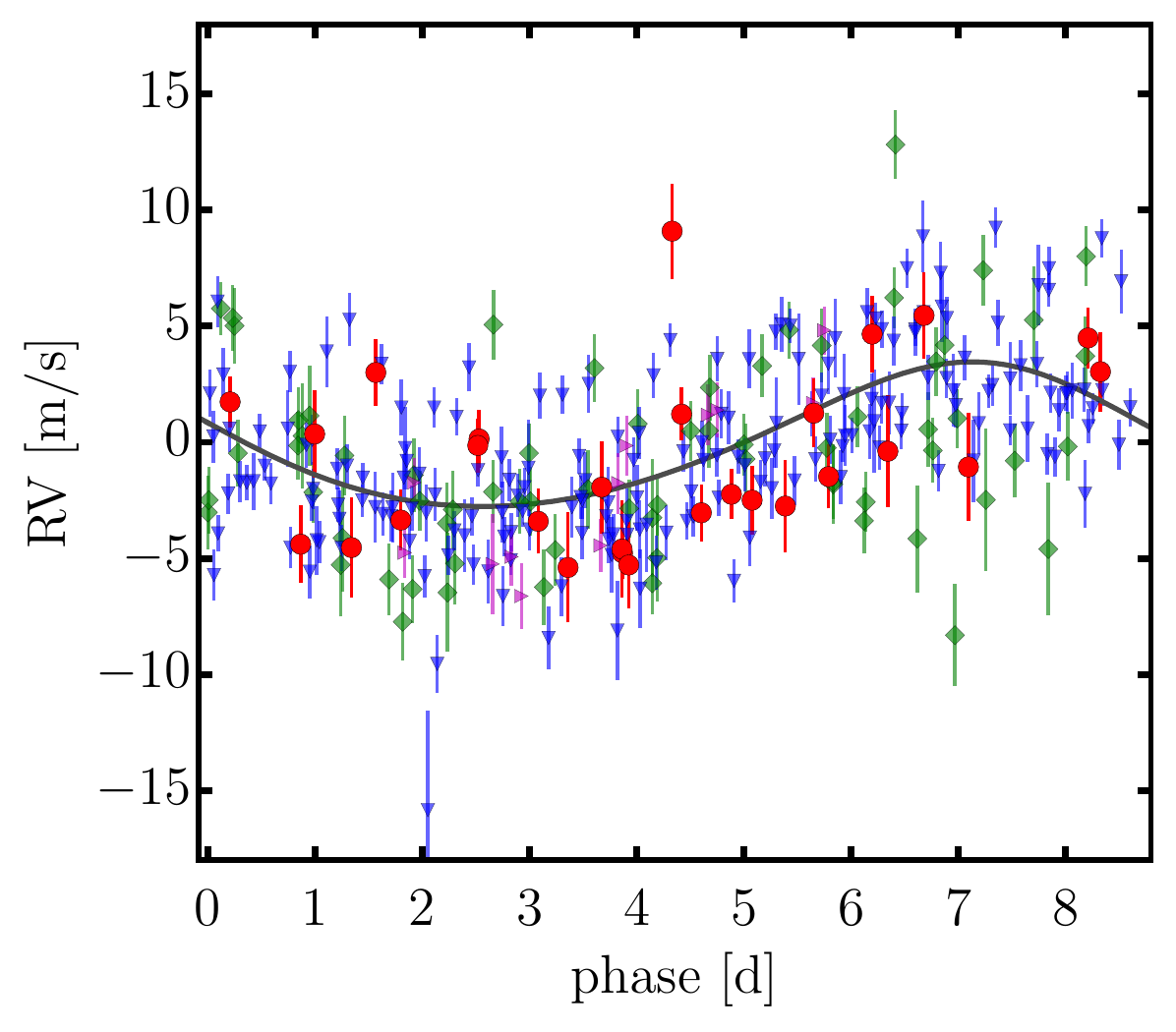}
\put(-130,130){GJ\,536\,b}  \put(-20,130){b)} 
\includegraphics[width=6cm]{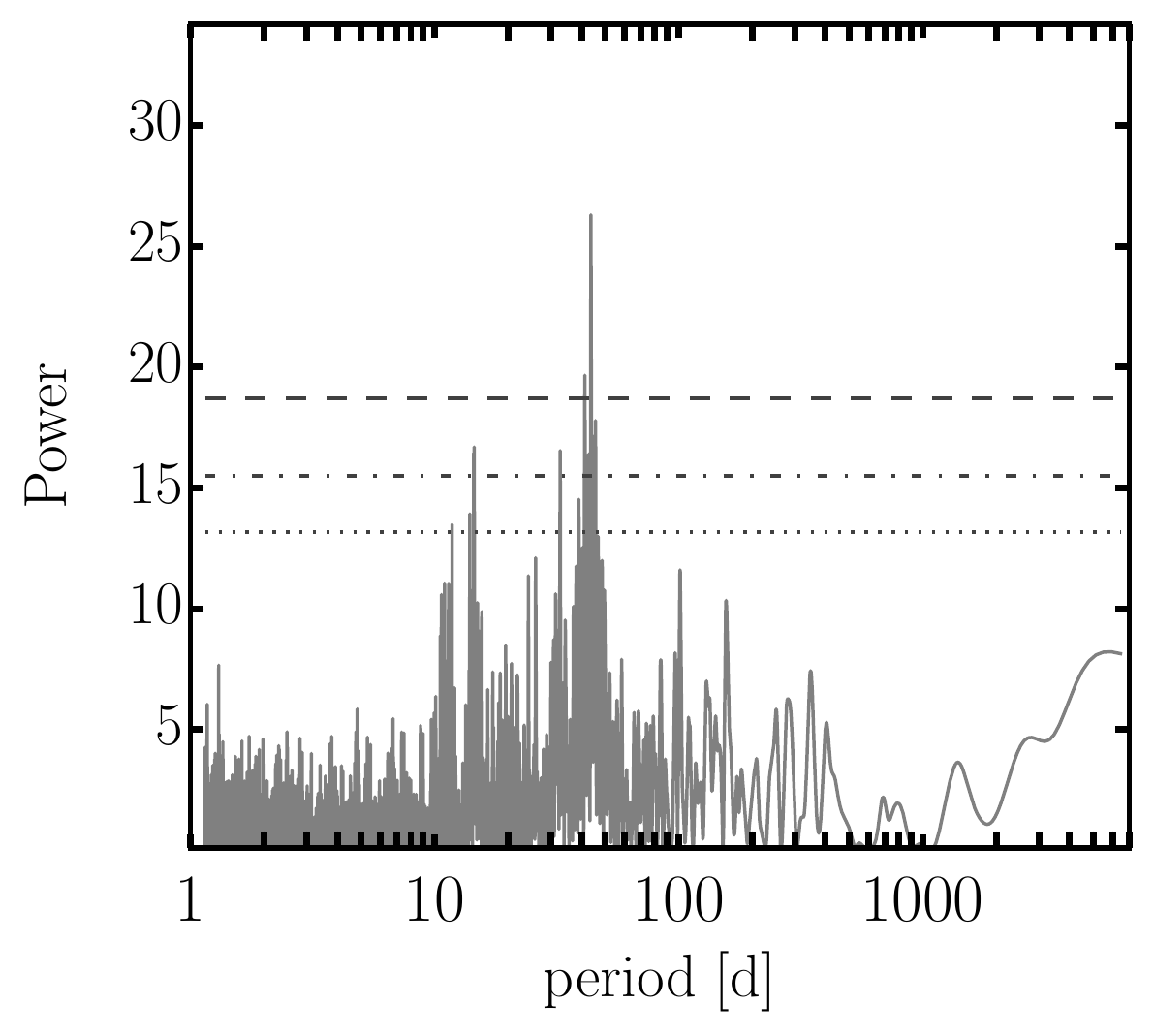} 
\put(-95,130){43.75d}  \put(-20,130){c)} 
\end{array} $
\end{center}

\caption{Same as Fig.~\ref{fig2} and Fig.~\ref{fig3}, but for GJ\,536. 
Panel a) GLS periodogram of the combined data for GJ\,536. 
The significant periods are at 8.71\,d (induced by GJ\,536\,b) and at 43.78\,d, respectively, the latter likely
due to the stellar rotational period ($P_{\rm rot}$  $\approx$ 43.3\,d).
The CARMENES, HARPS, HIRES and HARPS-N (magenta triangles) RV data for GJ\,536, and 
the phase folded best Keplerian fit are shown in panel b).
Panel c) GLS periodogram of the best fit residuals, revealing only the 43.75\,d activity peak.
}   

\label{fig4} 
\end{figure*}

\subsubsection{GJ\,436}

The 113 CARMENES RVs confirm GJ\,436\,b, showing full consistency with the 
356 HIRES RVs from \citet{Butler2017} and the 169 HARPS  RV measurements.
Our updated Keplerian parameters of GJ\,436\,b, based on the modeling 
of all 638 Doppler measurements has $\chi_{\nu}^2$~=~3.47, overall $rms$ = 3.27\,m\,s$^{-1}$, 
leading to a planet semi-amplitude $K_{\rm b}$ = 17.38\,m\,s$^{-1}$, 
period of $P_{\rm b}$~=~2.644 days, and eccentricity $e_{\rm b}$~=~0.152.
Our orbital period determination for GJ\,436\,b is consistent with the most precise transit 
time series photometry values of $P_{\rm b}$= 2.64388 $\pm$ 0.00006 days
performed with the {\em Hubble Space Telescope} \citep[][]{Bean2008}.
These parameters and the inclination constraints from the  transit ($i_{\rm b}$ = 85.80$_{-0.21}^{+0.25}$)
yield a planetary dynamical mass of $m_{\rm b} $ = 21.4$_{-0.2}^{+0.2}$ $M_{\oplus}$ and semi-major axis of 
$a_{\rm b}$ = 0.028$_{-0.001}^{+0.001}$ au. 
Detailed orbital parameters from our fit and their asymmetric 
bootstrap uncertainties are listed in Table~\ref{table:3}.

In Fig.~\ref{fig3} panel a), we show the GLS periodogram for the merged RV data,
which reveals a very strong peak at 2.644\,d and its one-day
alias at $\sim$1.6\,d,  while in panel b) in Fig.~\ref{fig3}
we show our best-fit Keplerian model together with the data phase-folded to the 2.644 day period of the planet.
We also inspected the GJ\,436 residuals after removing the Doppler 
contribution from the planet with a GLS periodogram.  
The right panel of Fig.~\ref{fig3} illustrates many peaks
above the 0.1\% FAP level, the most significant of them with a period of 
23.7\,d followed by peaks at 53.2\,d, 186.5\,d, 11.6\,d and others.
 We find that all three data sets on their own contain many 
significant GLS peaks in their residuals, which do not mutually agree. 
For example, all three data sets show a forest of residual periods in the range 42\,d--50\,d,
but with no clear match between the sets.
The 23.7\,d peak is seen in CARMENES and HARPS, but not seen in HIRES, which conversely presents the 11.6\,d peak.
Therefore, we do not associate any of these peaks with the signature of additional companions.
They could be due to  stellar activity, or potentially be related to the window function of the observations and its aliases.

Our best-fit orbital estimates for GJ\,436 are within the uncertainties from the literature,
but due to the large number of data from three independent high-precision instruments, 
they possibly represent  the most accurate  planetary orbit. 
We  also  quantify the CARMENES precision from the scatter around the orbital solution:
for GJ\,436 our data has a weighted $rms_{\rm CARMENES}$ = 2.56\,m\,s$^{-1}$, 
which is smaller than that of the \citet{Butler2017} data with
$rms_{\rm HIRES}$ = 4.37 m\,s$^{-1}$, but slightly higher than the one from HARPS  with 
$rms_{\rm HARPS}$ = 2.28 m\,s$^{-1}$.

It is worth noting that three of our CARMENES RVs were obtained during transit 
(JD~=~2457490.475, 2457511.606 and 2457511.617).
Similar to the HARPS transit time observations presented in \citet{Lanotte2014},
however, we did not detect any excursion potentially related to the Rossiter-McLaughlin effect on GJ\,436
due to the  expected low amplitude of $<$ 1\,m\,s$^{-1}$.

\subsubsection{GJ\,536}

In our initial CARMENES scheduling program, GJ\,536 was assigned moderate priority, 
and thus visited only nine times between January and June 2016, when the star was observable from Calar Alto.
After the planet announcement by \citet{Mascareno2017}, 
we secured 19 more Doppler measurements between January and February 
2017 in an attempt to cover as much of the planetary orbit as possible.
Currently, our 28 CARMENES RVs by themselves do not show any significant GLS peaks,
and only sparsely cover one full orbital phase when compared to the HARPS  and the 
literature velocities \citep{Mascareno2017,Butler2017}, which recover well the planetary signal.
We aim, however, at studying the individual performance of CARMENES for GJ\,536 
and check the agreement with the planet signal.

\begin{table}[tbp]
\caption{Best fit Keplerian parameters for the single planet systems GJ\,176, GJ\,436 and GJ\,536 based 
on the combined CARMENES and literature RVs.}  
\begin{adjustwidth}{-5.0cm}{} 

\centering  

\resizebox{0.75\textheight}{!}{\begin{minipage}{\textwidth}

\label{table:3}  

\centering      
\begin{tabular}{l  r r r r r r r r}     
\hline 
\noalign{\vskip 0.5mm}
\hline
\noalign{\vskip 0.5mm}

Orb. param.~~~~~~~~~~& GJ\,176\,b   &  GJ\,436\,b &  GJ\,536\,b \\     
\hline 
\noalign{\vskip 0.9mm}
%
$K$  [m\,s$^{-1}$]                & 4.49$_{-0.23}^{+1.00}$      &  17.38$_{-0.17}^{+0.17}$      &  3.12$_{-0.19}^{+0.36}$      \\  \noalign{\vskip 0.9mm}
$P$ [d]                           & 8.776$_{-0.002}^{+0.001}$   &  2.644$_{-0.001}^{+0.001}$    &  8.708$_{-0.001}^{+0.002}$    \\  \noalign{\vskip 0.9mm}
$e$                               & 0.148$_{-0.036}^{+0.249}$   &  0.152$_{-0.008}^{+0.009}$    &  0.119$_{-0.032}^{+0.125}$     \\  \noalign{\vskip 0.9mm}
$\varpi$ [deg]                    & 150.6$_{-104.5}^{+42.2}$    & 325.8$_{-5.7}^{+5.4}$         & 19.2$_{-42.8}^{+36.9}$        \\  \noalign{\vskip 0.9mm}
$M$ [deg]                         & 352.9$_{-36.6}^{+95.2}$     &  78.3$_{-5.4}^{+5.5}$         &  50.3$_{-43.4}^{+46.8}$      \\  \noalign{\vskip 0.9mm}

$a$ [au]                          & 0.066$_{-0.001}^{+0.001}$   &  0.028$_{-0.001}^{+0.001}$    &  0.067$_{-0.001}^{+0.001}$      \\  \noalign{\vskip 0.9mm}
$m_{\rm p} \sin i$  [$M_{\oplus}$]& 9.06$_{-0.70}^{+1.54}$      &  21.36$_{-0.21}^{+0.20}$      &  6.52$_{-0.40}^{+0.69}$   \\  \noalign{\vskip 3.9mm} 

 $\gamma_{\rm HIRES}$~[m\,s$^{-1}$]    & 0.03$_{-0.46}^{+0.50}$     &  0.57$_{-0.23}^{+0.23}$     &  0.72$_{-0.45}^{+0.46}$  \\ \noalign{\vskip 0.9mm}
 $\gamma_{\rm HARPS}$~[m\,s$^{-1}$]    & $-$2.44$_{-0.62}^{+0.52}$  &   13.02$_{-0.20}^{+0.21}$   & -1.42$_{-0.20}^{+0.21}$ \\ \noalign{\vskip 0.9mm}
 $\gamma_{\rm HARPS-N}$~[m\,s$^{-1}$]  & \dots                      &  \dots                      & 0.19$_{-0.71}^{+0.67}$  \\  \noalign{\vskip 0.9mm}
 $\gamma_{\rm CARM.}$~[m\,s$^{-1}$]    & $-$5.68$_{-0.84}^{+0.66}$  &  -21.09$_{-0.21}^{+0.21}$   & 9.92$_{-0.57}^{+0.58}$\\  \noalign{\vskip 3.9mm}
              \\
$rms$~[m\,s$^{-1}]$               & 4.33               &  3.27                 &  2.91                \\ \noalign{\vskip 0.9mm}

$rms_{\rm HIRES}$~[m\,s$^{-1}]$               & 4.81              &  4.37                 &  3.66                \\ \noalign{\vskip 0.9mm}
$rms_{\rm HARPS}$~[m\,s$^{-1}]$               & 4.16               &  2.28                 &  2.72                \\ \noalign{\vskip 0.9mm}
$rms_{\rm HARPS-N}$~[m\,s$^{-1}]$               & \dots               &  \dots                 &  2.17                \\ \noalign{\vskip 0.9mm}
$rms_{\rm CARM.}$~[m\,s$^{-1}]$               & 2.95              &  2.56                 &  3.08                \\ \noalign{\vskip 0.9mm}

$\chi_{\nu}^2$                    & 15.29              &  3.47                 &  6.68                 \\ \noalign{\vskip 0.9mm}
Valid for                                   &                  &     &     \\  \noalign{\vskip 0.9mm}

 $T_0$ [JD-2450000]                                  &  839.760                 &  1552.077   &   1410.730 \\  \noalign{\vskip 0.9mm}
  
\noalign{\vskip 0.5mm}
\hline

\end{tabular} 
\end{minipage}}
 
\end{adjustwidth}

\tablefoot{The HARPS-N data for GJ\,536 are taken from \citet{Mascareno2017}, 
but with subtracted absolute RV of 25\,620 m\,s$^{-1}$
to roughly match the RV offsets of HIRES, HARPS and CARMENES.
}

\end{table}

The  periodogram power spectrum of the combined HARPS, HARPS-N, HIRES and CARMENES data
for GJ\,536 in panel a) of Fig.~\ref{fig4}. 
A significant peak at $P$ = 8.71 days, is presumably induced by the planet, and another one at 43.78\,d, likely by activity,
since it is near the stellar rotational period $P_{\rm rot}$ $\approx$ 43.3\,d \citep{Mascareno2017}.
Similar to the case of GJ\,176, the $\sim$44\,d peak is only seen by 
HARPS, which seems to be more sensitive to activity induced RV signals than HIRES and CARMENES.
Our updated Keplerian orbital parameters for GJ\,536\,b  and statistics are listed in Table~\ref{table:3}, while 
 panel b) of Fig.~\ref{fig4} illustrates the phase folded best-fit Keplerian model and data.
Panel c) of Fig.~\ref{fig4} shows that the best-fit residuals yield a significant activity peak at 43.75\,d.
For GJ\,536\,b  we  determine an orbital period of $P_{\rm b}$ = 8.708 days, an eccentricity 
of $e_{\rm b}$ = 0.119, and a semi-amplitude of $K_{\rm b}$ = 3.12 m\,s$^{-1}$ 
implying a super-Earth planet with a minimum mass of 
$m_{\rm b} \sin i \approx$  6.5 $M_{\oplus}$  and a semi-major axis of $a_{\rm b}$ = 0.067 au. 
This fit has  an overall scatter $rms$ = 2.91 m\,s$^{-1}$, which is of the same order as the planetary signal.
Our RV data have a scatter around the best fit of $rms_{\rm CARMENES}$ = 3.08 m\,s$^{-1}$, which is lower than the 
one from HIRES with $rms_{\rm HIRES}$ = 3.66 m\,s$^{-1}$, but higher 
than HARPS and HARPS-N with $rms_{\rm HARPS}$ = 2.72 m\,s$^{-1}$ and $rms_{\rm HARPS-N}$ = 2.17 m\,s$^{-1}$.
The larger $rms_{\rm HIRES}$ may be the reason why our estimated value of the minimum mass of the planet GJ\,536 b is a bit 
larger than that in \citet{Mascareno2017}.

We fit a flat model with variable RV zero offset applied only to the CARMENES data alone 
and we find $rms_{\rm CARMENES}$ = 3.44 m\,s$^{-1}$.
An F-test shows that the improvement achieved by the one-planet model
for our 28 RVs ($rms_{\rm CARMENES}$ = 3.08 m\,s$^{-1}$) is still insignificant.
However, the 8.71-d periodogram peak increases its power and significance
when we combine the HIRES and the CARMENES data, meaning that all data sets seem to contain the same signal.
Similar to the case of GJ\,176, even though we cannot independently confirm the planet around GJ\,536, 
the CARMENES data support the presence of a planetary companion and follow the overall planet signature.

\begin{figure*}[tbp]
\begin{center}$
\begin{array}{cc} 
\includegraphics[width=6cm,height=5.2cm]{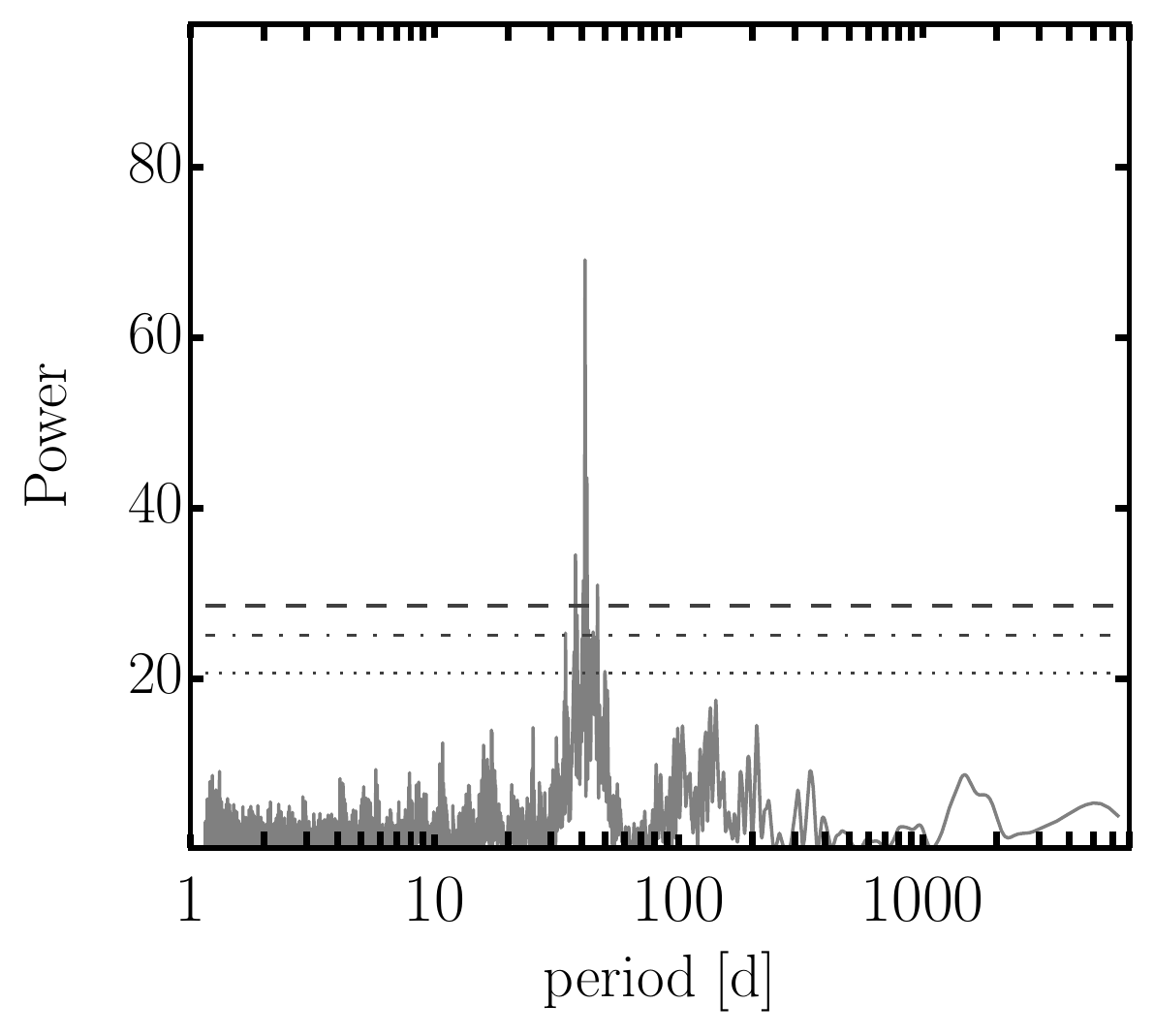} 
\put(-130,120){a)} \put(-95,115){41.4d} \put(-110,130){HIRES+CARMENES}
\includegraphics[width=6cm,height=5.2cm]{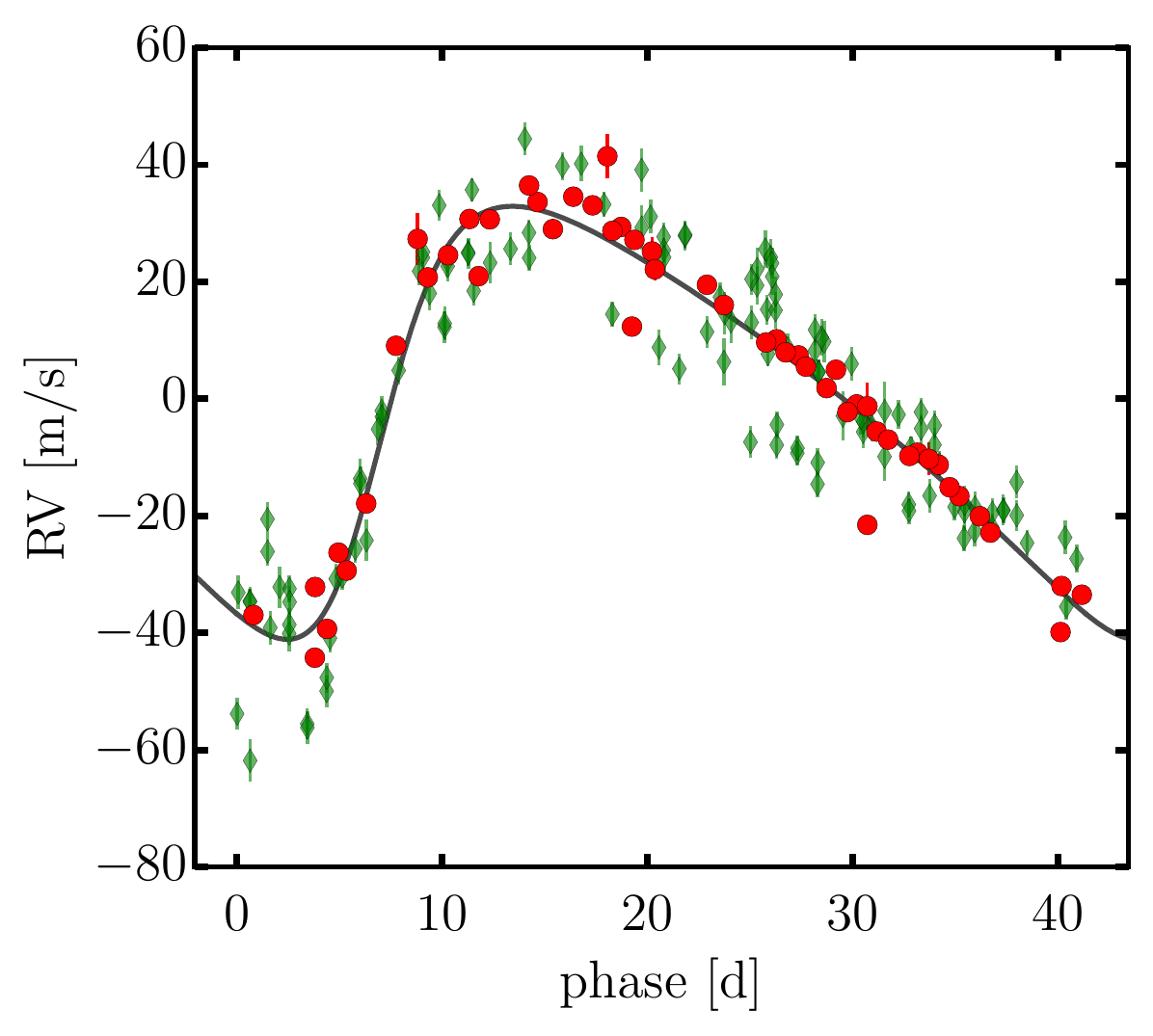} 
\put(-130,110){b)} \put(-55,130){GJ\,1148 b} \\

\includegraphics[width=6cm,height=5.2cm]{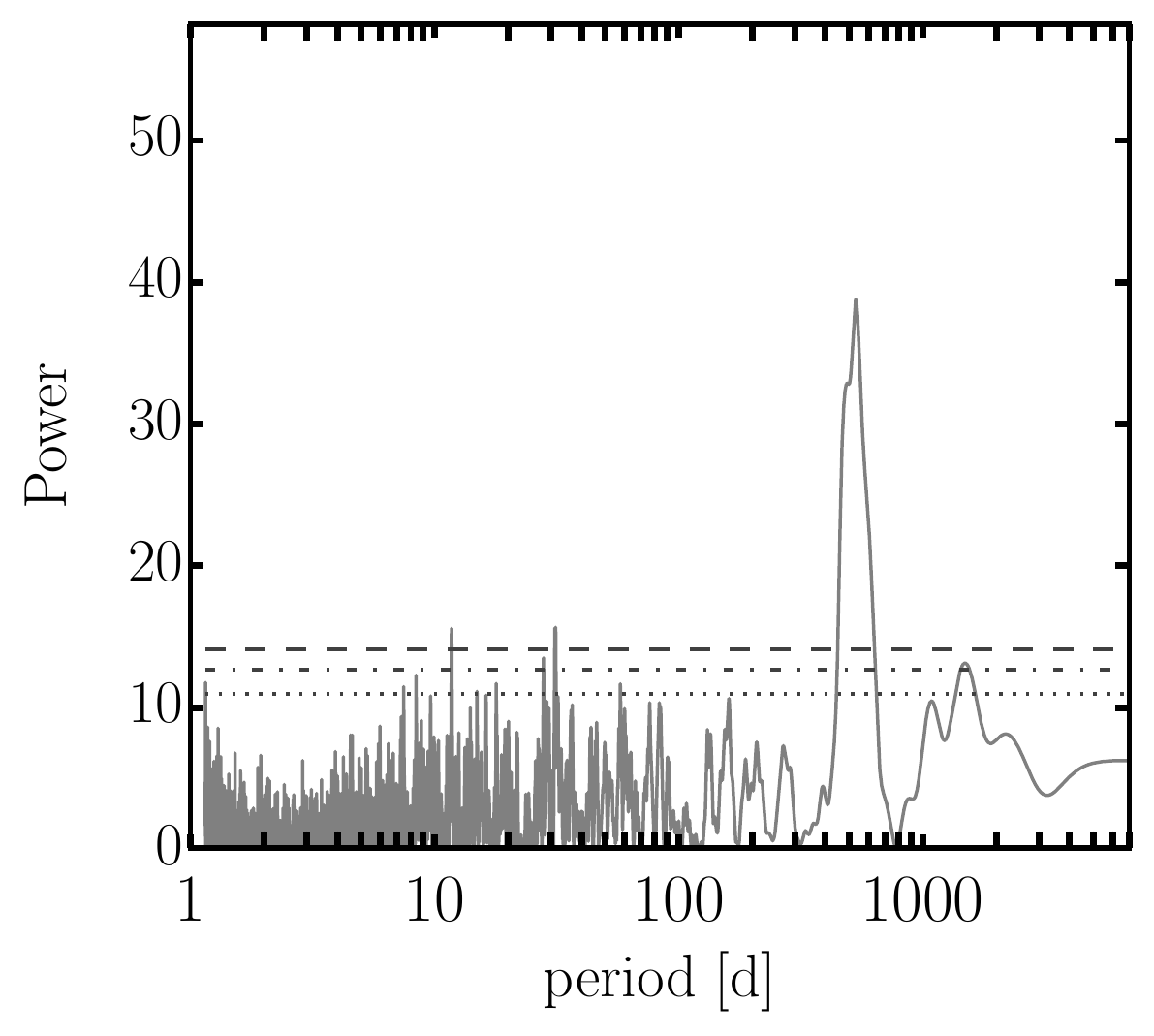}
\put(-130,120){c)} \put(-55,110){531.5d}   \put(-110,130){1pl o-c HIRES} 

\includegraphics[width=6cm,height=5.2cm]{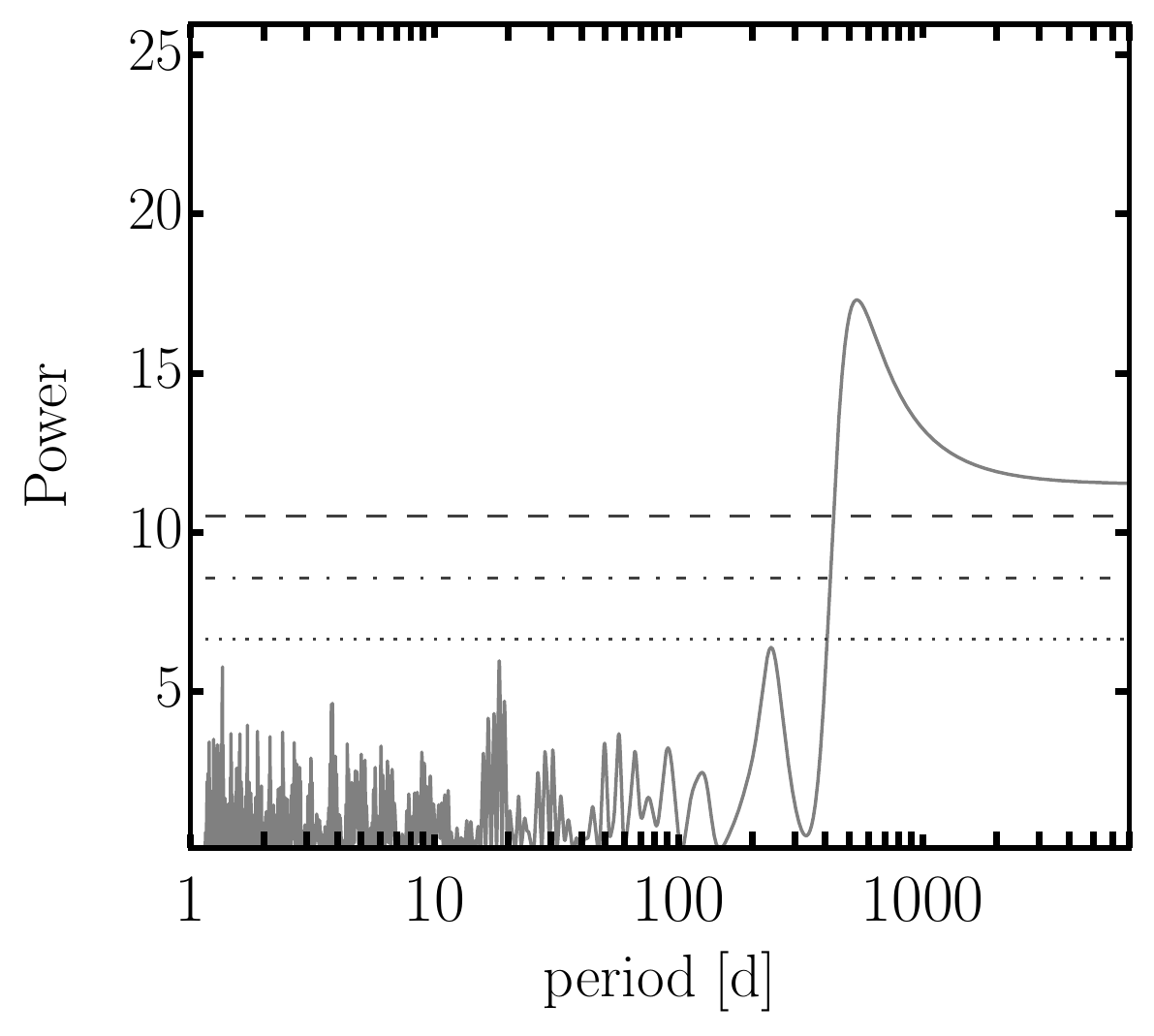}
\put(-130,120){d)} \put(-55,110){538.9d}   \put(-110,130){1pl o-c CARMENES} 

\includegraphics[width=6cm,height=5.2cm]{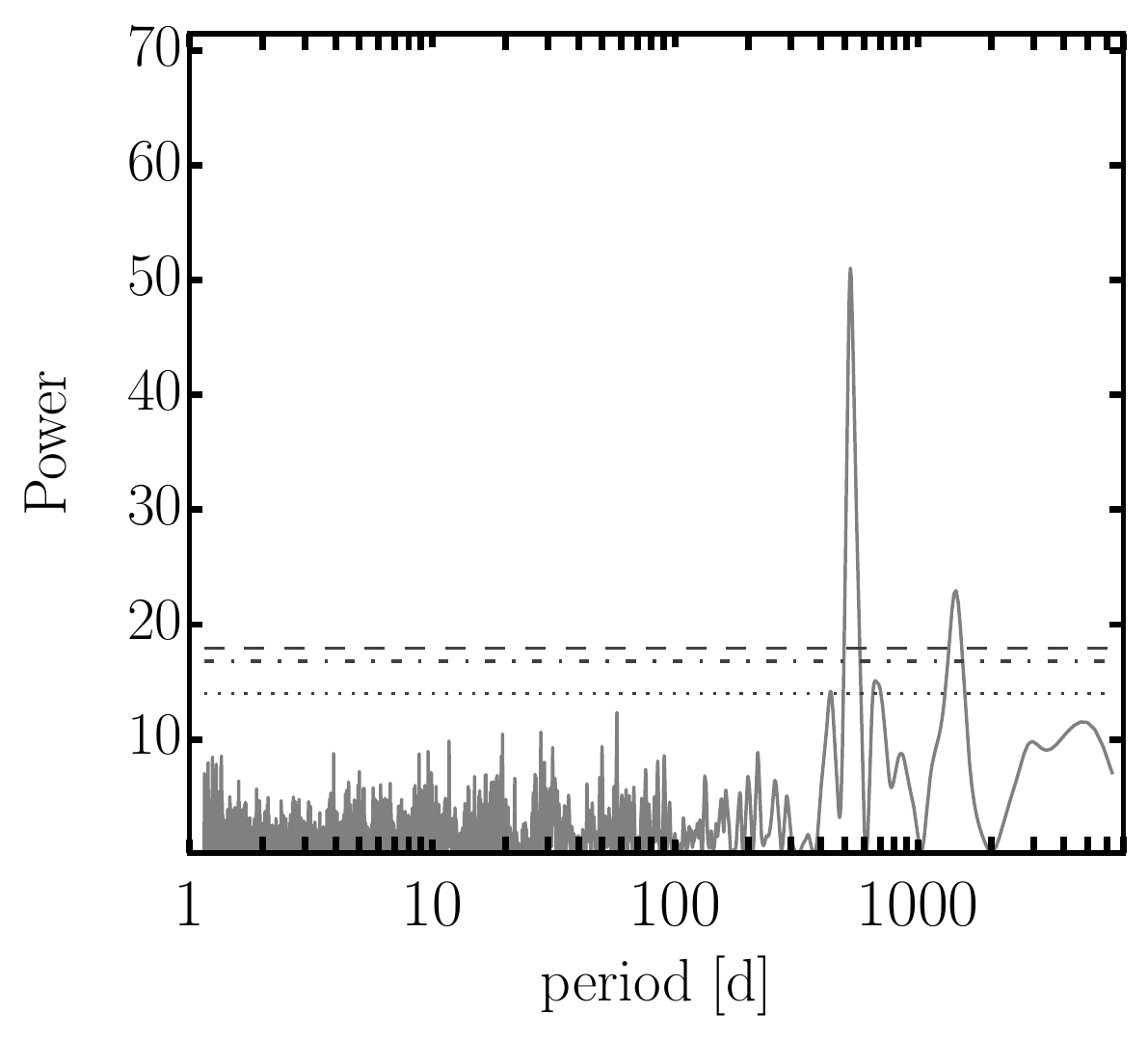}
\put(-130,120){e)} \put(-55,115){525.9d}  \put(-40,80){1434.3d}   \put(-130,130){1pl o-c HIRES+CARMENES}     \\

\includegraphics[width=6cm,height=5.2cm]{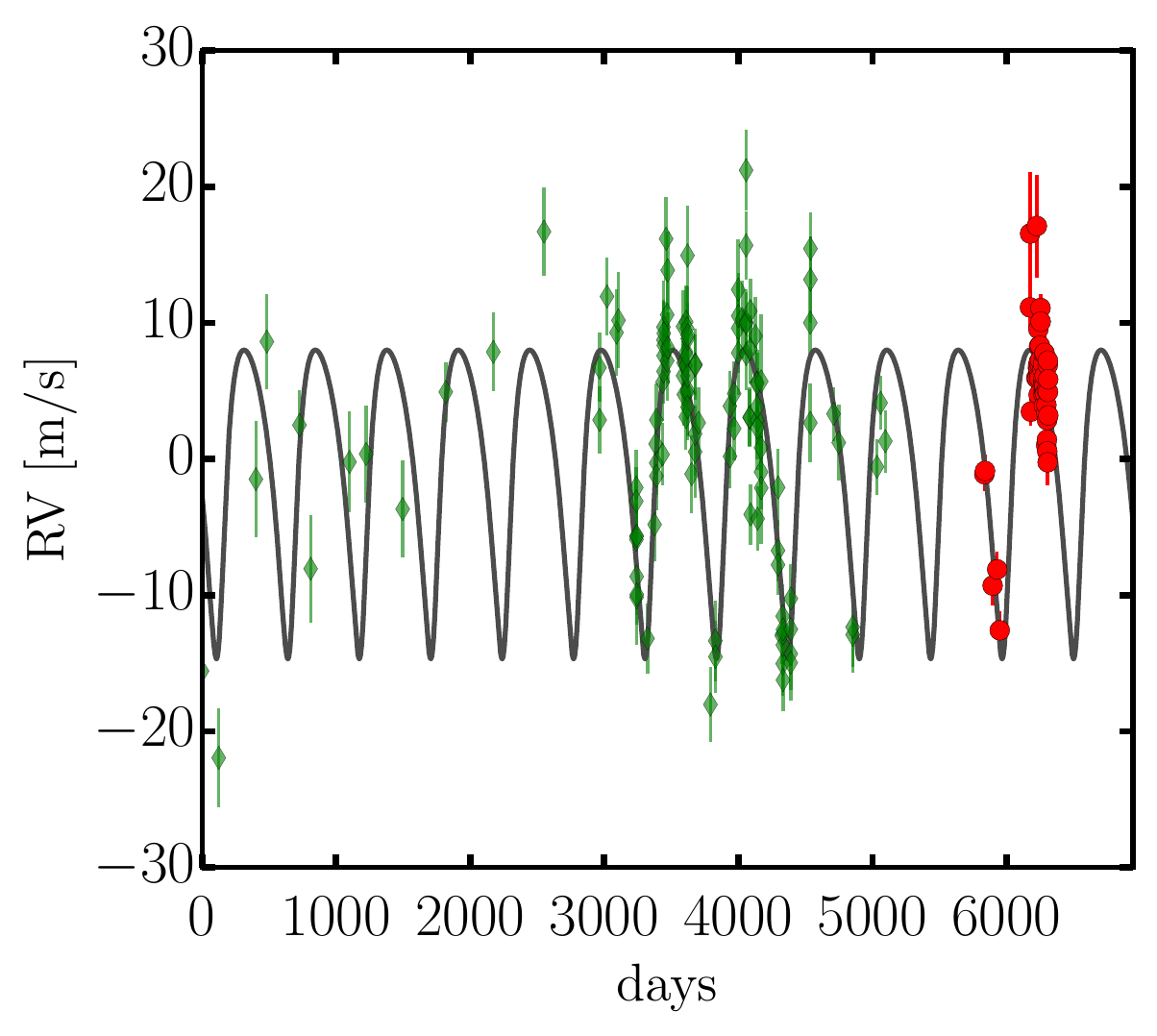} 
\put(-130,110){f)}\put(-55,130){GJ\,1148 c} 

\includegraphics[width=6cm,height=5.2cm]{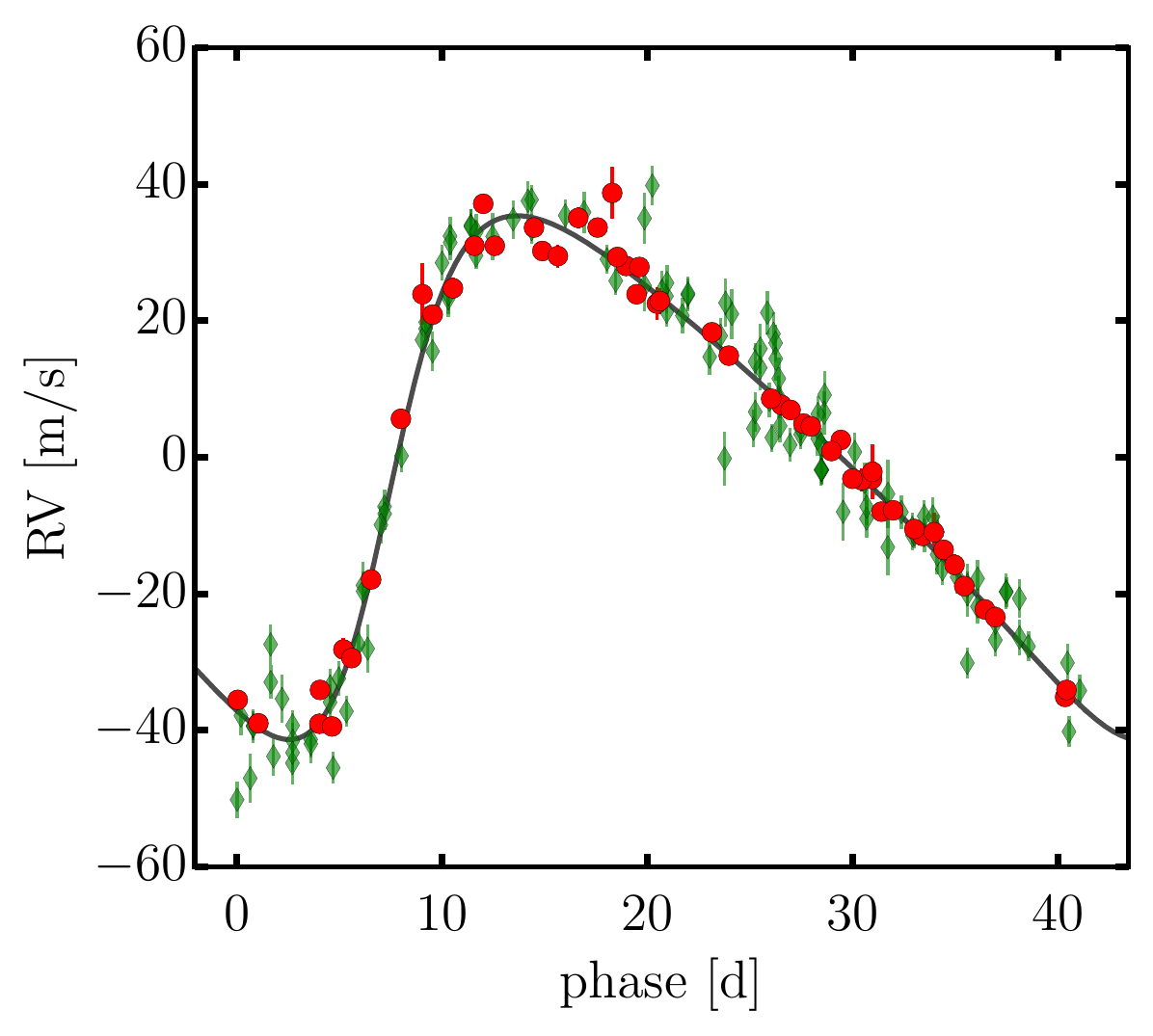} 
\put(-130,110){g)} \put(-55,130){GJ\,1148 b}  

\includegraphics[width=6cm,height=5.2cm]{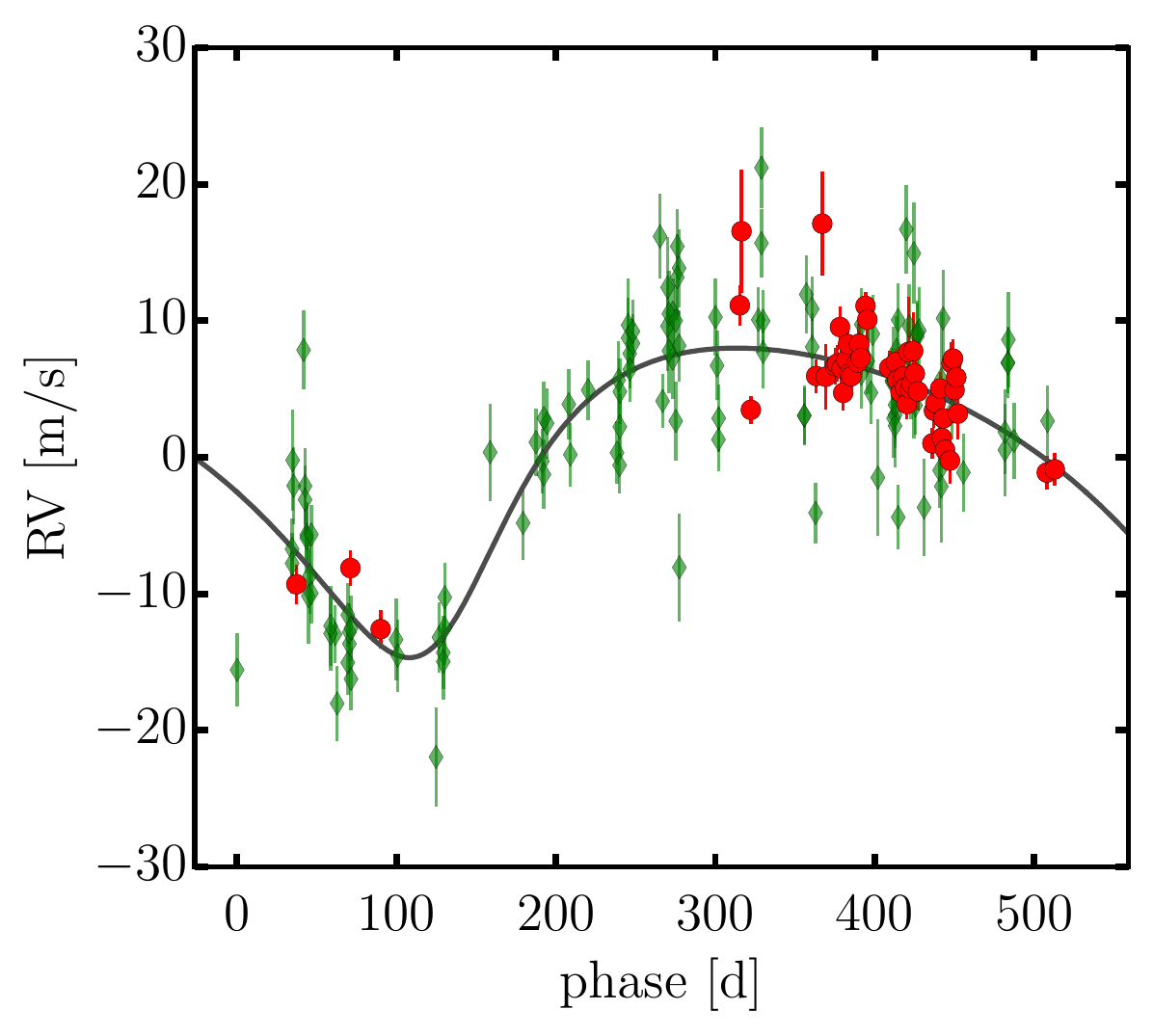}
\put(-130,110){h)} \put(-55,130){GJ\,1148 c}\\
\end{array} $
\end{center}

\caption{GJ\,1148  Doppler data obtained with HIRES (green diamonds) and  
CARMENES (red circles) show two distinct periodic signals consistent
with two eccentric Saturn mass planets with orbital periods of 41\,d and 526\,d.
Panels a) and b) show the GLS signal of the dominant planet  GJ\,1148 b  and its 
single planet Keplerian model phase folded at the best-fit period, respectively.
Panels c), d) and e) show the GLS  analysis of the HIRES, CARMENES and 
the combined data residuals after subtracting the signal from  GJ\,1148 b shown in Panel b). 
Both data sets reveal the existence of a second planet candidate with a period near 530\,d.
Panel f) shows the RV signal of the second planet as determined from the simultaneous two-planet fit,
while panels g) and h) show the individual Doppler signals of GJ\,1148 b  and GJ\,1148~c, 
respectively, phase folded at their best-fit periods.
}   

\label{fig1148_kep} 
\end{figure*}

\begin{figure*}[tp]
\begin{center}$
\begin{array}{cc} 

 \includegraphics[width=9cm,height=7.5cm]{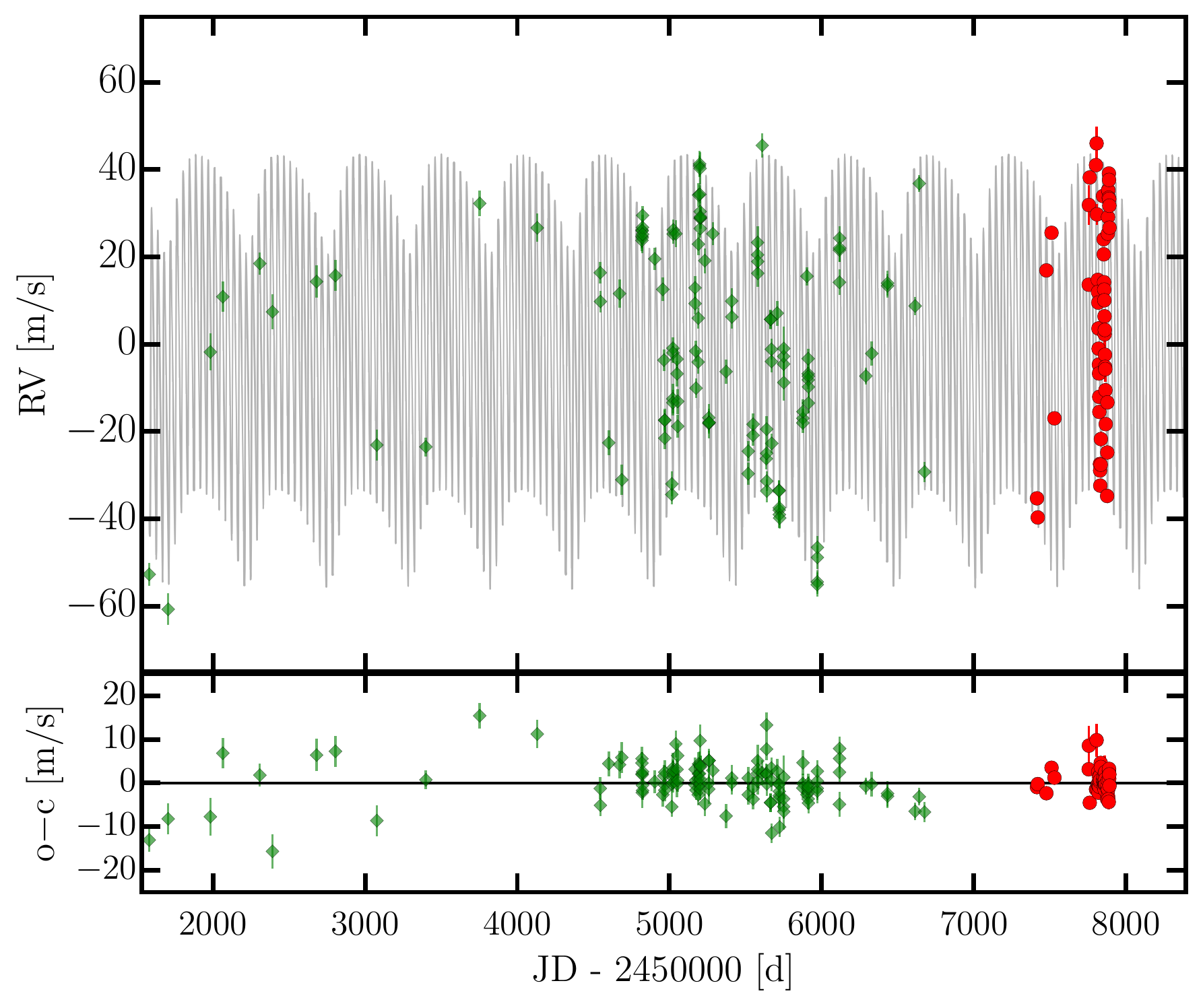} 
 \put(-140,190){GJ\,1148} \put(-210,185){a)}
 \includegraphics[width=9cm,height=7.5cm]{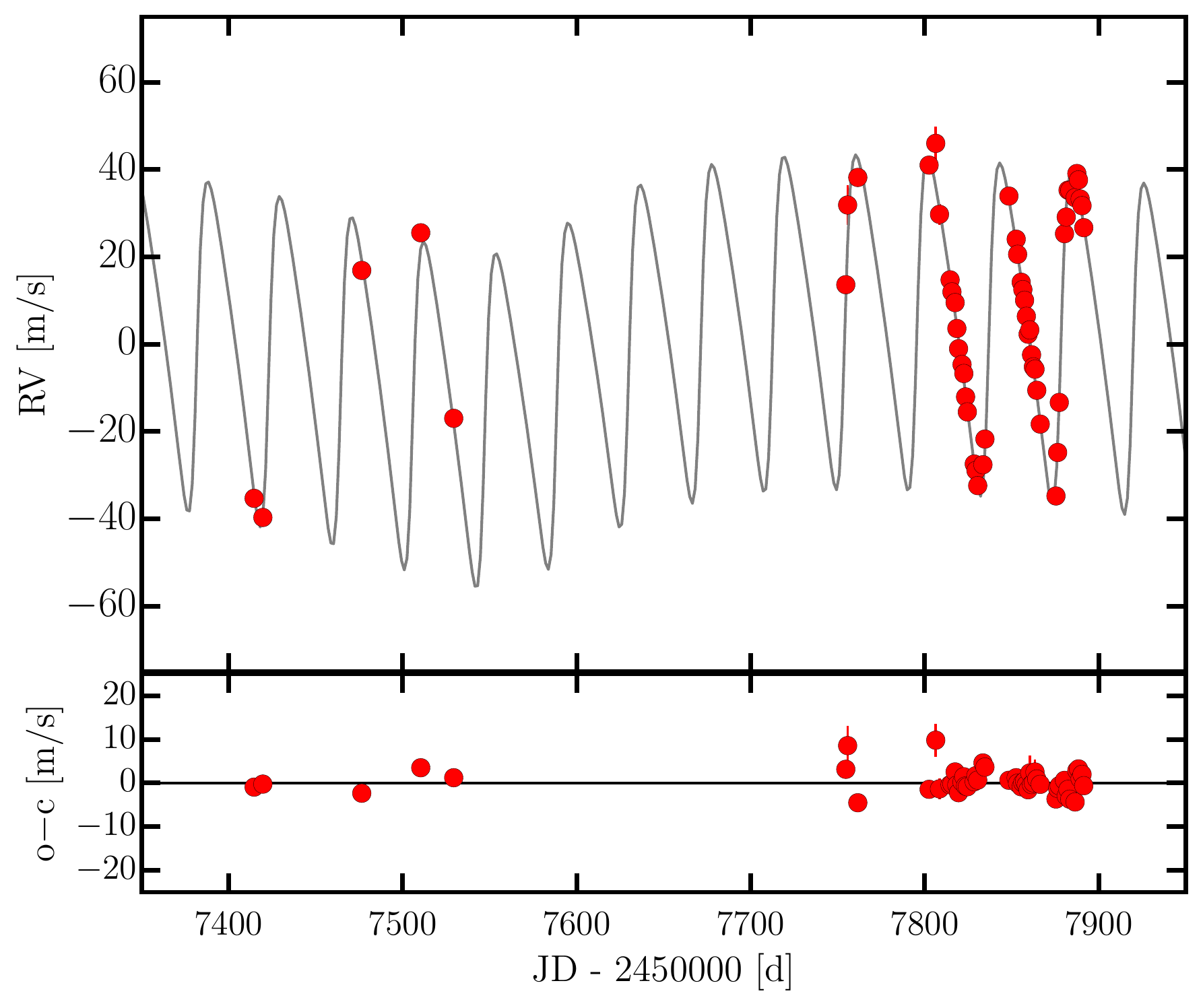} 
 \put(-160,190){CARMENES data}  \put(-210,185){b)} \\
  \includegraphics[width=9cm,height=7.5cm]{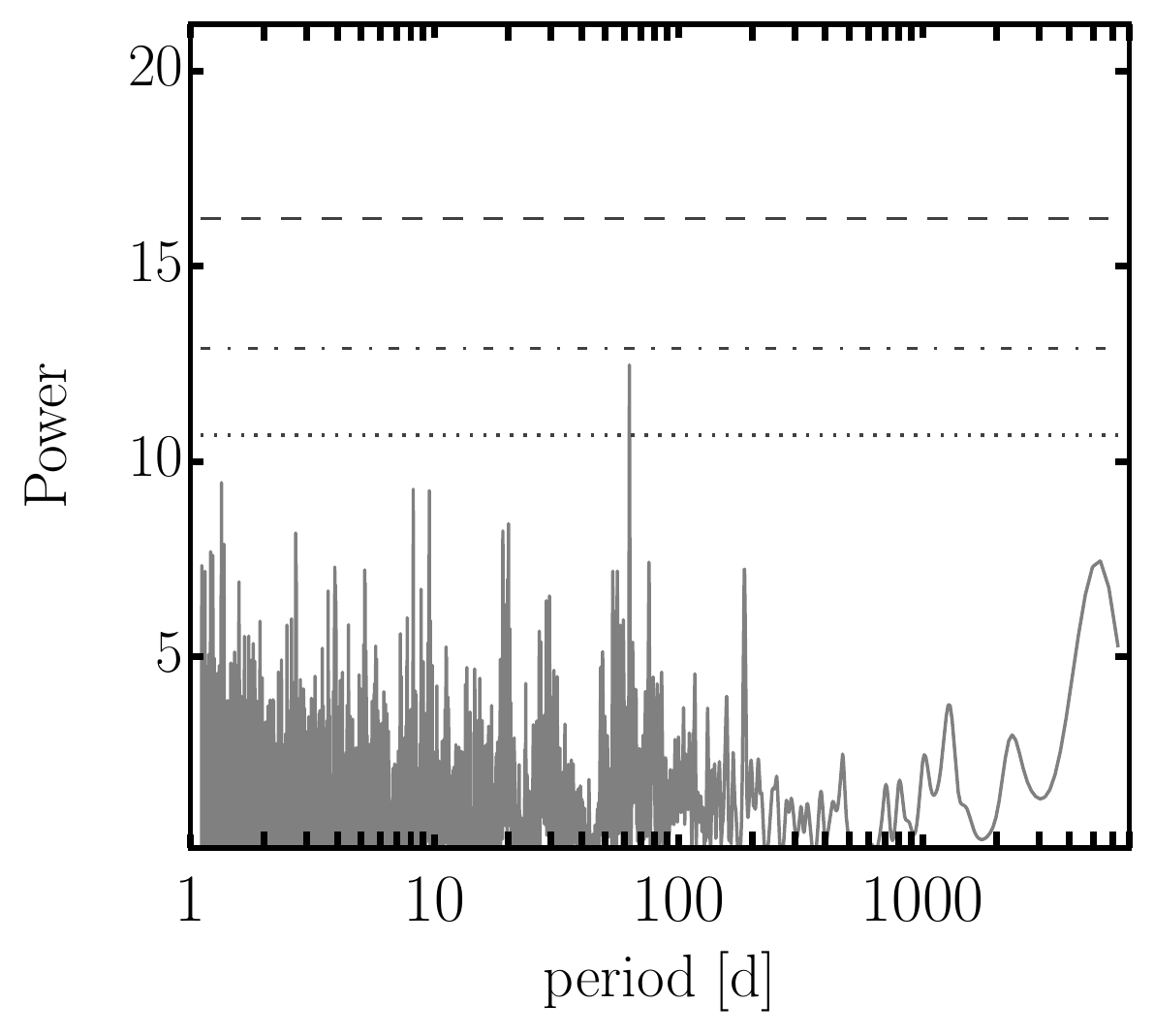}
  \put(-190,190){c)}\put(-125,150){62.9d}
 \includegraphics[width=9cm,height=7.5cm]{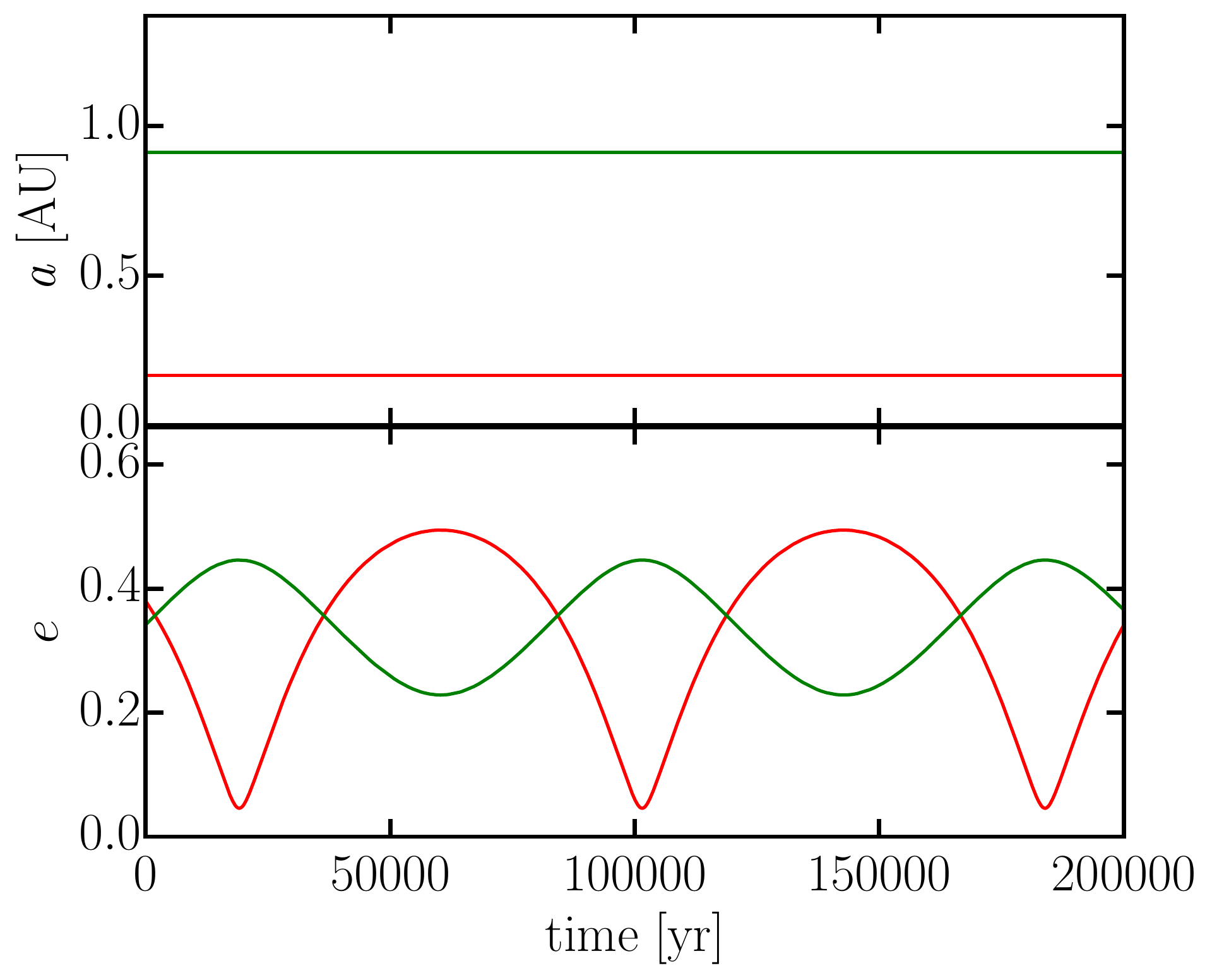} 
 \put(-210,190){d)} \put(-145,135){GJ\,1148 b} \put(-145,185){GJ\,1148 c}    

\end{array} $
\end{center}

\caption{Panel a) Available Doppler measurements for GJ\,1148 obtained with HIRES (green diamonds)
and CARMENES (red circles) fitted with a two-planet N-body model.
Panel b) Zoom of the CARMENES time series together with the same model. 
Panel c) No significant signal is left in the residuals of the two-planet dynamical model.
A peak at 62.9\,d presents an interesting possibility for a third lower-mass planetary companion 
that might be locked in a 3:2 MMR with GJ\,1148\,b, but currently this peak is still below the 1\% FAP level.
Panel d) Results from a stability analysis of the GJ\,1148 system composed of planets b and c. 
This two-planet fit is stable for at least 10 Myr, but for 
illustrative purposes we plot only  a 200\,000 yr extent of the orbital evolution 
of the planetary semi-major axes and eccentricities for the best two-planet dynamical fit. 
In this configuration the two planets exhibit large secular oscillations of $e_{\rm b}$ and
$e_{\rm c}$ with a secular period of $\sim$80\,000 years.
}

\label{dyn_1148} 
\end{figure*}

\subsection{The multiple planet systems}
\label{Sec3.2}

\subsubsection{GJ\,1148}
\label{GJ1148}

In Section~\ref{Sec2} this target was introduced as a known single-planet host 
harboring a $\sim$41-d Saturn-mass planet designated as GJ\,1148~b  \citep{Haghighipour2010} and 
a possible second planetary companion with a period of $\sim$530\,d \citep{Butler2017}.
In this section, we confirm the existence of a second eccentric Saturn-mass planet, hereafter GJ\,1148~c, 
with a period of $P_{\rm c}$ = 532.6\,d, making GJ\,1148 a multiple planet system.
For the first time, we present its full two-planet orbital configuration.
The GJ\,1148~c planet discovery is based on the combined 125 literature HIRES RVs 
presented in \citet{Butler2017} and the additional 52 precise Doppler measurements  that we secured with CARMENES.
Both data sets independently contain the GJ\,1148~b and GJ\,1148~c planetary signals, 
and thus further strengthen the two-planet hypothesis.

We now introduce the RV analysis sequence leading to the detection of GJ\,1148~c.
In Fig.~\ref{fig1148_kep}, panel a) we show the GLS power spectrum for the available Doppler data, which
reveals a strong peak at 41.4\,d, attributed to the presence of GJ\,1148\,b.
A single-planet Keplerian model to the combined HIRES and CARMENES data suggests a planetary period of 
$P_{\rm b}$ = 41.4 days, a moderately large eccentricity of $e_{\rm b}$ = 0.392,
and a semi-amplitude $K_{\rm b}$ = 37.0 m\,s$^{-1}$ 
from which we derive  a minimum mass of $m_{\rm b} \sin i$ = 92.8 $M_{\oplus}$ and 
a semi-major axis of $a_{\rm b}$ = 0.166 au. 
More detailed one planet best-fit parameters and their uncertainties are shown in Table~\ref{table:1148}, 
while the phase-folded single-planet fit is shown in Fig.~\ref{fig1148_kep}, panel b.

Similar to the GJ\,1148  best-fit presented in \citet{Haghighipour2010}
our one-planet fit has a large overall scatter of $rms$ = 7.05 m\,s$^{-1}$, 
leading to a poor $\chi_{\nu}^2$ = 11.05. Based on the 37 HIRES discovery RVs, \citet{Haghighipour2010} 
found that including a linear trend of 2.465 $\pm$ 1.205 m\,s$^{-1}$\,yr$^{-1}$ led to a better fit, reducing the 
$rms$ from 9.23 m\,s$^{-1}$ to 8.06 m\,s$^{-1}$.
However, introducing a linear trend in our combined data set of HIRES and CARMENES 
did not lead to a model improvement, thus we did not fit a linear trend in our analysis.
When we analyze the one-planet best-fit residuals, however, we find 
that both data sets exhibit a significant periodicity around 530\,d, 
which we attribute to the possible second planet GJ\,1148\,c.
The GLS periodograms of the one-planet model residuals for CARMENES, HIRES and the combined data
are shown in Fig.~\ref{fig1148_kep}, panels c), d) and e), respectively.
The CARMENES data residuals reveal a significant GLS peak at 538.9\,d, while for the HIRES data 
this peak is even stronger and better resolved (due to the higher number of measurements  and longer temporal
baseline of the observations) at around 531.5\,d. 
The combined data set residuals reveal two significant peaks at 525.9\,d and 1434.3\,d. 
The broad 1434.3\,d peak is very likely related to the 1196\,d alias of GJ\,1148\,c and the one sideral year.

We investigated the possibility of the 525.9\,d signal being caused by stellar activity.
A rotational modulation of star spots can be excluded, since the observed 525.9\,d RV signal
is much longer than the estimated rotational period for GJ\,1148 of $P_{\rm rot}$ = 73.5\,d 
sugested by \citep{Hartman2011} or the somewhat longer period of $P_{\rm rot}$ = 98.1\,d given in \citet{Haghighipour2010}.
However, long-period magnetic cycles in M dwarfs cannot be easily excluded.
As we showed in Fig.\ref{fig0}, our CARMENES H$\alpha$ index measurements for GJ\,1148 do not exhibit any significant
peaks that could be associated with activity, which supports the GJ\,1148\,c planet hypothesis.
However, even though insignificant, the highest peak in the CARMENES H$\alpha$ index 
power spectrum is consistent with signals beyond 500\,d, and thus deserves a note of caution.
Unfortunately, because of the low significance and low frequency resolution 
(note the short time baseline in Fig.~\ref{dyn_1148} and the 
large observational gap between June 2016 and January 2017)
the available CARMENES H$\alpha$ index time series does not allow us to verify 
whether this activity power is related to the significant $\sim$530\,d RV peak.

\begin{figure}[tp]
\begin{center}$
\begin{array}{cc} 

\includegraphics[width=9cm]{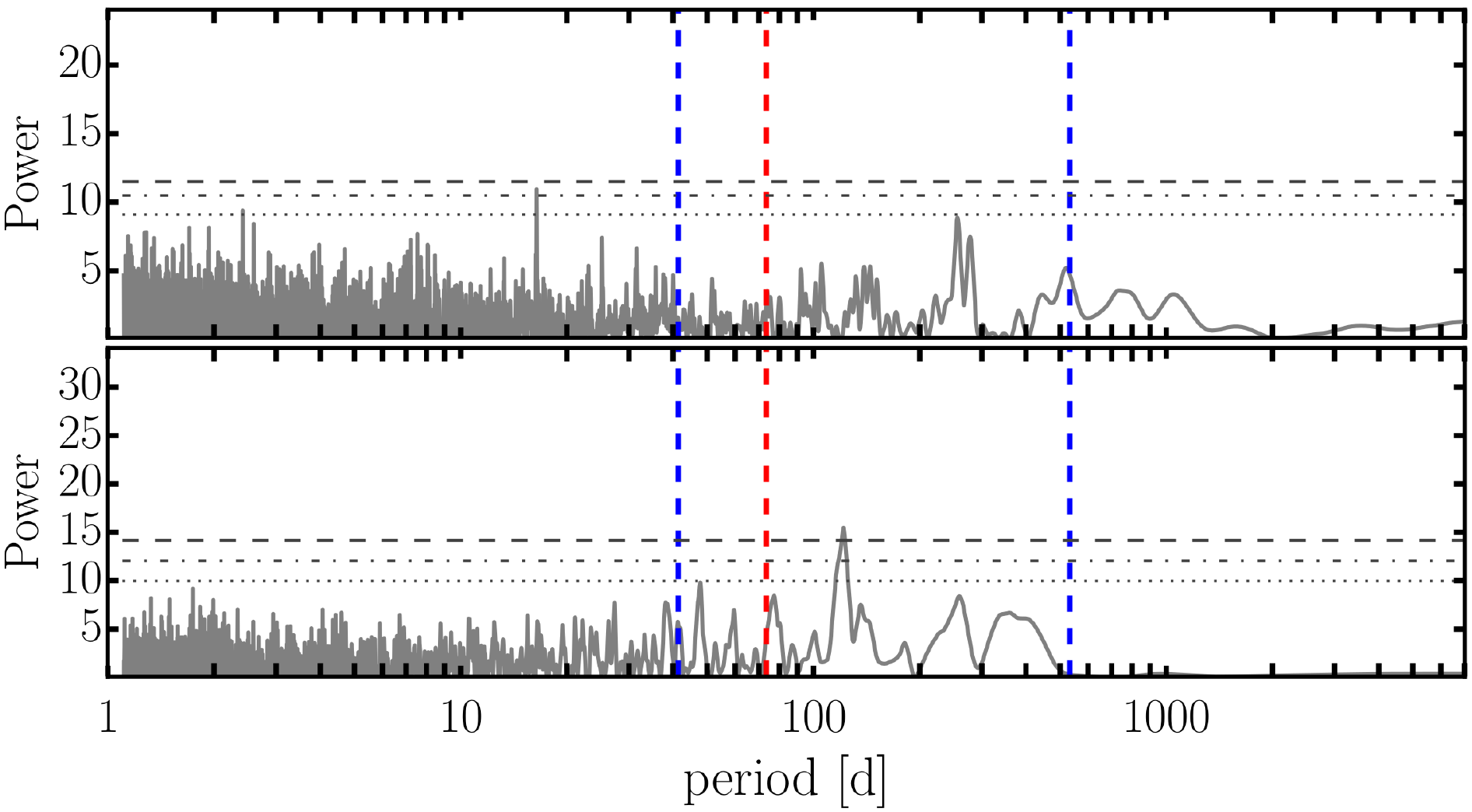} 
\put(-230,130){HIRES~~~S-index} \put(-230,70){HIRES~~~H-index} \put(-120,60){\tiny 121.7d} \\   

\end{array} $
\end{center}

\caption{ GLS periodograms of the S- and H-index measurements from the HIRES data for GJ\,1148.
Only the H-index periodogram reveals a significant peak at 121,7\,d, likely related to activity,
but not associated with the stellar rotation for GJ\,1148 $P_{\rm rot}$ = 73.5\,d (red dashed line), or 
with either of the planetary periods $P_{\rm b}$ = 41.4\,d and $P_{\rm c}$ = 532.6\,d  (blue dashed lines) seen in the HIRES RVs.
}

\label{s_h_index} 
\end{figure}

The HIRES data for GJ\,1148 from \citet{Butler2017} contain S- and H-index activity indicator measurements with a
much longer temporal baseline than our CARMENES data, and are therefore more suitable to search for long-period activity.
The HIRES S-index activity indicator is measured in the Ca~{\sc ii} H\&K
wavelength region, while the H-index measures the H$\alpha$ 
flux variations with respect to the local continuum \citep[for more details see][]{Butler2017}.
Fig.~\ref{s_h_index} shows the GLS periodograms of the HIRES activity indicators.
The S-index data do not show significant peaks, while the H-index measurements
reveal a marginally significant peak at 121.7\,d, which cannot be associated with the planetary signals.
Therefore, we conclude that the CARMENES and the HIRES activity indicators 
so far do not show any evidence of a long-period activity cycle, which could mimic
a planet. Thus, the most plausible interpretation for the observed $\sim$530\,d RV signal
is a second eccentric Saturn-mass planet in orbit around GJ\,1148.

\begin{table}[tbp]

\centering  

\caption{One-planet Keplerian and a coplanar edge-on two-planet dynamical best-fit parameters for GJ\,1148 based 
on the combined CARMENES and HIRES literature RVs.}  

\resizebox{0.70\textheight}{!}{\begin{minipage}{\textwidth}

\label{table:1148}      
  
\begin{tabular}{l r r r r r r }     

\hline \hline \noalign{\vskip 0.7mm}  

\makebox[0.1\textwidth][l]{\hspace{29 mm}One-planet\hspace{21 mm}Two-planet} \\
\makebox[0.1\textwidth][l]{\hspace{28 mm}Keplerian fit\hspace{18 mm}dynamical fit} \\

\cline{2-2}\cline{4-5}\noalign{\vskip 0.9mm} 

Orb. param. & GJ\,1148 b && GJ\,1148 b  &  GJ\,1148~c       \\     
\hline\noalign{\vskip 0.5mm} 
\noalign{\vskip 0.9mm}

$K$  [m\,s$^{-1}$]       & 37.02$_{-0.90}^{+0.92}$    && 38.37$_{-0.49}^{+0.59}$     &  11.34$_{-0.36}^{+0.79}$         \\  \noalign{\vskip 0.9mm}
$P$ [d]                  & 41.382$_{-0.002}^{+0.003}$ && 41.380$_{-0.001}^{+0.002}$  &  532.58$_{-2.52}^{+4.14}$     \\  \noalign{\vskip 0.9mm} 
$e$                      & 0.392$_{-0.022}^{+0.019}$  && 0.380$_{-0.012}^{+0.010}$   &  0.342$_{-0.062}^{+0.050}$       \\  \noalign{\vskip 0.9mm}
$\varpi$ [deg]           & 253.6$_{-3.0}^{+3.1}$      && 258.1$_{-1.8}^{+2.0}$       &  210.4$_{-9.1}^{+12.0}$          \\  \noalign{\vskip 0.9mm}
$M$ [deg]                & 303.9$_{-3.0}^{+3.0}$      && 299.0$_{-2.0}^{+3.1}$       &  272.6$_{-10.7}^{+15.9}$          \\  \noalign{\vskip 0.9mm}

$a$ [au]                 & 0.166$_{-0.001}^{+0.002}$  && 0.166$_{-0.001}^{+0.001}$   & 0.912$_{-0.002}^{+0.005}$       \\  \noalign{\vskip 0.9mm}
$m_{\rm p} \sin i$  [$M_{\oplus}$]   & 92.77$_{-2.00}^{+2.10}$     && 96.70$_{-1.02}^{+1.41}$   &  68.06$_{-2.19}^{+4.91}$       \\  \noalign{\vskip 3.9mm}

$\gamma_{\rm HIRES}$~[m\,s$^{-1}$]   & 2.89$_{-0.82}^{+0.78}$      &&  \makebox[\dimexpr(\width-3em)][l]{1.78$_{-0.44}^{+0.37}$}                 \\ \noalign{\vskip 0.9mm}
$\gamma_{\rm CARM.}$~[m\,s$^{-1}$]   & $-$30.36$_{-0.62}^{+0.58}$  &&  \makebox[\dimexpr(\width-3em)][l]{-34.92$_{-1.42}^{+0.83}$} &                 \\  \noalign{\vskip 3.9mm}

$rms$~[m\,s$^{-1}]$      &  7.05                      && \makebox[\dimexpr(\width-3em)][l]{3.71}                  \\ \noalign{\vskip 0.9mm}
$rms_{\rm HIRES}$~[m\,s$^{-1}]$      &  8.62                      && \makebox[\dimexpr(\width-3em)][l]{4.59}                  \\ \noalign{\vskip 0.9mm}
$rms_{\rm CARM.}$~[m\,s$^{-1}]$      &  4.49                      && \makebox[\dimexpr(\width-3em)][l]{2.23}                  \\ \noalign{\vskip 0.9mm}

$\chi_{\nu}^2$           &  11.05                     && \makebox[\dimexpr(\width-3em)][l]{2.97}                \\ \noalign{\vskip 0.9mm}
Valid for                &                           &&     \\  \noalign{\vskip 0.9mm}

 $T_0$ [JD-2450000]      & \makebox[\dimexpr(\width-6em)][l]{1581.046}                  &&     \\  \noalign{\vskip 0.9mm}

\hline

\end{tabular} 
\end{minipage}}

\end{table}

A simultaneous double Keplerian model fitting two-planets on initially 41.4 and 527\,d-period orbits
converged to a best fit with significantly improved $\chi_{\nu}^2$ = 2.97 and $rms$ = 3.71 m\,s$^{-1}$ 
when compared to the single-planet fit. 
Based on our two-planet best fit we derive updated orbital parameters for GJ\,1148 b:
$K_{\rm b}$ = 38.37 m\,s$^{-1}$,
$P_{\rm b}$ = 41.380  days,
$e_{\rm b}$ = 0.379,  
and for the new planet GJ\,1148~c:  
$K_{\rm c}$ = 11.34 m\,s$^{-1}$,
$P_{\rm c}$ = 532.6 days,
$e_{\rm c}$ = 0.341,  
from which we derive minimum planetary masses of  
$m_b \sin i$  = 0.304 $M_{\rm Jup}$, $m_c \sin i$ = 0.214 $M_{\rm Jup}$ (96.7 and 68.1 $M_\oplus$),
and semi-major axes $a_{\rm b}$ = 0.166 au, $a_{\rm b}$ = 0.913 au, respectively.
The phase-folded Keplerian planetary signals for GJ\,1148 b and~c are shown 
in Fig.~\ref{fig1148_kep}, panels g) and h), respectively. 
No significant GLS peaks are left in the two-planet model residuals, confirming that 
the 1434.3-day peak is indeed related to the lower frequency alias of the GJ\,1148~c planetary signal.

According to an F-test, the double Keplerian best-fit represents a significant improvement 
over the one-planet model with an extremely convincing false-alarm probability of 2.8x$10^{-46}$.
The CARMENES RV scatter for the two-planet model is $rms_{\rm CARMENES}$ = 2.23 m\,s$^{-1}$, 
which is better than the scatter from HIRES data of $rms_{\rm HIRES}$ = 4.60 m\,s$^{-1}$.
From panels f), g) and h) in Fig.~\ref{fig1148_kep} it can be seen 
that the scatter around the two-planet fit is significantly
reduced when compared to the one-planet best-fit solution shown in panel b). 
Both the HIRES and the CARMENES data follow very well the two-planet model 
providing supporting evidence for the multiple planet system architecture of GJ\,1148.

As a next step we adopted the two-planet Keplerian best-fit parameters as an initial guess for 
our more accurate N-body dynamical model.
The two-planet dynamical fit parameters and uncertainties are provided in Table~\ref{table:1148}.
The actual fit to the HIRES and the CARMENES data and 
their residuals are shown in  Fig.~\ref{dyn_1148} panel a), while panel b) 
shows a zoom of the fit and the CARMENES time series.
In Fig.~\ref{dyn_1148} panel c), the best-fit residuals do not show significant peaks above the 0.1\% FAP level.
We note, however, an interesting GLS peak at 62.9\,d near the 1\% FAP level, which could be due 
to an additional $\sim$ 7 $M_\oplus$ mass planet potentially locked in a 3:2 MMR with GJ\,1148\,b.
However, due to the still insignificant power of this peak and the close proximity to the eccentric GJ\,1148\,b
our confidence in the putative third planet is currently low.

We find that the best-fit parameters from the two-planet N-body fit
are practically the same as those from our two-planet Keplerian model.
The small difference between our unperturbed Keplerian and our N-body dynamical model
is a result of the relatively large separation between the GJ\,1148 b  and~c  planets,
leading to negligible dynamical interactions during the observational time baseline.

The long-term dynamical interactions for the GJ\,1148 system, however, are not negligible.
The best dynamical fit is stable for 10\,Myr, showing strong long-term secular
dynamical interactions due to the large planetary eccentricities.
While the planetary semi-major axes are practically constant at 
$a_{\rm b}$ = 0.166 au and $a_{\rm c}$ = 0.912 au, 
the orbital eccentricities exhibit large variations
in the range of $e_{\rm b}$ = 0.05 to 0.49 and 
$e_{\rm c}$ = 0.22 to 0.44 with secular time scales of $\sim$80\,000 years.
This can be seen from the 200\,000 year extent of the GJ\,1148's 
best-fit evolution, which is shown in Fig.~\ref{dyn_1148}, panel d).  
With these large secular eccentricity oscillations, the minimum 
pericenter distance $q_{\rm min} = a(1-e_{\rm max})$
and maximum apocenter distance  $p_{\rm max} = a(1+e_{\rm max})$
for the planets are $q_{\rm b}$ $\approx$ 0.08 au,  $p_{\rm b}$ $\approx$ 
0.25 au and $q_{\rm c}$ $\approx$ 0.50 au,  $p_{\rm c}$ $\approx$ 1.32 au,
which makes it unlikely that additional low-mass planets in close proximity to GJ\,1148 b 
and~c would be able to survive on stable orbits. 
Most likely the two Saturn-mass planets are the only companions to GJ\,1148 at least up to $\sim$1.4 au.

\subsubsection{GJ\,581}

For GJ\,581 we secured 20  precise CARMENES Doppler measurements between January 2016 and February 2017. 
In addition, we found 251 publicly available ESO HARPS spectra, which we re-processed with SERVAL,
and 413 HIRES literature RVs \citep{Butler2017}.
The large number of precise HIRES and HARPS data is an excellent opportunity for 
a comparative analysis with the newly obtained CARMENES data,
and a subsequent update of the orbital architecture of the GJ\,581 system.
GJ\,581 is currently known to have three bona-fide planets, which 
when listed in ascending order by orbital period are designated as GJ\,581\,e, GJ\,581\,b and GJ\,581\,c. 
Our RV analysis for GJ\,581 consisted of several standard consecutive 
steps of GLS period search and Keplerian fitting.
First we identified the strongest GLS peak of the combined data at 5.37\,d, corresponding to GJ\,581\,b.
Then using this period as an initial guess we fit a full Keplerian model for GJ\,581\,b whose residuals  
revealed a significant GLS peak at 12.9\,d, which is due to GJ\,581\,c. 
We added an additional Keplerian term and we fit the combined data simultaneously for GJ\,581\,b and GJ\,581\,c.
The residuals of the two-planet model revealed another strong GLS peak at 66.7\,d, which in the past was 
designated as GJ\,581\,d \citep{Udry2007}, but is now believed 
to be due to stellar activity \citep{Baluev2013, Robertson2014,Mascareno2015, Hatzes2016, Mascareno2017b}. 
We skipped this peak and adopted the next strongest peak at 3.15\,d, which is actually induced by GJ\,581\,e.
Finally, we obtained a simultaneous three-planet Keplerian model for GJ\,581.
We used the Keplerian three-planet best-fit parameters  as an initial guess for our more 
accurate three-planet dynamical model, which 
takes into account the gravitational interactions between planets GJ\,581\,e, b, and c, while fitting the RVs.

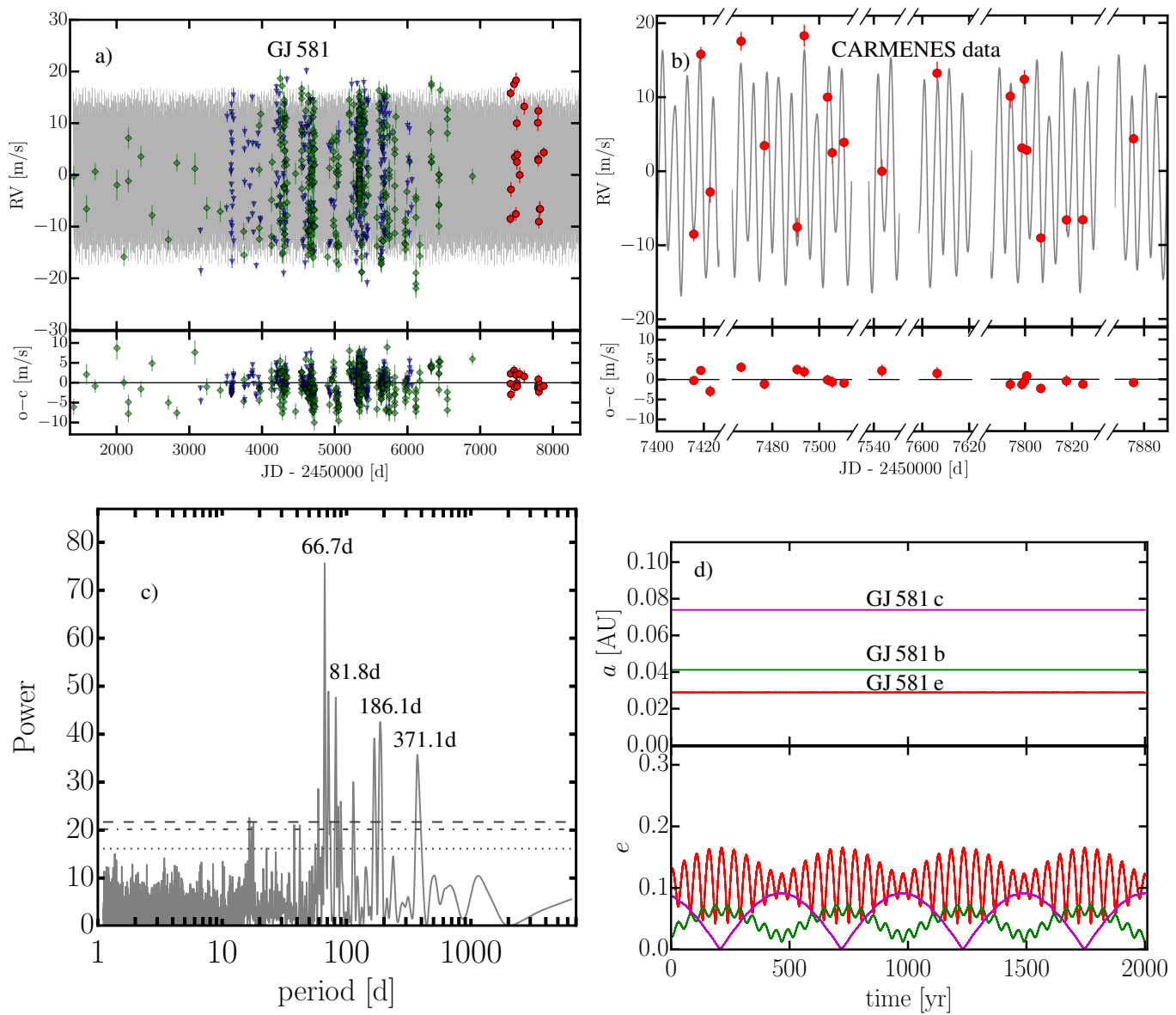
\begin{figure*}[tp]
\begin{center}$
\begin{array}{cc} 

 \includegraphics[width=9cm,height=7.5cm]{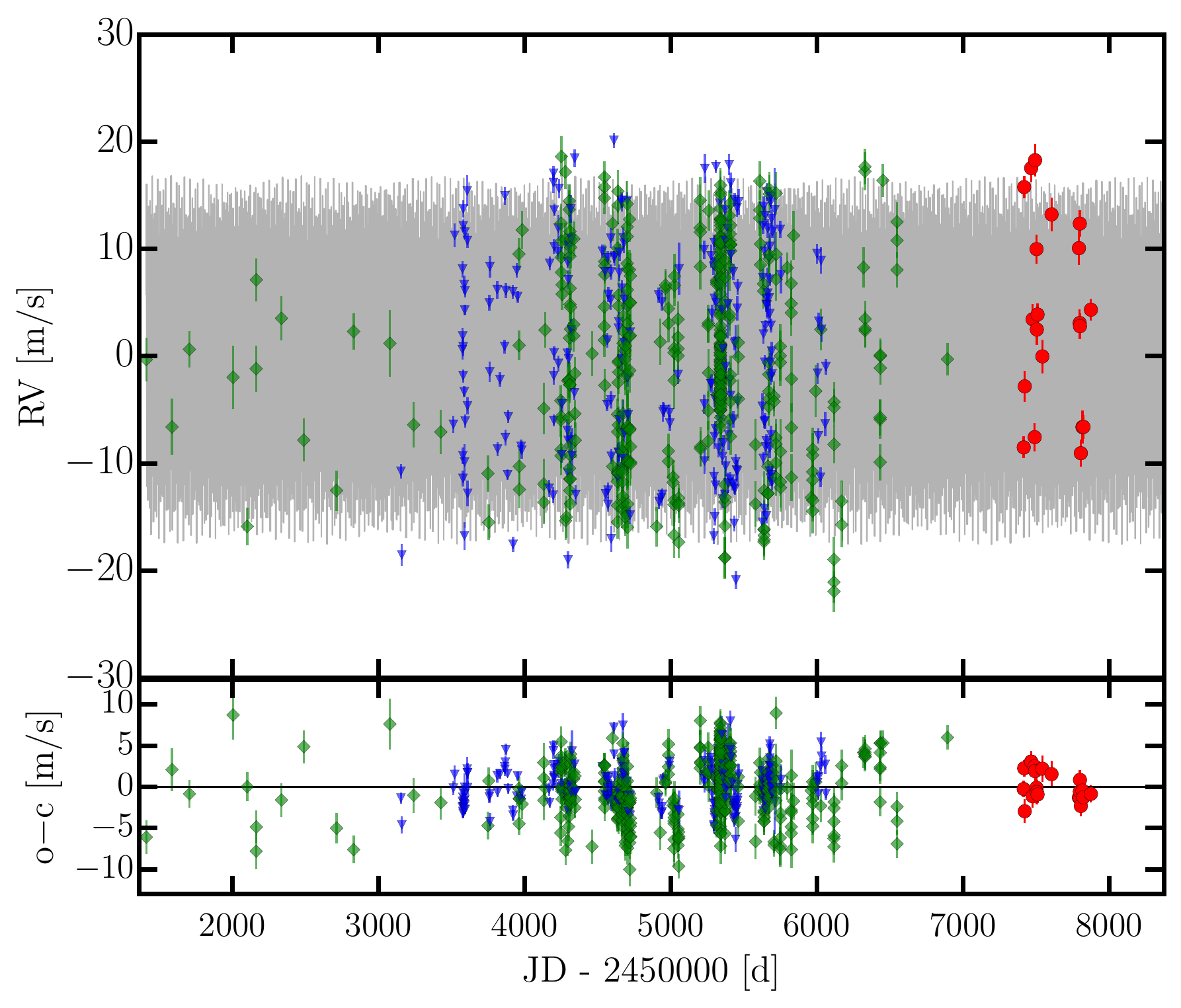} 
 \put(-140,190){GJ\,581}  \put(-215,187){a)}
 \includegraphics[width=9cm,height=7.5cm]{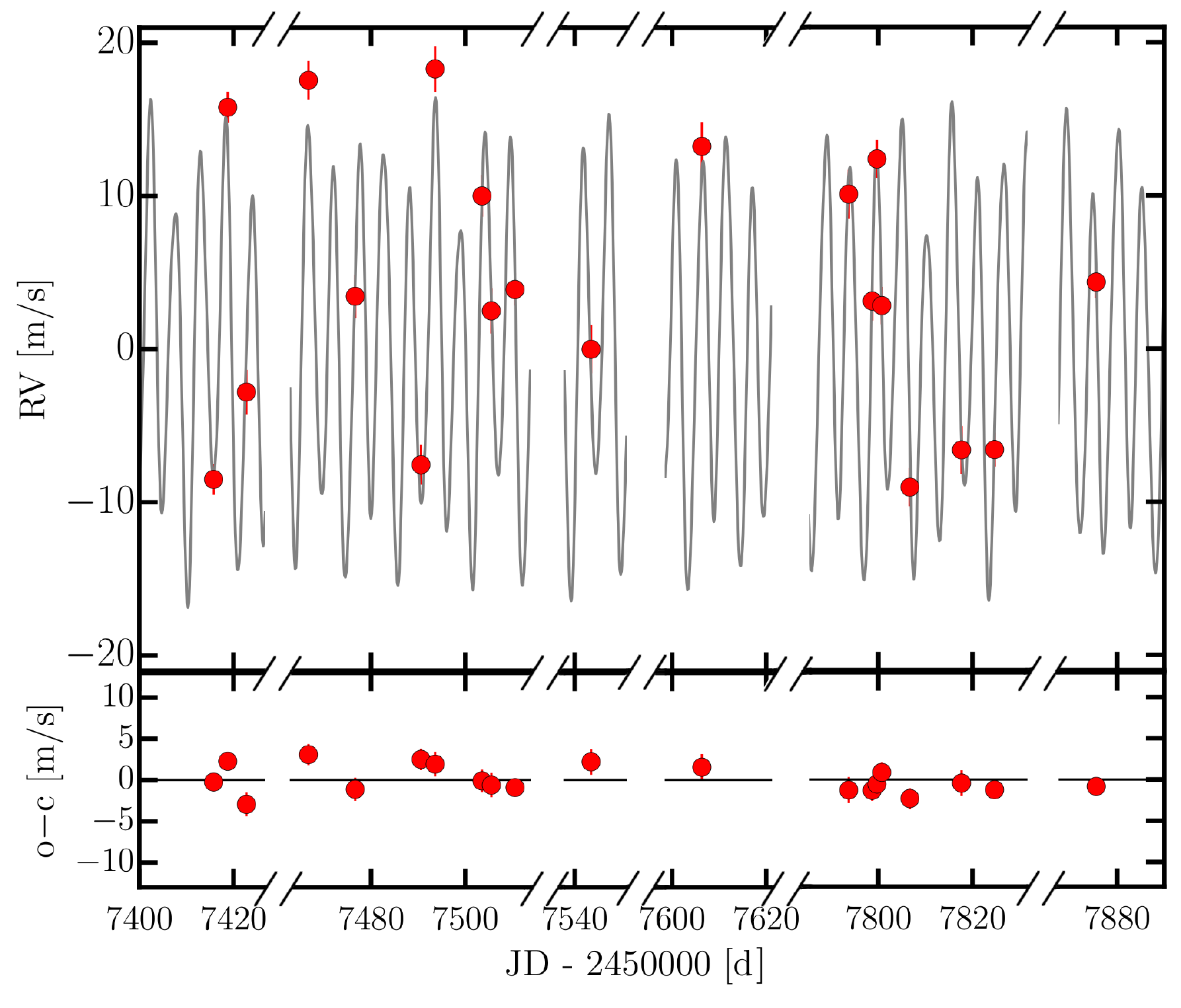}
 \put(-150,190){CARMENES data}  \put(-220,185){b)}\\
\includegraphics[width=9cm]{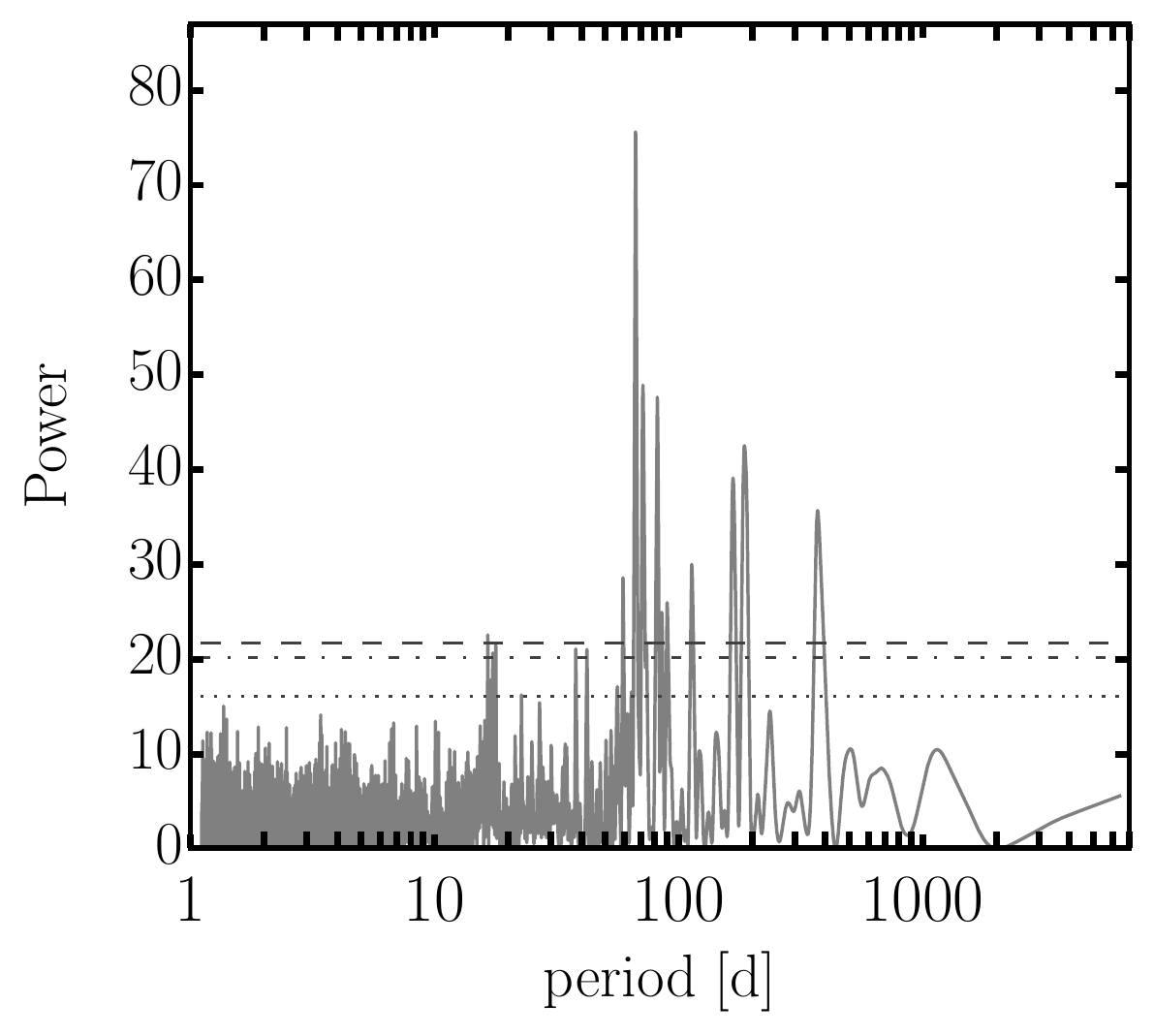}
\put(-125,205){66.7d} \put(-113,150){81.8d} \put(-100,135){186.1d} \put(-85,120){371.1d}  \put(-195,185){c)}
 \includegraphics[width=9cm,height=7.5cm]{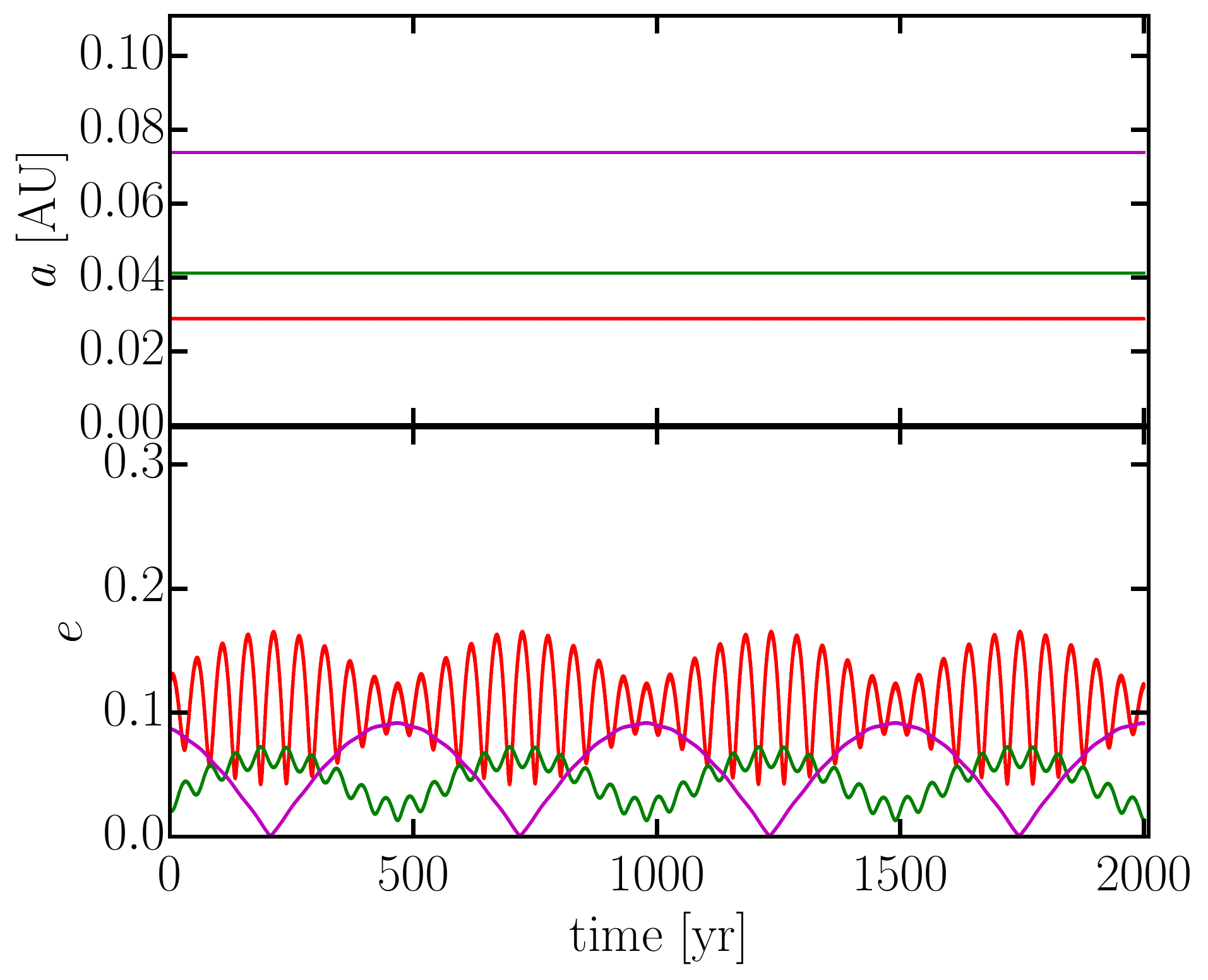} 
\put(-210,195){d)} \put(-135,145){GJ\,581\,e} \put(-135,158){GJ\,581\,b} \put(-135,182){GJ\,581\,c}  

\end{array} $
\end{center}

\caption{Panel a) Available Doppler measurements for GJ\,581 obtained with HIRES (green diamonds), HARPS (blue triangles)
and CARMENES (red circles) mutually agree when fitted with a three-planet self-consistent N-body model.
Panel b) Time series plot for only the CARMENES data show that they are fully consistent with 
the best-fit three-planet dynamical model yielding an $rms$ = 1.64 m\,s$^{-1}$, the lowest among the three data sets.  
Panel c) GLS power spectrum of the residuals from the three-planet dynamical model for GJ\,581.
The strong peak at 66.7\,d  was earlier attributed to a planet designated GJ\,581\,d, but is now believed to be induced by stellar activity.
Several other peaks with periods near 81.8\,d, 186.1\,d and 371.2\,d are significant, but unlikely of planetary nature
and most likely also related to activity and the window function. 
Panel d) Dynamical evolution of the best three-planet fit for the GJ\,581 planetary system.
This fit is stable for at least 10 Myr, but for illustrative purposes we show only a 2000 yr extent of the 
N-body integration, which clearly shows the perfectly synchronized orbital evolution. 
In this best-fit configuration the eccentricities of the three planets 
are oscillating with a period of $\sim$500 yr
in addition to the $\sim$50 yr secular perturbations between GJ\,581\,e and GJ\,581\,b.
}

\label{581_dyn} 
\end{figure*}

We converged to a three-planet best-fit solution leading to: 
$K_{\rm e}$ = 1.55 m\,s$^{-1}$,
$P_{\rm e}$ = 3.153 days,
$e_{\rm e}$ = 0.125,
$K_{\rm b}$ = 12.35 m\,s$^{-1}$,
$P_{\rm b}$ = 5.368 days,
$e_{\rm b}$ = 0.022, 
and 
$K_{\rm c}$ = 3.28 m\,s$^{-1}$,
$P_{\rm c}$ = 12.919 days,
$e_{\rm c}$ = 0.087.
We derived planetary masses and semi-major axes, respectively, as:
$m_{\rm e,b,c}$  =  1.66, 15.20, 5.65 $M_{\oplus}$ (0.005, 0.050, 0.018 $M_{\rm Jup}$) and $a_{\rm e,b,c}$ = 0.029, 0.041 and 0.074 au.
This fit has $\chi_{\nu}^2$ = 5.85 and overall scatter $rms$ = 2.91 m\,s$^{-1}$.
Detailed orbital parameter estimates and their bootstrap uncertainties are provided in Table~\ref{table_gj581},
while in Fig.~\ref{581_dyn} panel a), we show the HIRES, HARPS and CARMENES time series data plot fitted with the best 
three-planet dynamical  model. In Fig.~\ref{581_dyn} panel b) we show only the time series from our CARMENES data,
which clearly follow the best-fit model that is heavily influenced by the HIRES and HARPS data.
The individual data set scatter around the best-fit is lowest for 
CARMENES with $rms_{\rm CARMENES}$ = 1.64 m\,s$^{-1}$, 
followed by $rms_{\rm HARPS}$ = 2.32 m\,s$^{-1}$ and $rms_{\rm HIRES}$ = 3.60 m\,s$^{-1}$
showing good consistency between the data sets.
The CARMENES data supports the current understanding of the GJ\,581 system, while 
our RV analysis presents an update of its three-planet configuration
based on dynamical modeling of all available data.

\begin{table}[tp]

\centering  


\caption{Coplanar edge-on best dynamical fit parameters for the multiple planet system GJ\,581 based 
on the combined CARMENES and literature RVs.}   
\label{table_gj581}      

\begin{tabular}{l r r r r r r  }     
\hline \hline \noalign{\vskip 0.7mm}  
\noalign{\vskip 0.9mm}
Orb. param.    & GJ\,581\,e  &  GJ\,581\,b   &  GJ\,581\,c        \\     
\hline \noalign{\vskip 0.5mm} 
\noalign{\vskip 0.9mm}

$K$  [m\,s$^{-1}$]                 & 1.55$_{-0.13}^{+0.22}$     &  12.35$_{-0.20}^{+0.18}$   &  3.28$_{-0.12}^{+0.22}$      \\ \noalign{\vskip 0.9mm}
$P$ [d]                            & 3.153$_{-0.006}^{+0.001}$  &  5.368$_{-0.001}^{+0.001}$ &  12.919$_{-0.002}^{+0.003}$  \\ \noalign{\vskip 0.9mm} 
$e$                                & 0.125$_{-0.015}^{+0.078}$  &  0.022$_{-0.005}^{+0.027}$ &   0.087$_{-0.016}^{+0.150}$  \\ \noalign{\vskip 0.9mm}
$\varpi$ [deg]                     & 77.4$_{-43.6}^{+23.0}$     &  118.3$_{-22.9}^{+27.4}$   &  148.7$_{-33.0}^{+71.5}$     \\ \noalign{\vskip 0.9mm} 
$M$ [deg]                          & 203.7$_{-21.4}^{+56.6}$    &  163.4$_{-23.9}^{+22.9}$   &  218.0$_{-68.4}^{+37.3}$   \\ \noalign{\vskip 0.9mm}

$a$ [au]                           & 0.029$_{-0.001}^{+0.001}$  &  0.041$_{-0.001}^{+0.001}$ &  0.074$_{-0.001}^{+0.001}$   \\ \noalign{\vskip 0.9mm}
$m$  [$M_{\oplus}$]        & 1.657$_{-0.161}^{+0.240}$  &  15.20$_{-0.27}^{+0.22}$   &  5.652$_{-0.239}^{+0.386}$   \\ \noalign{\vskip 3.9mm}  

$\gamma_{\rm HIRES}$~[m\,s$^{-1}$]    & &0.61$_{-0.15}^{+0.15}$                             &                \\ \noalign{\vskip 0.9mm}
$\gamma_{\rm HARPS}$~[m\,s$^{-1}$]    & &12.19$_{-0.10}^{+0.12}$                          &                  \\ \noalign{\vskip 0.9mm}
$\gamma_{\rm CARM.}$~[m\,s$^{-1}$]    & &$-$6.83$_{-0.29}^{+0.28}$                         &                   \\  \noalign{\vskip 3.9mm}

$rms$~[m\,s$^{-1}]$                & &2.91                                      &                          \\ \noalign{\vskip 0.9mm}  
$rms_{\rm HIRES}$~[m\,s$^{-1}]$                & &3.60                                      &                          \\ \noalign{\vskip 0.9mm}  
$rms_{\rm HARPS}$~[m\,s$^{-1}]$                & &2.32                                      &                          \\ \noalign{\vskip 0.9mm}  
$rms_{\rm CARM.}$~[m\,s$^{-1}]$                & &1.64                                      &                          \\ \noalign{\vskip 0.9mm}  

$\chi_{\nu}^2$        & &5.85               &                                    \\ \noalign{\vskip 0.9mm}  
 
Valid for                &                           &       \\  \noalign{\vskip 0.9mm}

$T_0$ [JD-2450000]      & &1409.762                         \\  \noalign{\vskip 0.9mm}

\hline

\end{tabular} 
\end{table}

A GLS periodogram for the best-fit residuals is shown  panel c) of Fig.~\ref{581_dyn}. 
We find several significant residual periodic signals,
the strongest of which is at 66.7\,d, which was originally believed to be due to an 
additional planet GJ\,581\,d,
followed by 71.5\,d, 81.8\,d, 186.1\,d, 371.1\,d, etc., most of which
are likely also due to activity, in particular aliases of the dominant peak
and the one year observational window.
For example, the second strongest peak at 81.8\,d is  a 365.25 d alias of 66.7\,d. 
The 371.1\,d peak is close to one year period and likely comes from the observational window function,
while the 186.1\,d is close to half a year which is approximately one observing season, 
or alternatively this peak might be related to an alias of 371.1\,d and 365.25\,d.

We find our best fit to be stable for at least 10 Myr, demonstrating that the three-planet system 
has a perfectly synchronized orbital evolution.
In Fig.~\ref{581_dyn} panel d) we show a 2000 yr extent of the orbital evolution.
While the planetary semi-major axes are nearly constant, the 
planetary eccentricities are oscillating with moderate amplitudes and 
a secular period of $\sim$500 yr in addition to the shorter term $\sim$50-year 
secular perturbations between planets~e and b.

\begin{figure*}[tp]
\begin{center}$
\begin{array}{cc} 

\includegraphics[width=9cm,height=7.5cm]{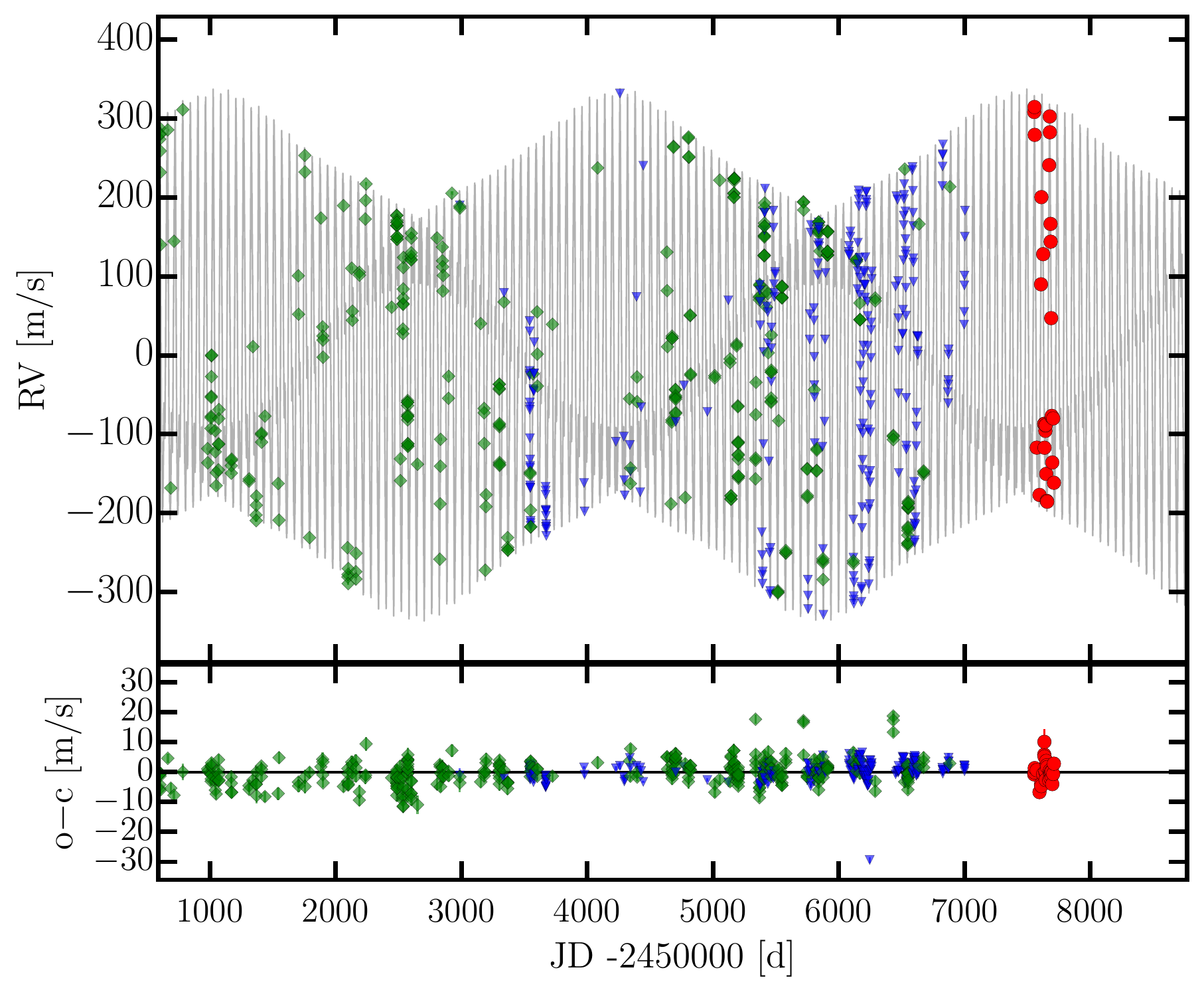} 
\put(-140,195){GJ\,876}  \put(-210,195){a)}
\includegraphics[width=9cm,height=7.5cm]{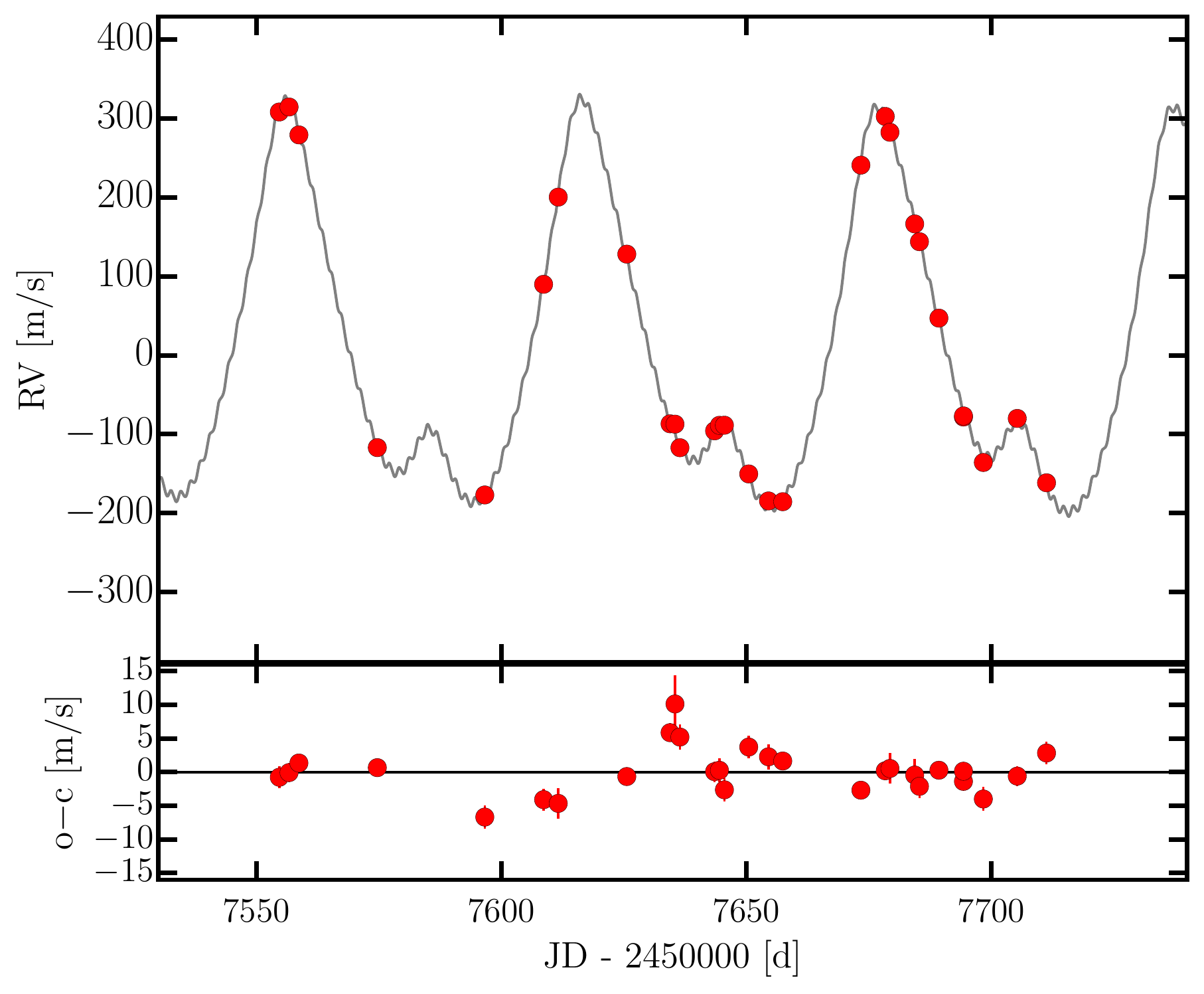}  
\put(-160,195){CARMENES data}  \put(-210,195){b)}\\
\includegraphics[width=9cm]{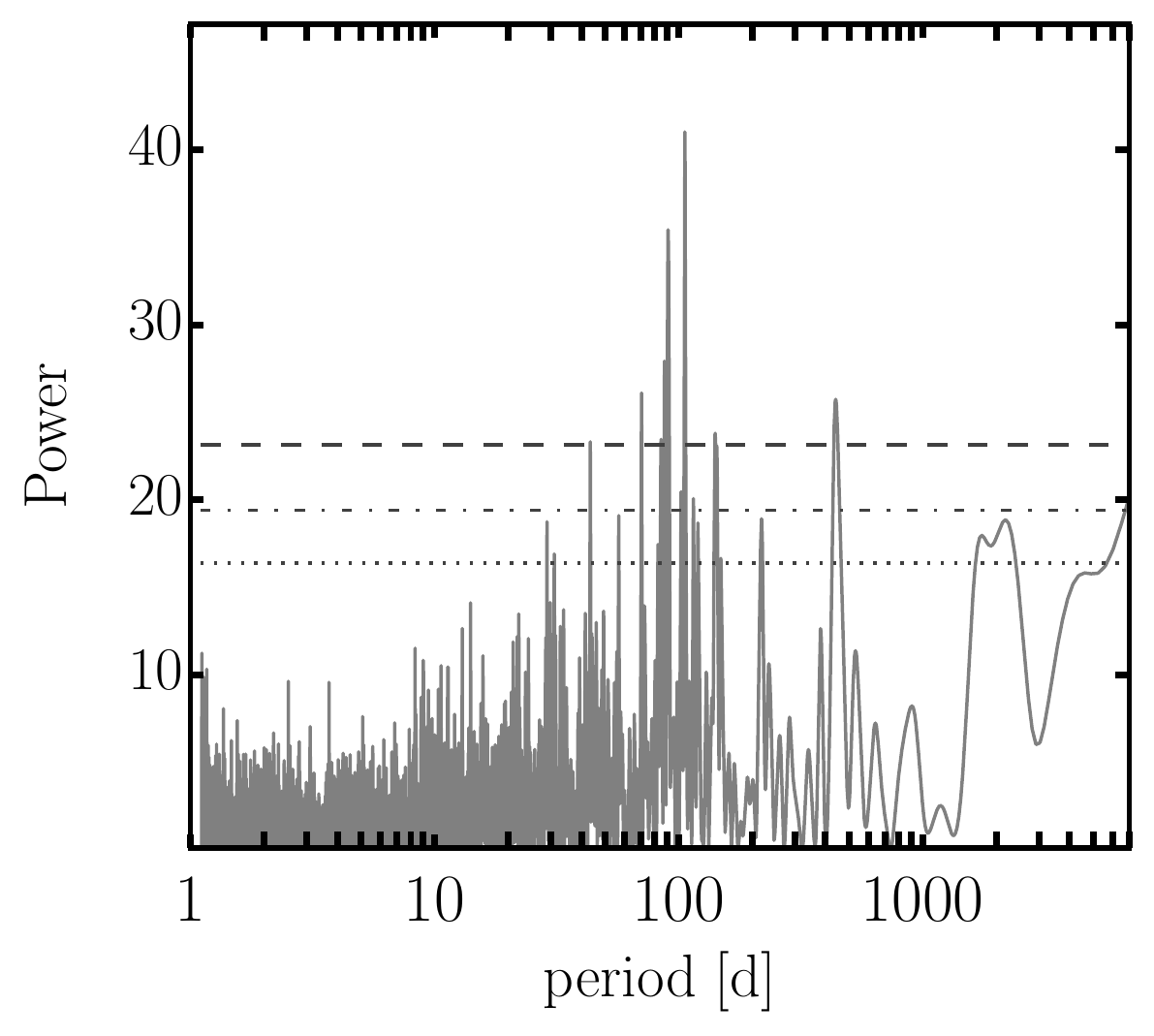}
\put(-115,205){106.0d} \put(-125,180){90.6d}  \put(-80,150){441.1d}  \put(-190,195){c)}
\includegraphics[width=9cm,height=7.5cm]{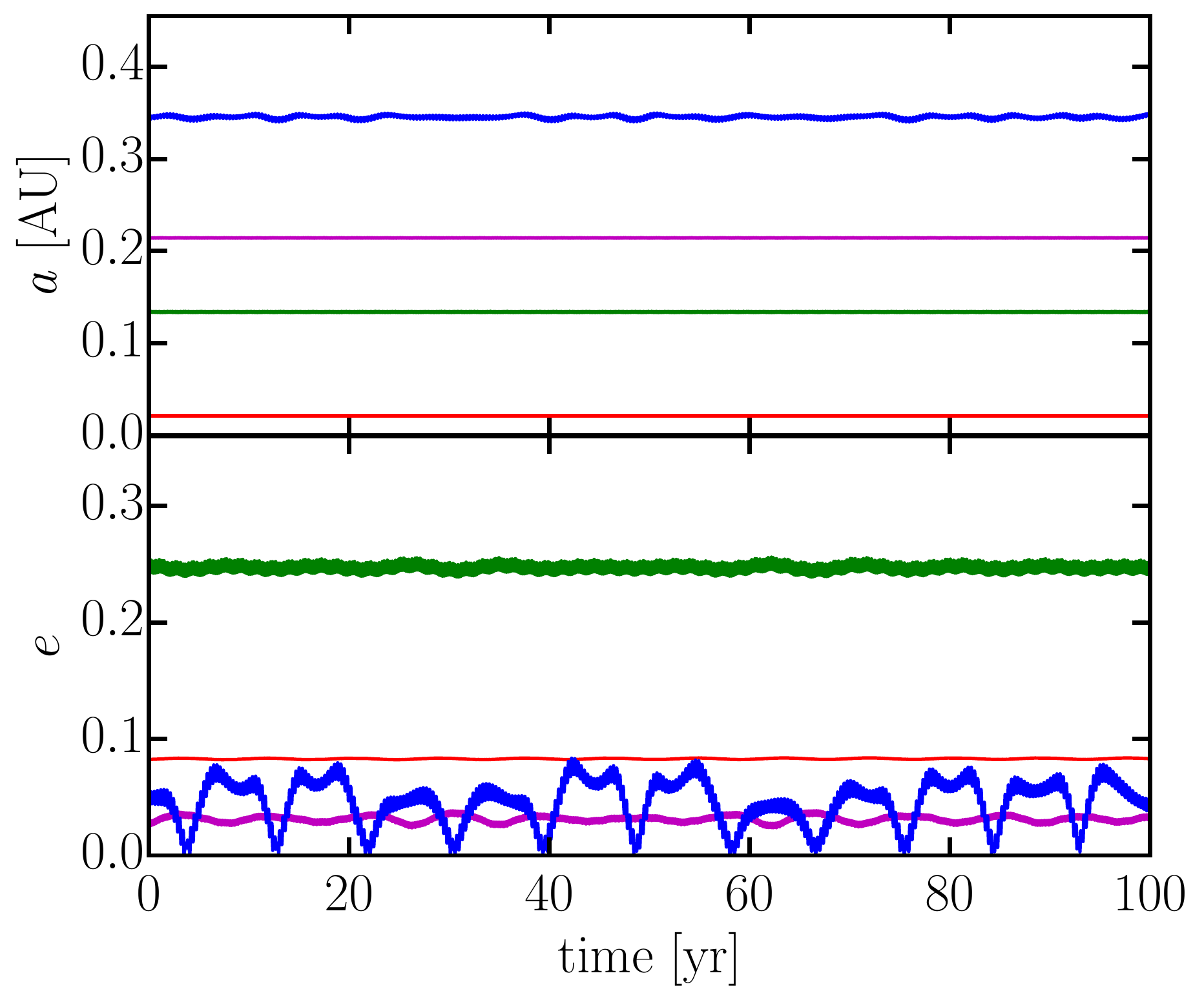} 
\put(-210,195){d)}  \put(-135,130){GJ\,876\,d} \put(-135,149){GJ\,876\,c} \put(-135,167){GJ\,876\,b} \put(-135,192){GJ\,876\,e}

\end{array} $
\end{center}

\caption{Data colors and symbols are the same as in Fig.~\ref{581_dyn}, but for GJ\,876. 
Panel a) The total amount of 622 precise RVs from HARPS, HIRES and CARMENES are fitted with a four-planet Newtonian model.
Panel b) Zoomed extent of the CARMENES data, which clearly follow the four planet model 
with a very low $rms$ = 2.97 m\,s$^{-1}$,  very similar to the HARPS data with $rms$ = 2.95 m\,s$^{-1}$, 
but better than the HIRES data whose $rms$ scatter is 4.35 m\,s$^{-1}$. 
Panel c) GLS power spectrum of the residuals from the four-planet dynamical model for GJ\,876 
showing several peaks above FAP = 0.1\% near 106.0\,d, 90.6\,d and 441.1\,d. 
However, these are unlikely planetary signals (see text for details). 
Panel d) Our updated four-planet dynamical fit is stable for 10 M yrs and is 
consistent with a chaotic Laplace 1:2:4 MMR orbital evolution between planets b, c and~e, already known for this system.
Our best fit and orbital evolution of the planetary system, however, suggests smaller (compared to the literature)
eccentricity ($e_{\rm d}$ = 0.082) of the innermost planet GJ\,876\,d.
}

\label{fig876} 
\end{figure*}

\begin{table*}[tb]

\centering  


\caption{Coplanar best dynamical fit parameters for the multiple planet system GJ\,876 based 
on the combined CARMENES and literature RVs.  }   
\label{table:multi876}      
  
\begin{tabular}{l r r r r r r r }     
\hline\noalign{\vskip 0.5mm}\hline
\noalign{\vskip 0.9mm}
Orb. param.~~~~~~~~~~~~~~~~~~~~~~~~& GJ\,876\,d  & & GJ\,876\,c  & & GJ\,876\,b  & & GJ\,876\,e     \\     
\hline\noalign{\vskip 0.5mm} 
\noalign{\vskip 0.9mm}

$K$  [m\,s$^{-1}$]                 & 6.14$_{-0.22}^{+0.23}$     & & 88.34$_{-0.25}^{+0.23}$     & & 212.07$_{-0.26}^{+0.27}$   & & 3.39$_{-0.28}^{+0.29}$        \\ \noalign{\vskip 0.9mm}
$P$ [d]                            & 1.938$_{-0.001}^{+0.001}$  & & 30.126$_{-0.003}^{+0.011}$  & & 61.082$_{-0.010}^{+0.006}$ & & 124.4$_{-0.7}^{+0.3}$   \\ \noalign{\vskip 0.9mm}
$e$                                & 0.082$_{-0.025}^{+0.043}$  & & 0.250$_{-0.002}^{+0.001}$   & & 0.027$_{-0.002}^{+0.002}$  & & 0.040$_{-0.004}^{+0.021}$     \\ \noalign{\vskip 0.9mm}
$\varpi$ [deg]                     & 272.8$_{-29.5}^{+21.8}$    & & 51.6$_{-1.0}^{+0.4}$        & & 35.1$_{-1.9}^{+6.7}$       & & 263.6$_{-46.0}^{+28.3}$       \\ \noalign{\vskip 0.9mm}
$M$ [deg]                          & 316.7$_{-20.0}^{+28.6}$    & & 293.3$_{-0.4}^{+1.1}$       & & 341.1$_{-6.8}^{+2.0}$      & & 310.3$_{-29.2}^{+46.7}$       \\ \noalign{\vskip 0.9mm}

$a$ [au]                           & 0.021$_{-0.001}^{+0.001}$  & & 0.134$_{-0.001}^{+0.001}$   & & 0.214$_{-0.001}^{+0.001}$  & & 0.345$_{-0.002}^{+0.001}$     \\ \noalign{\vskip 0.9mm}
$m$        [$M_\oplus$]            & 6.910$_{-0.270}^{+0.220}$  & & 241.5$_{-0.6}^{+0.7}$       & & 760.9$_{-1.0}^{+1.0}$      & & 15.43$_{-1.27}^{+1.29}$       \\  \noalign{\vskip 0.9mm} 
$i$ [deg]                          & 59.0 (fixed)               & & 59.0 (fixed)                & & 59.0 (fixed)               & & 59.0 (fixed)                  \\ \noalign{\vskip 0.9mm}
$\Omega$ [deg]                     & 0.0 (fixed)                & & 0.0 (fixed)                 & & 0.0 (fixed)                & & 0.0 (fixed)                  \\ \noalign{\vskip 3.9mm}

$\gamma_{\rm HIRES}$~[m\,s$^{-1}$]    & \makebox[\dimexpr(\width-10em)][l]{27.50$_{-0.30}^{+0.32}$}      & &                       & &                 & &  \\ \noalign{\vskip 0.9mm}
$\gamma_{\rm HARPS}$~[m\,s$^{-1}$]    & \makebox[\dimexpr(\width-10em)][l]{138.09$_{-0.10}^{+0.12}$}  & &                       & &                 & &  \\ \noalign{\vskip 0.9mm}
$\gamma_{\rm CARM.}$~[m\,s$^{-1}$]    & \makebox[\dimexpr(\width-10em)][l]{-260.24$_{-0.65}^{+0.60}$}  & &                       & &                 & & \\  \noalign{\vskip 3.9mm}

$rms$~[m\,s$^{-1}]$                & \makebox[\dimexpr(\width-10em)][l]{3.49}               & &                       & &                 & &                            \\ \noalign{\vskip 0.9mm}
$rms_{\rm HIRES}$~[m\,s$^{-1}]$                & \makebox[\dimexpr(\width-10em)][l]{4.35}               & &                       & &                 & &                            \\ \noalign{\vskip 0.9mm}
$rms_{\rm HARPS}$~[m\,s$^{-1}]$                & \makebox[\dimexpr(\width-10em)][l]{2.95}               & &                       & &                 & &                            \\ \noalign{\vskip 0.9mm}
$rms_{\rm CARM.}$~[m\,s$^{-1}]$                & \makebox[\dimexpr(\width-10em)][l]{2.97}               & &                       & &                 & &                            \\ \noalign{\vskip 0.9mm}

$\chi_{\nu}^2$                     & \makebox[\dimexpr(\width-10em)][l]{9.75}               & &                       & &                 & &                            \\ \noalign{\vskip 0.9mm}

Valid for                &                           &       \\  \noalign{\vskip 0.9mm}

$T_0$ [JD-2450000]      & \makebox[\dimexpr(\width-10em)][l]{602.093}                  &       \\  \noalign{\vskip 0.9mm}

\hline

\end{tabular} 
\end{table*}

\subsubsection{GJ\,876}

For GJ\,876 we obtained 28 precise CARMENES Doppler measurements between June 2016 and December 2016. 
We find 256 publicly available ESO HARPS spectra, which we re-processed with SERVAL,
and 338 HIRES literature RVs \citep{Butler2017}.
GJ\,876 is known to host four planets, namely GJ\,876\,d,\,c,\,b and e,
the last three of which  are locked in a strongly interacting 1:2:4 Laplace MMR \citep{Rivera2010}.
For our RV analysis we simply combined all available RV data and we 
applied a four-planet dynamical model starting with coplanar
orbital parameters taken from \citep{Rivera2010}. 
Following their coplanar test we also fixed the line of sight inclination at $i$~=~59$^\circ$. 
We did not make further attempts to constrain the coplanar or mutual inclinations for this system, although 
we are aware that this might lead to an additional model improvement \citep{Nelson2016}.

Our four-planet dynamical fit  takes into account 622 precise Doppler 
measurements taken over twenty years,  which is by far the most complete set of 
high-precision RV data  that has been analyzed for this star.
Our updated orbital four-planet best-fit configuration suggests: 
$K_{\rm d}$ = 6.14 m\,s$^{-1}$,
$P_{\rm d}$ = 1.938 days,
$e_{\rm d}$ = 0.082, 
for the inner resonant pair we obtain:
$K_{\rm c}$ = 88.34 m\,s$^{-1}$,
$P_{\rm c}$ = 30.126 days,
$e_{\rm c}$ = 0.250  
and 
$K_{\rm b}$ = 212.07  m\,s$^{-1}$,
$P_{\rm b}$ = 61.082  days,
$e_{\rm b}$ = 0.027,
and for the outermost planet~e we obtain:
$K_{\rm e}$ = 3.39 m\,s$^{-1}$,
$P_{\rm e}$ = 124.4 days,
$e_{\rm e}$ = 0.040,
valid at the epoch JD = 2450602.093, the same as in \citet{Rivera2010}.
We derive planetary masses and semi-major axes as follows:
$m_{\rm d,c,b,e}$  = 6.91, 241.5, 760.9, 15.43 $M_\oplus$ (0.021, 0.760, 2.394 and 0.049~$M_{\rm Jup}$),
and $a_{\rm  d,c,b,e}$ = 0.021, 0.134, 0.214 and 0.345 au, respectively.
Detailed best-fit orbital parameter estimates and their bootstrap uncertainties
are provided in Table~\ref{table:multi876}.

In Fig.~\ref{fig876} panel a) we show the HIRES, HARPS and 
CARMENES data time series fitted with our best four-planet dynamical fit,
while  panel b) shows a zoom only to our CARMENES data.
All data sets yield very good agreement with the four-planet best-fit prediction
and show similar RV scatter residuals.
As in the GJ\,581 case, this fit is dominated by the HIRES and the HARPS data, 
which have much more extensive data sets when compared with the CARMENES RVs.
The CARMENES scatter around the best-fit is $rms_{\rm CARMENES}$ = 2.97 m\,s$^{-1}$, close to the 
one from HARPS with $rms_{\rm HARPS}$ = 2.95 m\,s$^{-1}$, and better 
than HIRES which has $rms_{\rm HIRES}$ = 4.35\,m\,s$^{-1}$.

Figure~\ref{fig876} panel c) shows a GLS periodogram of the best-fit residuals.
We find several significant GLS peaks 
that, when sorted by significance, appear at 106.0\,d, 90.6\,d, 87.6\,d, 441.1\,d, 70.5\,d, etc.
The origin of these periodic signals is likely stellar activity
induced by stellar rotation, magnetic cycles, the window function and their aliases.
For example \citet{Nelson2016} also found a dominant periodicity around 95\,d
in their four-planet best-fit residuals for GJ\,876 and attributed this signal (and aliases) to stellar activity. 
They found the same periodicity in the H$\alpha$ line when analyzing the publicly available HARPS spectra. 
In the $Carmencita$ catalog \citep{Caballero2016} the period for GJ\,876 (listed in Table~\ref{table:1}) 
is $\sim$81\,d (D\'{\i}ez-Alonso et al., in prep.), which is of the same order of magnitude, 
but slightly shorter than the 95\,d from \citet{Nelson2016} and the $\sim$87\,d estimate from \citet{Mascareno2015}. 
Thus, the true rotation period of GJ\,876 likely lies somewhere between these estimates.
We do not retrieve a 95\,d  peak in our  combined data residuals,  nor we did obtain 
a significant peak detection in our CARMENES H$\alpha$-index measurements.
However, the 90\,d peak in our four-planet model residuals is close to the current $P_{\rm rot}$ estimates for GJ\,876. 
By fitting a sine model with period of 90.6\,d to the residuals 
we also remove the 70.5\,d, 87.6\,d and 441.1\,d peaks, except the peak at 106.0\,d.
The~106.0\,d period,  could be related to differential stellar rotation and spots at two latitudes
creating peaks at 90.0\,d and 106.0\,d. The 106.0\,d signal could also be induced by model degeneracy
(i.e., imperfect dynamical modeling) since we do not fit for the mutual inclinations of the planets.

When compared to the previous four-planet solution of \citet{Rivera2010},
our fit is largely consistent, in terms of overall stability, orbital 
parameters and evolution of planets GJ\,876\,b, c and e.
For clarity, in  Fig.~\ref{fig876} panel d) we show a 100-yr 
section of our best fit dynamical evolution of GJ\,876, which we find stable for at least 10 Myr.
The orbital evolution of our best-fit is consistent with a chaotic 1:2:4 Laplace MMR in agreement with 
earlier dynamical studies for this system \citep{Rivera2010, Batygin2015, Nelson2016}. 
This is evident from the quasi-periodic pattern of the eccentricity evolution of the outermost planet 
GJ\,876\,e \citep[first described by][]{Rivera2010}.  
It should be noted that the time scale for the eccentricity evolution of planet GJ\,876\,e 
is only about 7.5 years and therefore shorter than the observational time baseline of the system. 
This result, combined with the chaotic nature, imposes an intrinsic limit of the fit quality since small changes 
well within the uncertainties can change the eccentricity evolution of GJ\,876\,e significantly.
Despite the chaotic behavior, \citet{Rivera2010} found stability for the GJ\,876 system over hundreds to thousand Myr.

\begin{figure*}[htp]
\begin{center}$
\begin{array}{cccc}
\includegraphics[width=6cm,height=5.2cm]{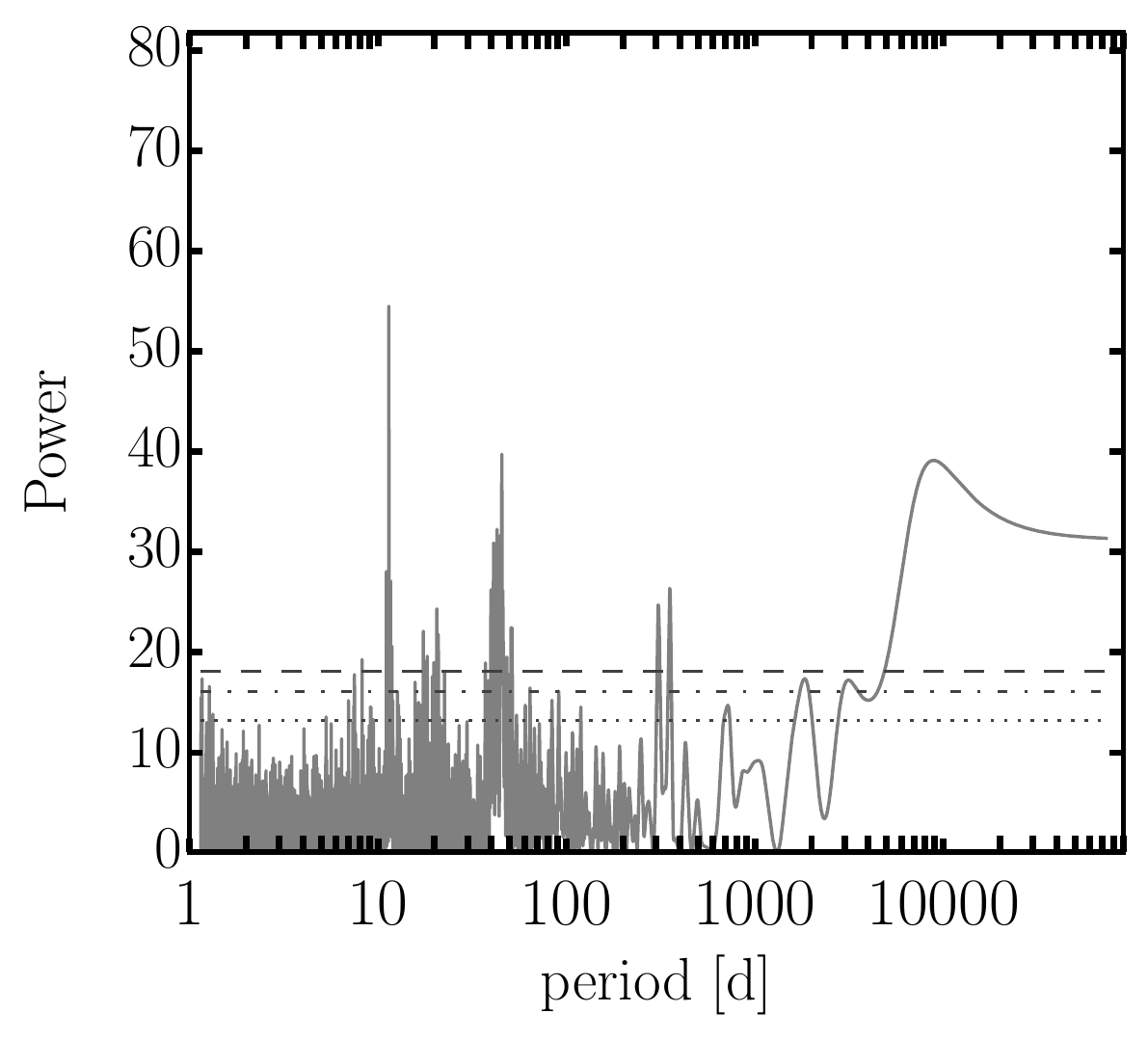} 
\put(-130,120){a)} \put(-110,125){HIRES+CARMENES} \put(-125,108){11.44d} \put(-110,90){45.46d} \put(-80,70){307d} \put(-75,80){354d} \put(-40,90){8892d} 
\includegraphics[width=6cm,height=5.2cm]{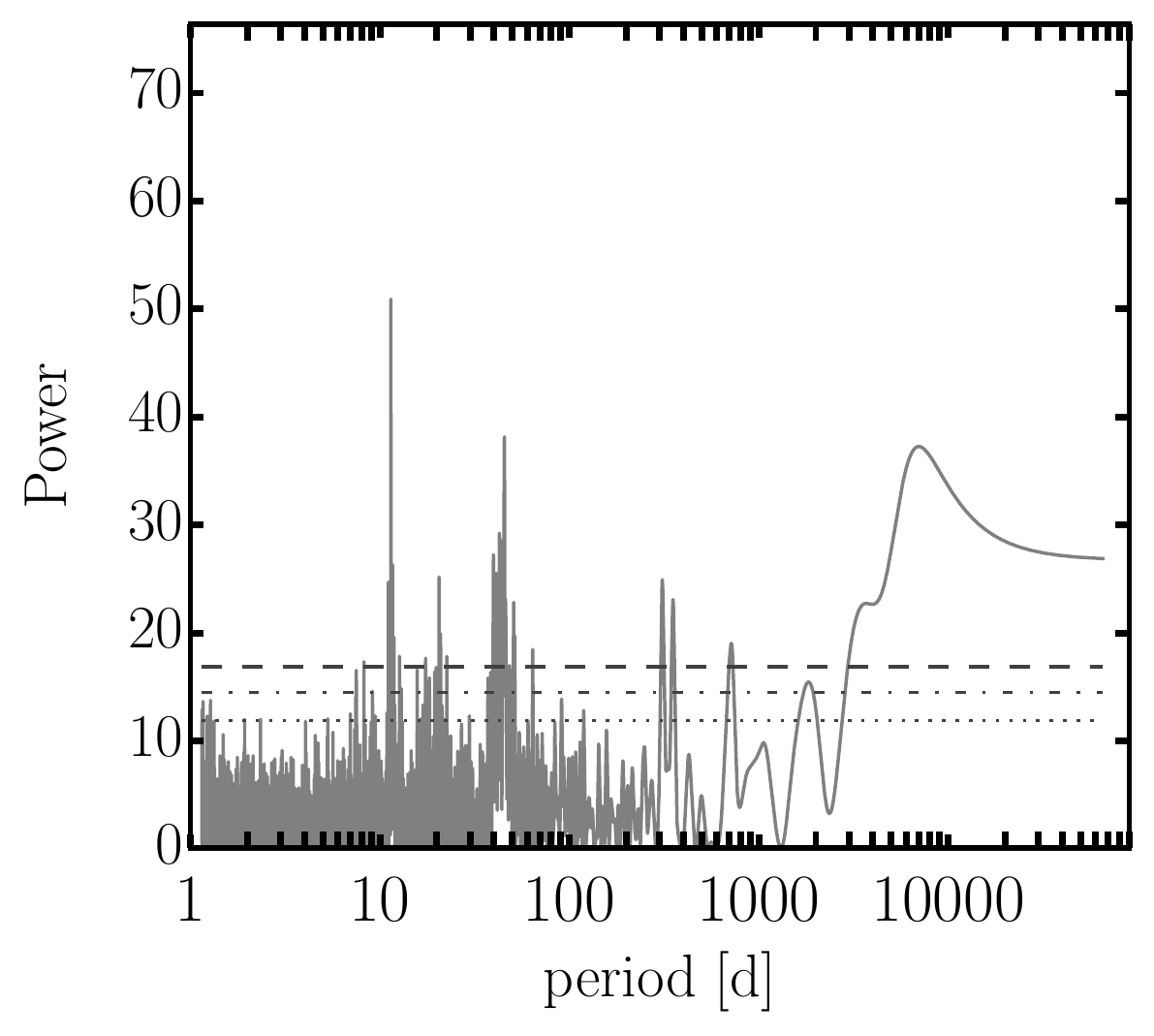} 
\put(-125,108){11.44d} \put(-110,90){45.42d}  \put(-130,120){b)} \put(-90,125){HIRES} \put(-80,70){309d} \put(-75,80){352d} \put(-45,90){6997d} 
\includegraphics[width=6cm,height=5.2cm]{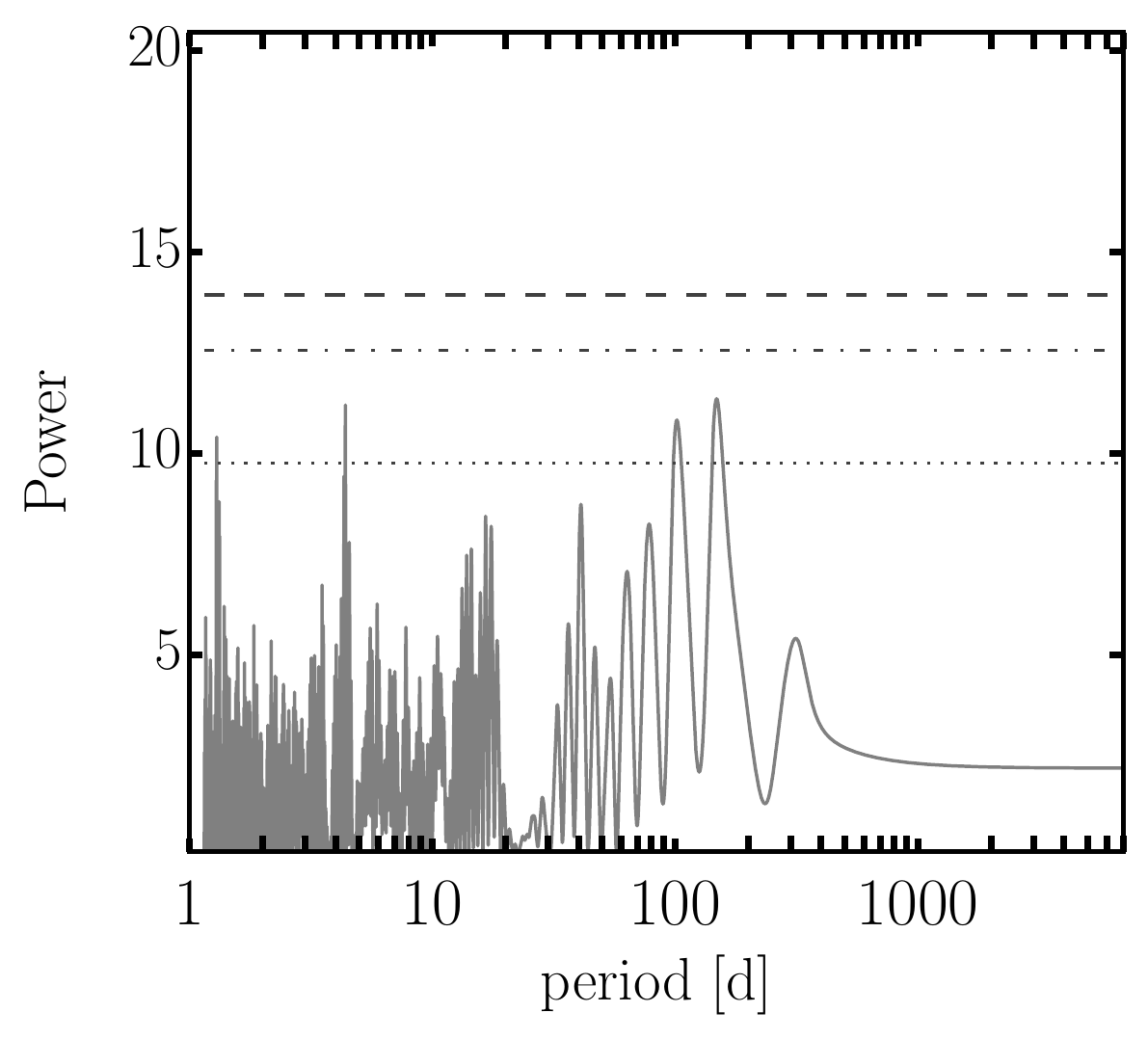} 
\put(-130,120){c)} 
\put(-100,125){CARMENES} \\

\includegraphics[width=6cm,height=5.2cm]{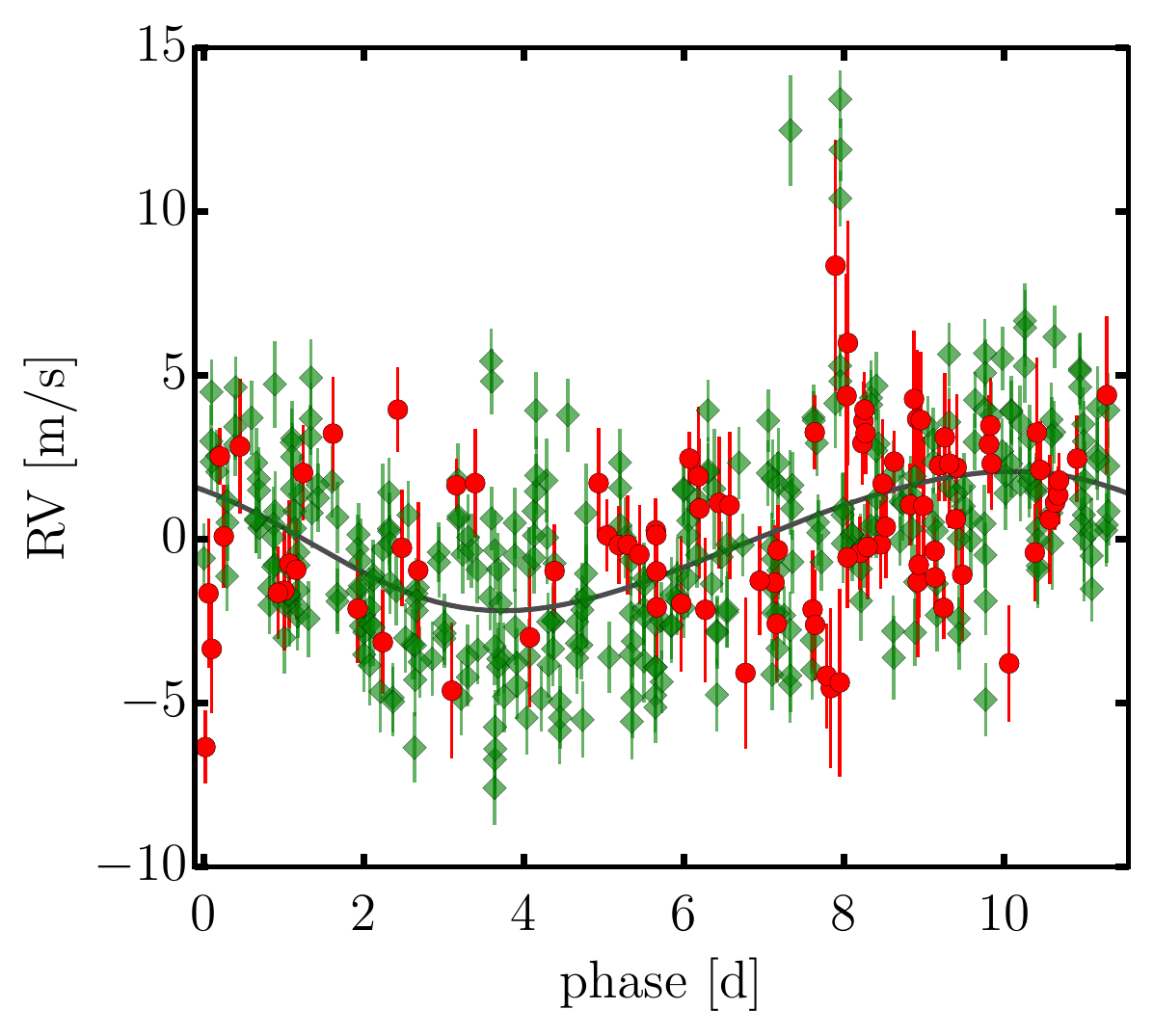} 
\put(-135,105){d)} \put(-120,110){\tiny GJ\,15\,Ab + lin. trend}  \put(-135,130){\tiny HIRES+CARMENES}
\includegraphics[width=6cm,height=5.2cm]{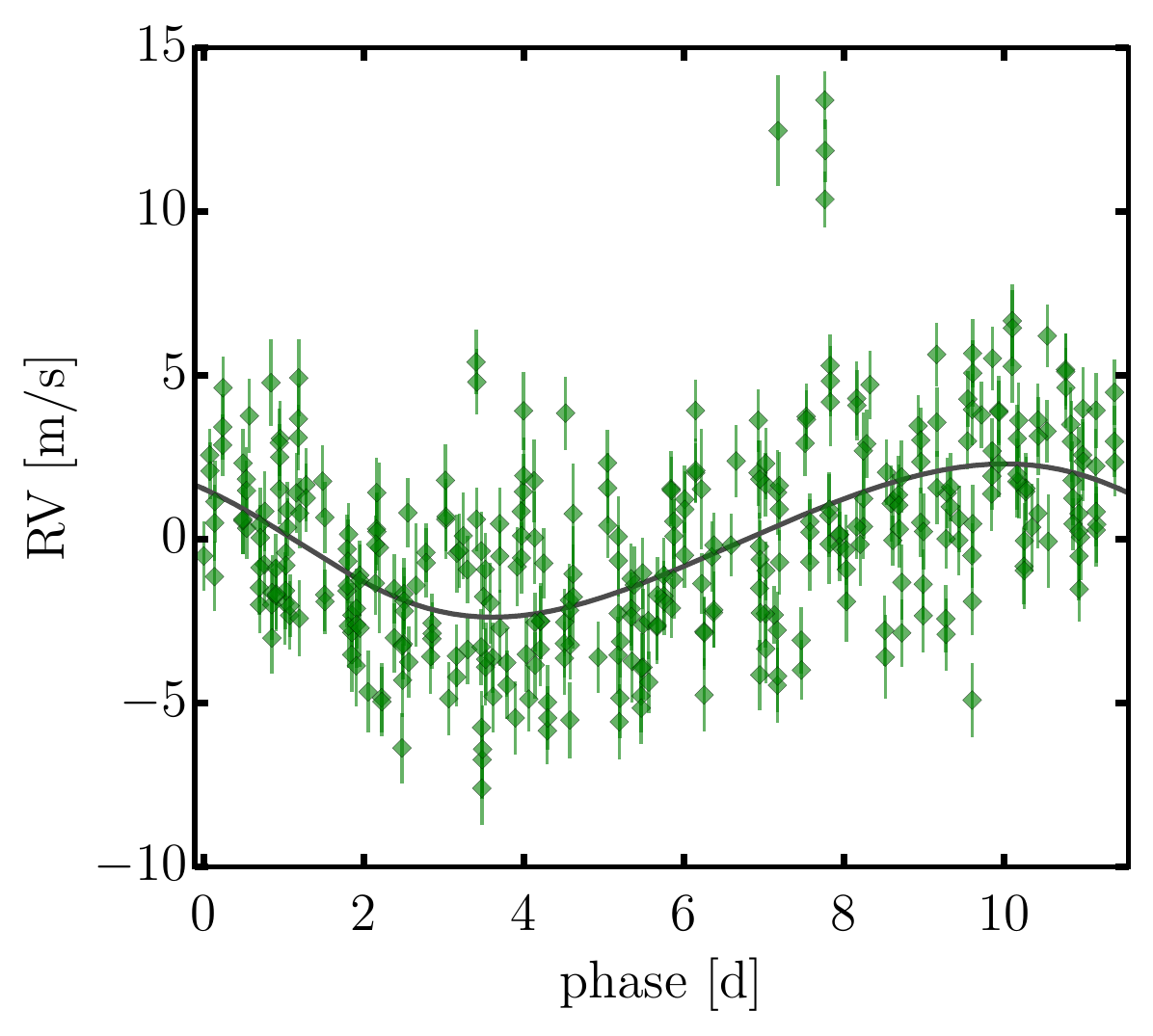} 
\put(-135,105){e)} \put(-120,110){\tiny GJ\,15\,Ab + lin. trend}  \put(-135,130){\tiny HIRES}
\includegraphics[width=6cm,height=5.2cm]{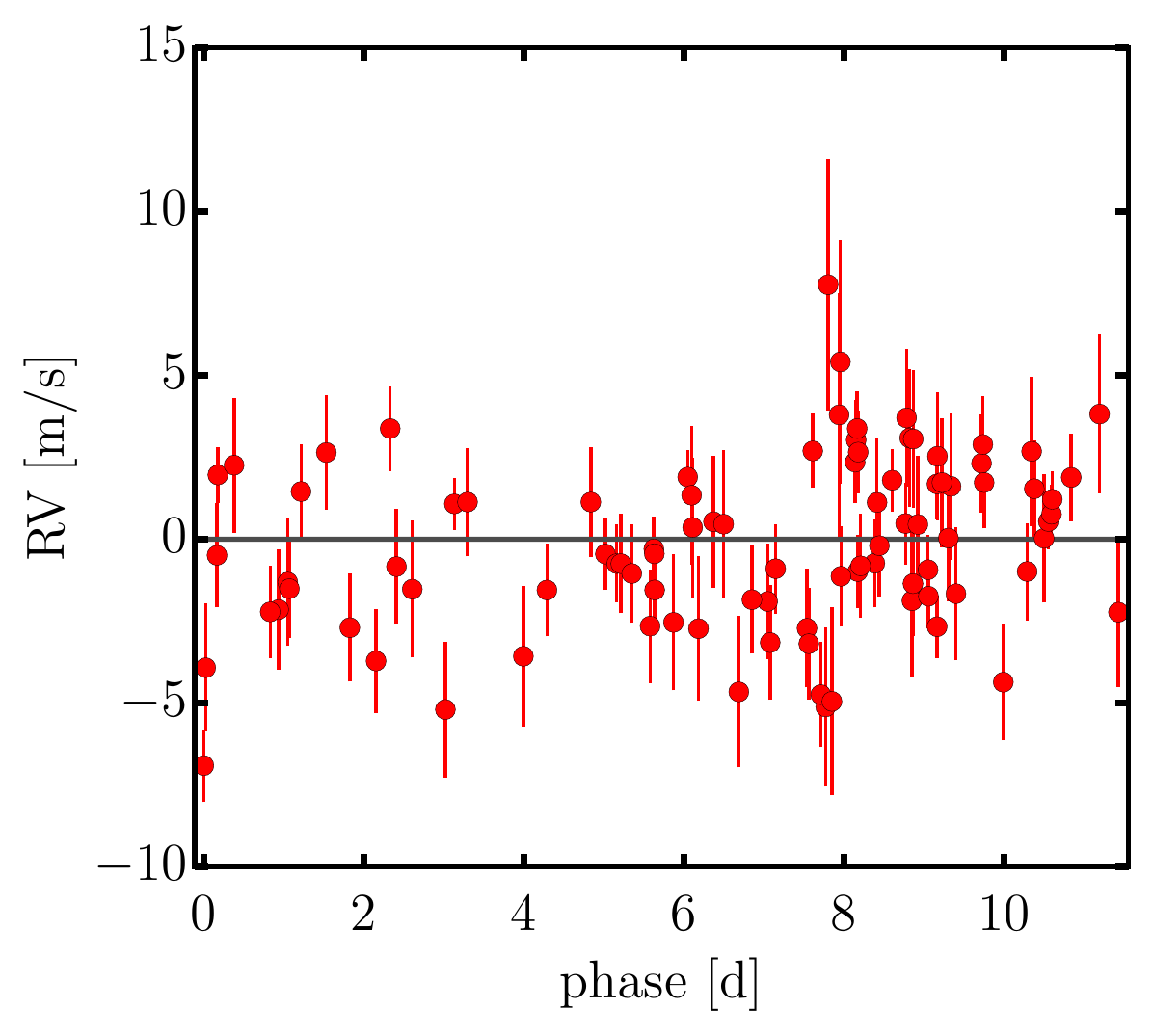} 
\put(-135,105){f)} \put(-120,110){\tiny no signal at P = 11.44d} \put(-135,130){\tiny CARMENES} \\

\end{array} $
\end{center}

\caption{
Panel a) show GLS periodogram of the combined HIRES (green diamonds) and CARMENES (red circles) data for GJ\,15\,A,
while panels b) and c) show separate GLS periodograms for the HIRES, and CARMENES data, respectively.
Panels d), e) and f) show the best-fit models to the combined and the separate data sets.
The HIRES data strongly suggest several periodicities, most notable at a period of 11.44\,d, 
which is associated with the putative planet GJ\,15\,Ab,
a period of 45.42\,d, attributed to stellar activity, and a few long-period peaks, which disappear when we fit a linear trend to the data.
A GLS power spectrum to the 92 CARMENES RVs, however, does not show significant peaks, 
which is also the case for more recent HIRES RVs (see Fig.~\ref{fig1.3}, bottom panel).
No valid Keplerian model with a period of 11.44\,d can be fitted to the post-discovery 
HIRES and CARMENES data, casting doubts on the existence of GJ\,15\,Ab.
The available data for GJ\,15\,A are dominated by the early HIRES RVs, and thus the putative planetary
signal is still detected in the combined data set.
}

\label{fig1} 
\end{figure*}

\begin{figure}[tp]
\begin{center}$
\begin{array}{cc} 
\includegraphics[width=4.5cm,height=3.8cm]{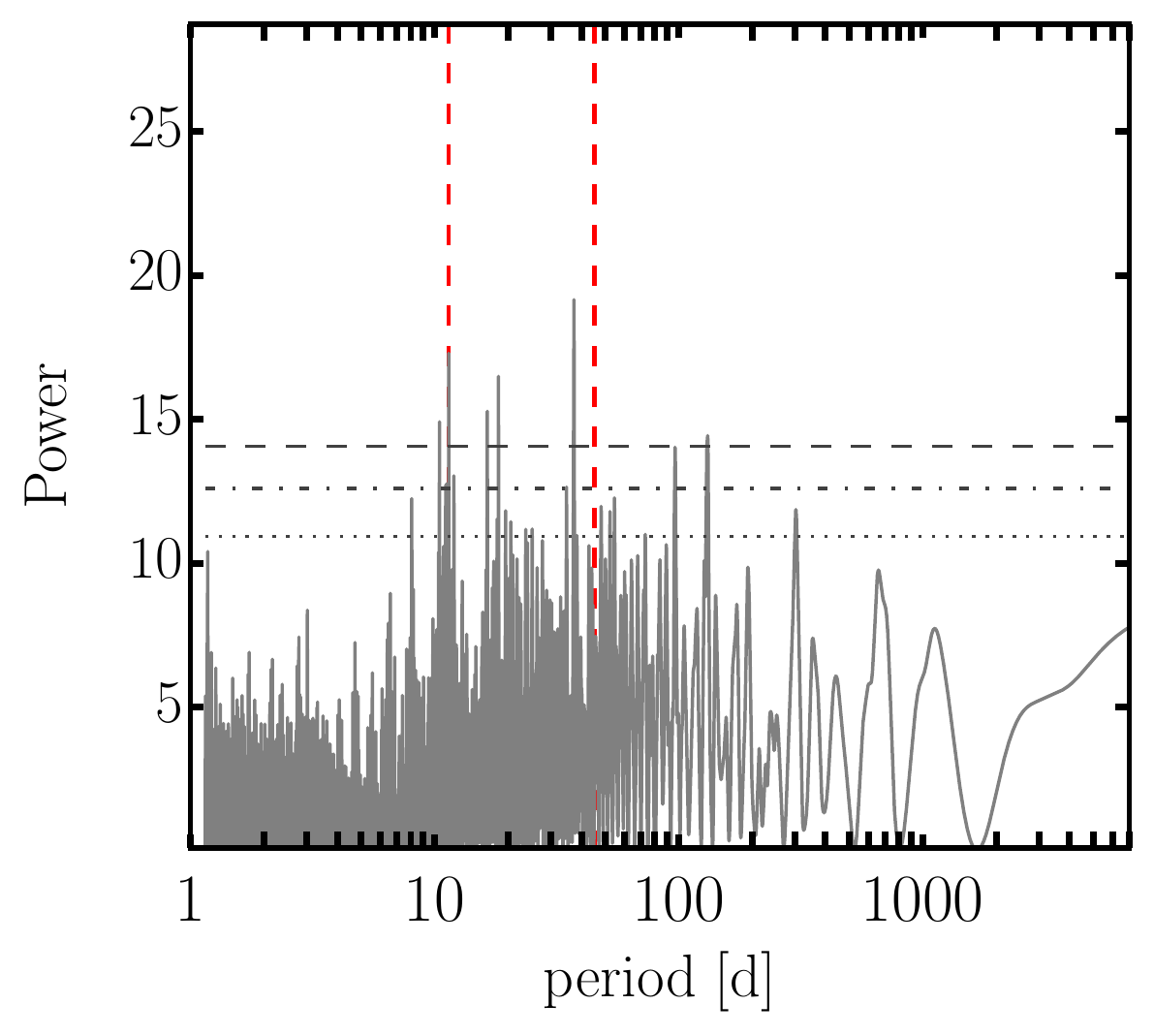}  
\put(-98,92){11.44 d} \put(-60,92){45.46 d}
\includegraphics[width=4.5cm,height=3.8cm]{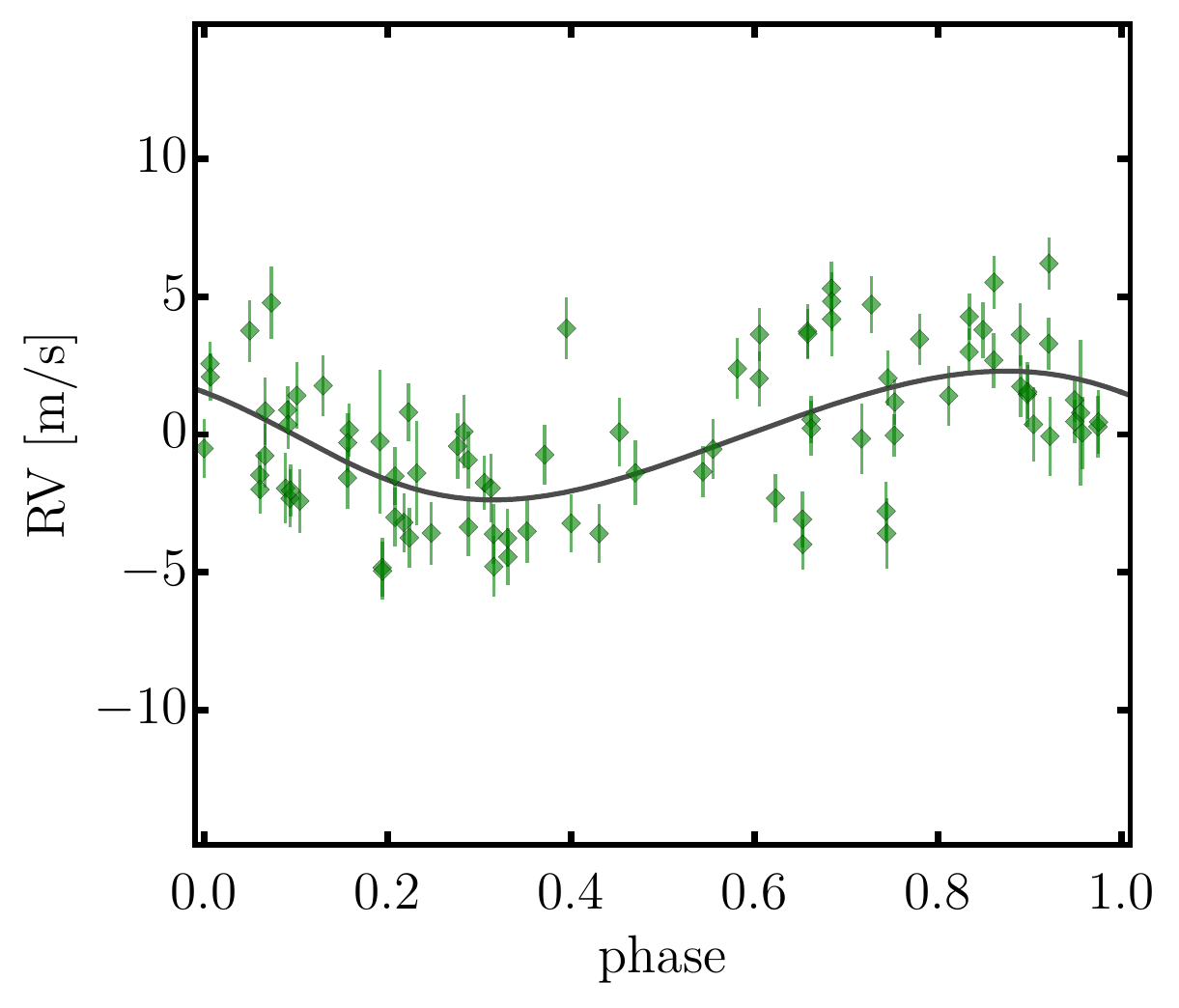} 
\put(-95,92){Jan 1997 - Dec 2009}\\
\includegraphics[width=4.5cm,height=3.8cm]{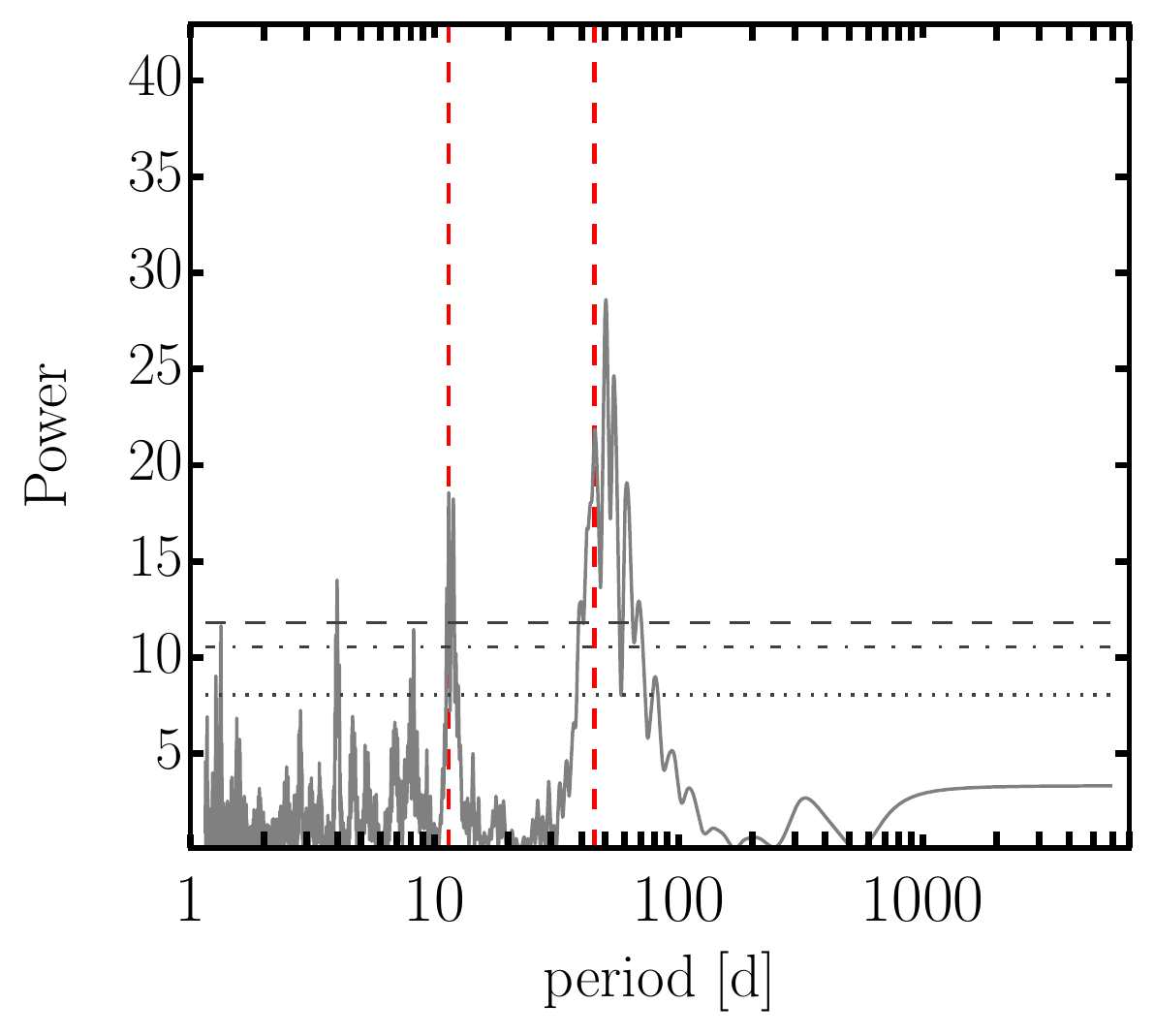}
\includegraphics[width=4.5cm,height=3.8cm]{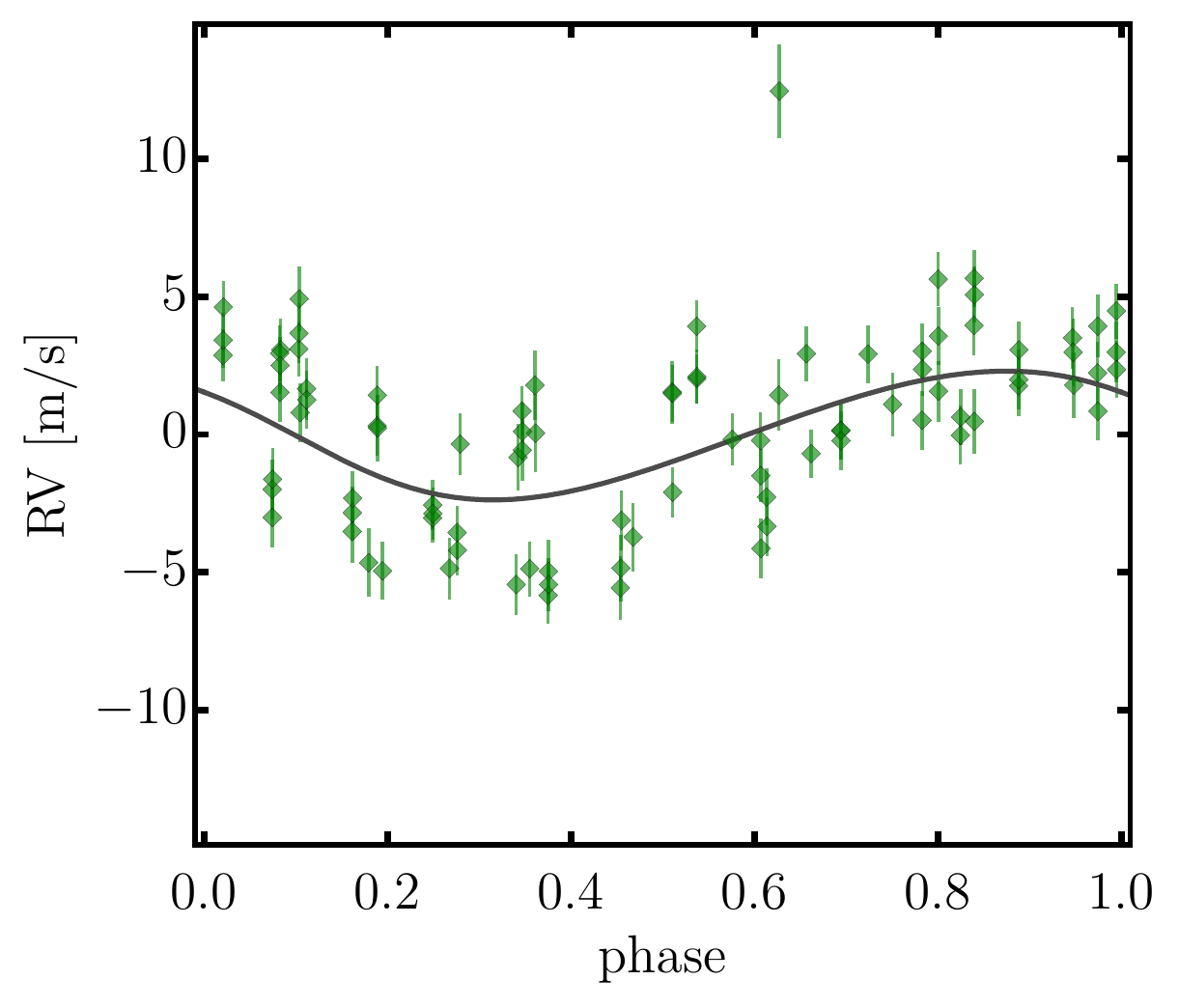} 
\put(-95,92){Jan 2010 - Sep 2011}\\
\includegraphics[width=4.5cm,height=3.8cm]{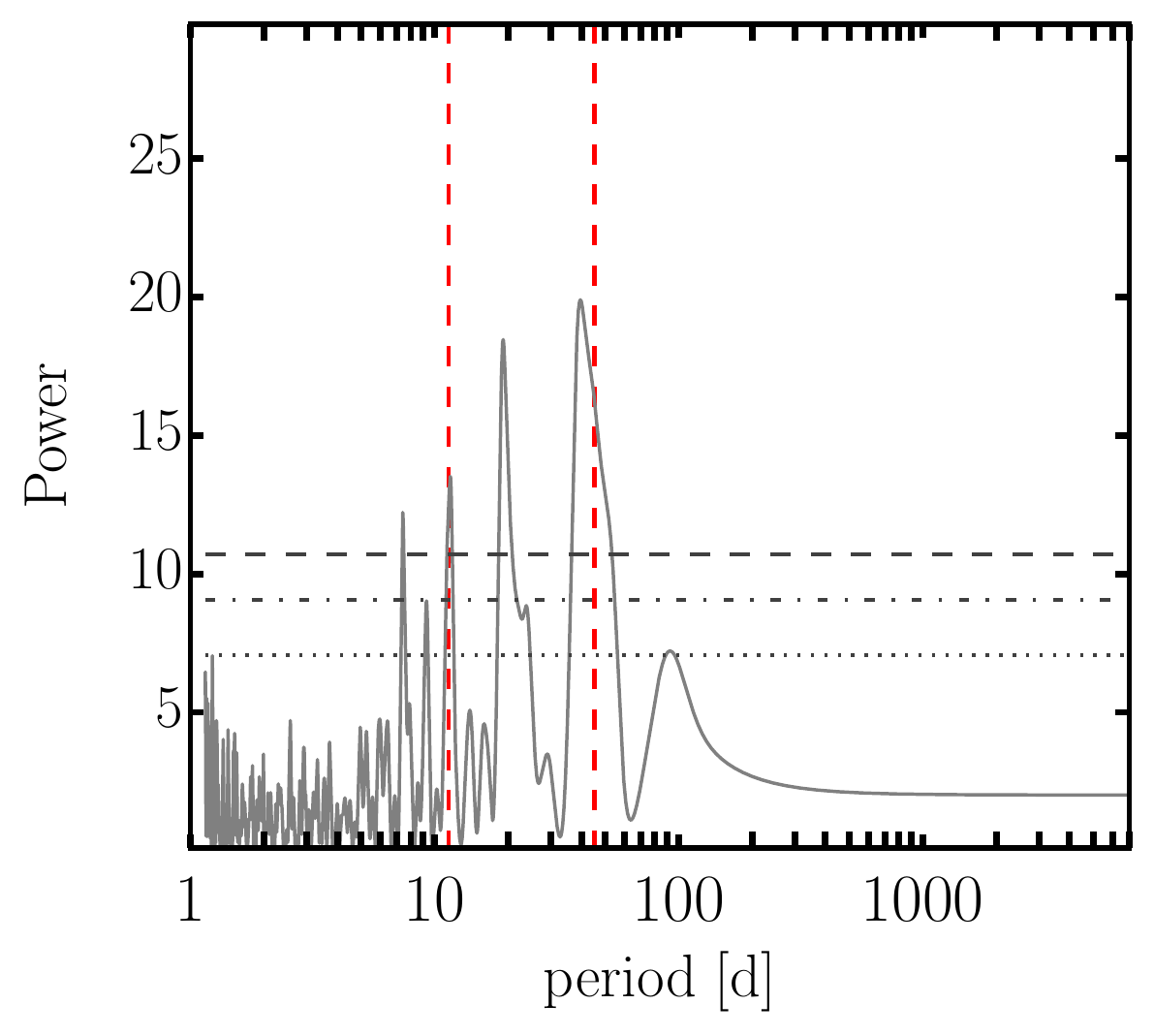}
\includegraphics[width=4.5cm,height=3.8cm]{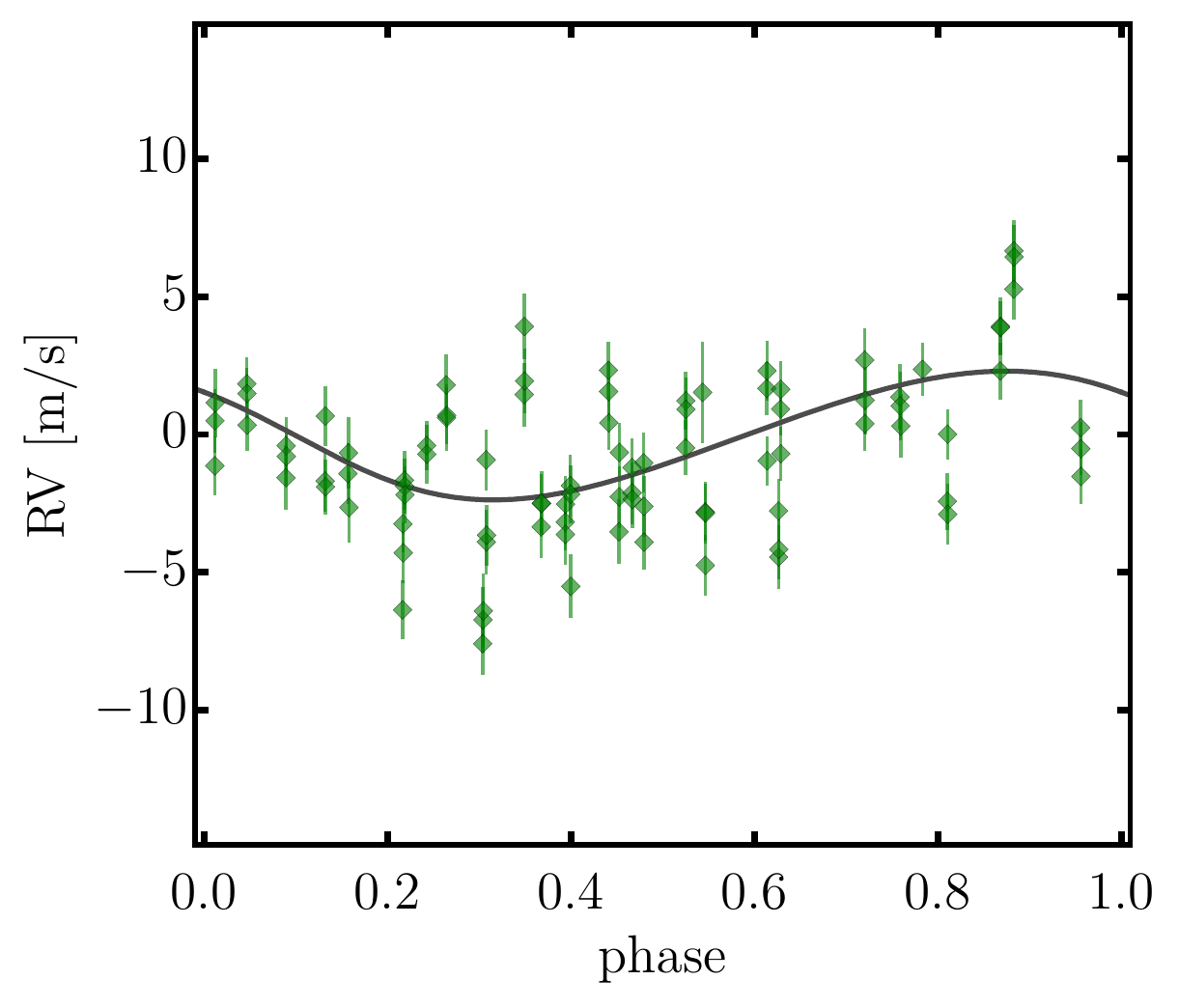} 
\put(-95,92){Aug 2011 - Dec 2011}\\
\includegraphics[width=4.5cm,height=3.8cm]{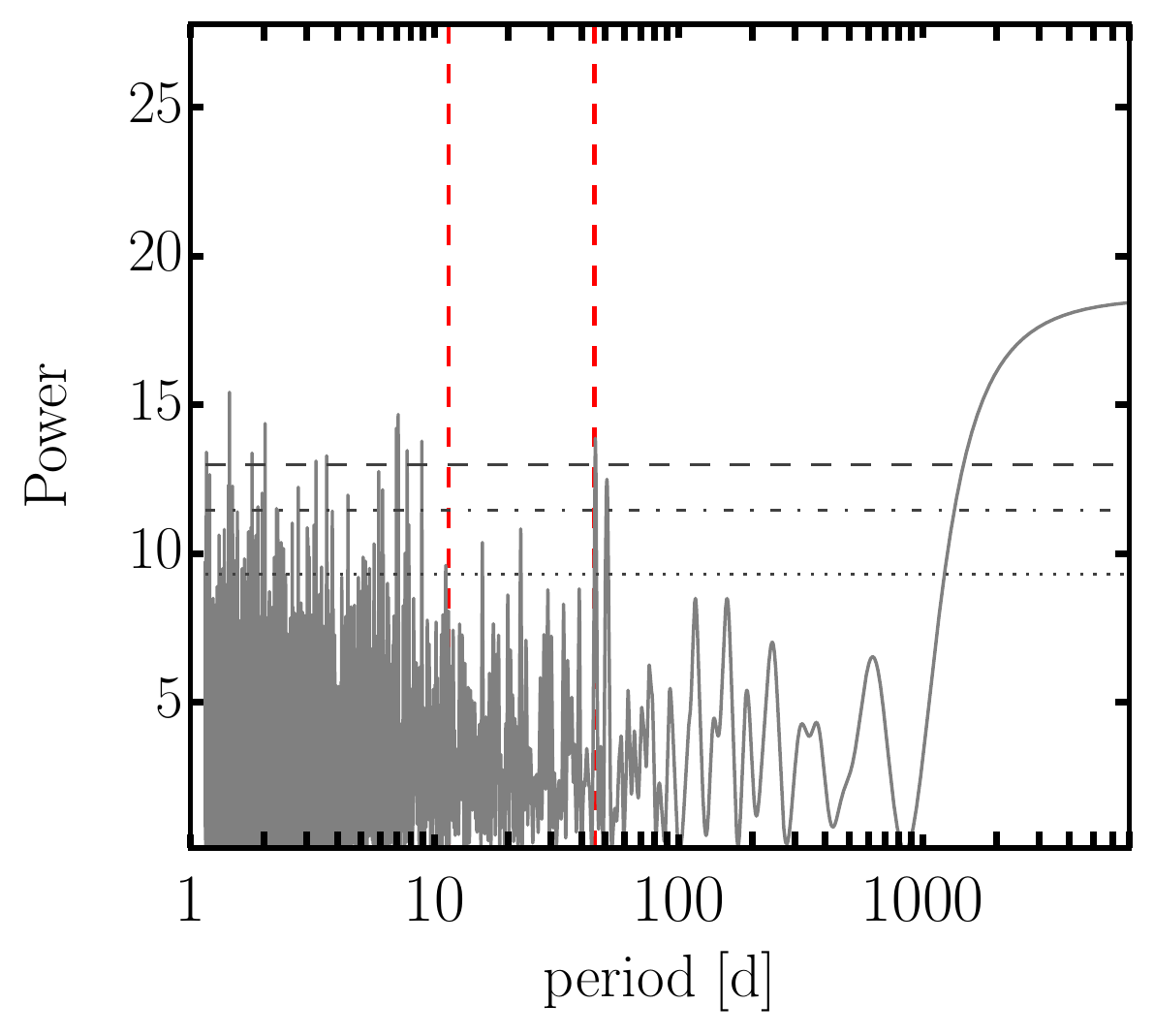}
\includegraphics[width=4.5cm,height=3.8cm]{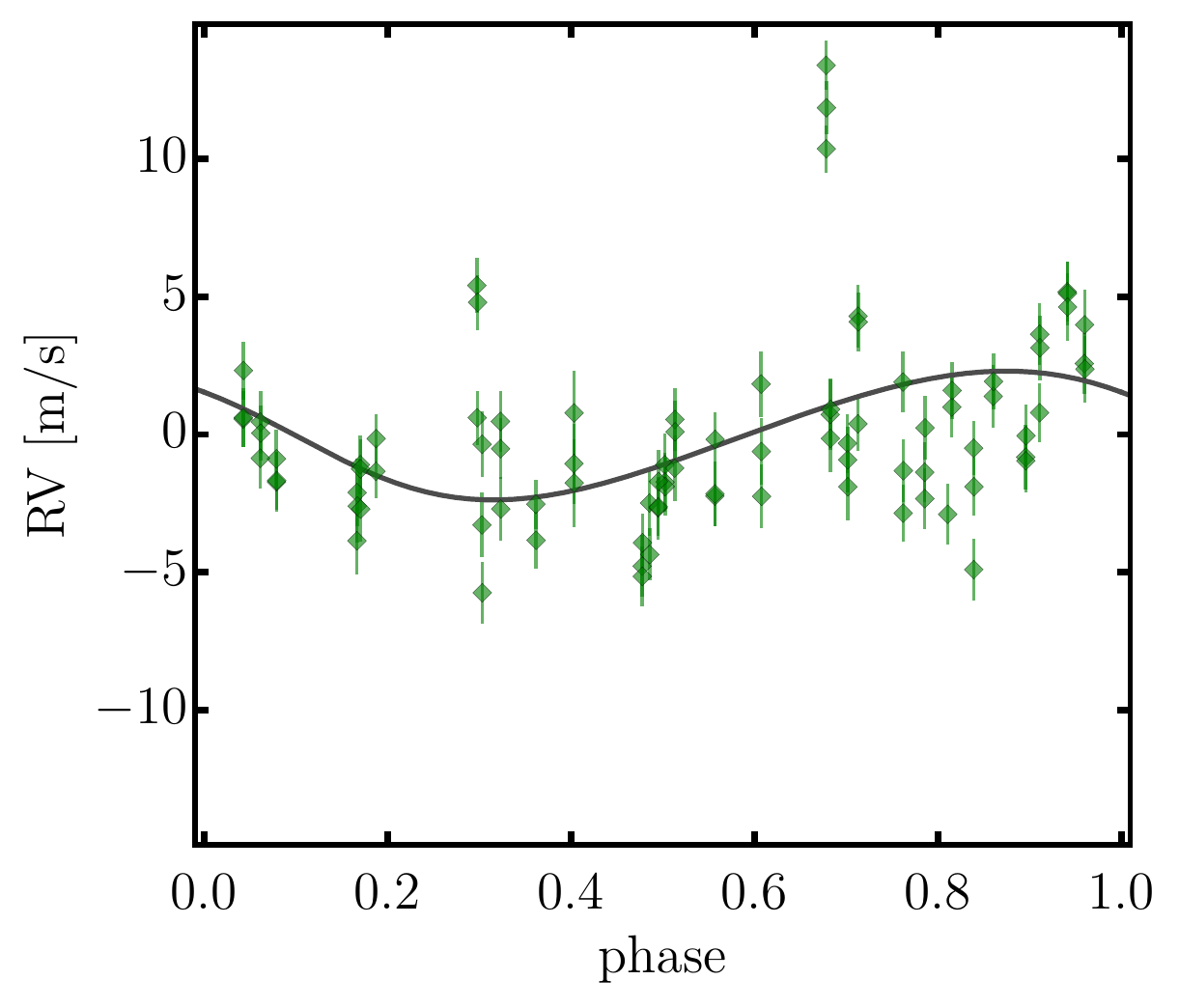} 
\put(-95,92){Jan 2012 - Dec 2014}\\
\end{array} $
\end{center}

\caption{The 358 HIRES RVs for GJ\,15\,A,  
separated into four subsets of 90, 90, 90, and 88 RVs, respectively, and analyzed individually.
The left panels show the GLS periodogram analysis for each of the four subsets,
while the right panels show the RV data phase-folded to the best Keplerian fit + a linear trend calculated using
the full set of HIRES data (see Fig~\ref{fig1}, panels a) and b)).
The first 90 RVs are taken over a 10 yr period and are consistent with a 11.44\,d,
a $\sim$40-day and other significant periodic signals. 
Between Jan. 2010 and Sep. 2011, GJ\,15\,A likely experienced a period of intensive activity 
leading to strong periodicity at $\sim$45\,d probably causing also the signal around 11.5\,d. 
The same is true for the 90 RVs taken between Aug 2011 and Dec 2011, which show 
periodicities at $\sim$40\,d, 20\,d, 11\,d and 8\,d.
The last three years of HIRES data, however, are not consistent with signal at 11.44\,d.
As shown in Fig~\ref{fig1} panels~c) and~d), at later 
epochs CARMENES is also not showing any periodic signal near~11\,d.
}   

\label{fig1.3} 
\end{figure}

\begin{figure}[htp]
\begin{center}$
\begin{array}{ccc}
\includegraphics[width=9cm]{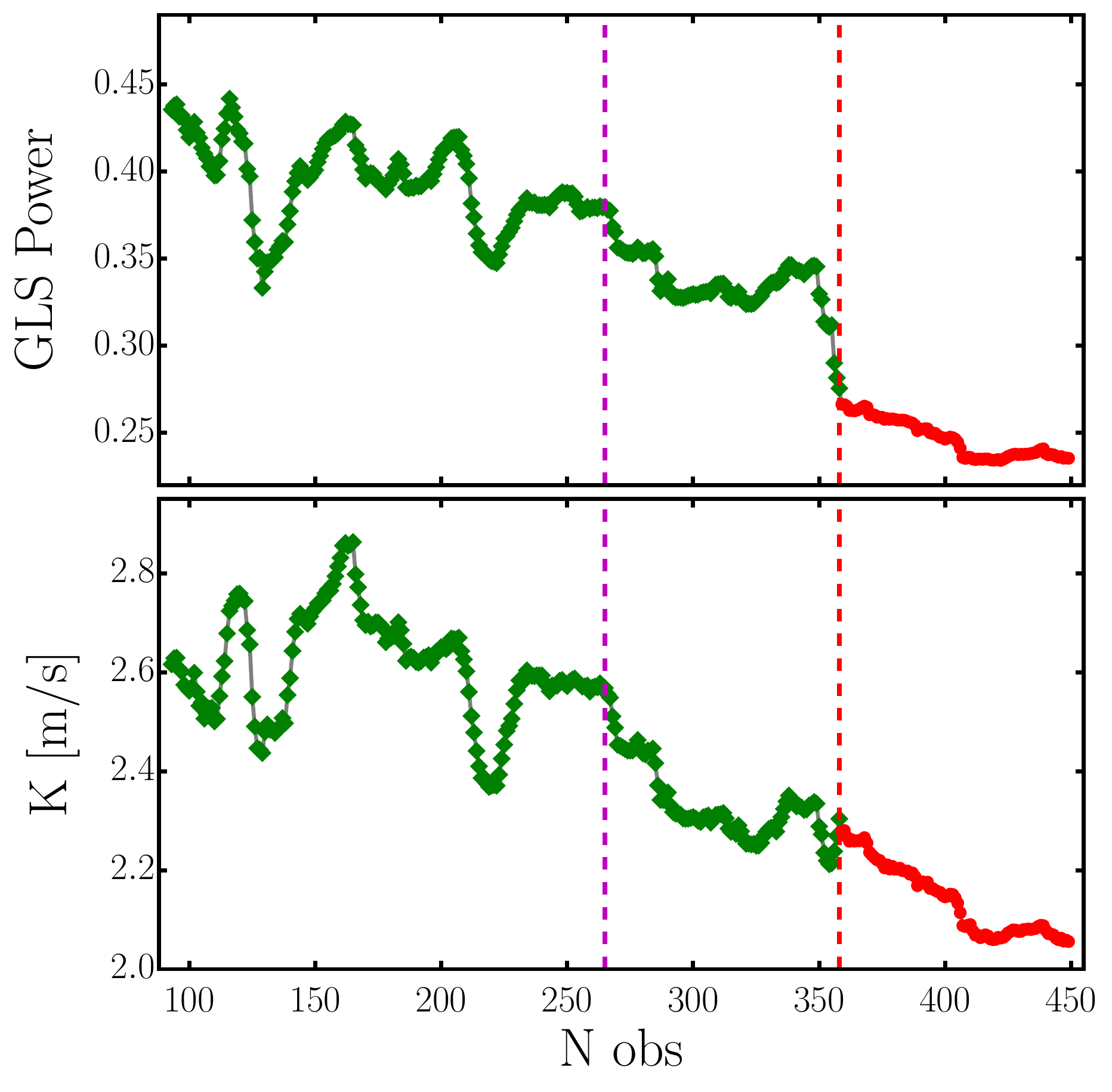} 
\put(-185,235){{\tiny HIRES}} \put(-195,228){{\tiny (pre-discovery)}} 
\put(-100,235){{\tiny HIRES}} \put(-115,228){{\tiny (post-discovery)}} \put(-55,235){{\tiny CARMENES}}

\end{array} $
\end{center}

\caption{GLS power and signal semi-amplitude as a function of  the number of data for the 
11.44\,d periodicity, supposed to be induced by a close-in super-Earth planet GJ\,15\,Ab. 
The magenta vertical line separates the pre- and post-discovery HIRES data for GJ\,15\,A,
while the red line indicates the beginning of the CARMENES data.
Even before the claimed planet discovery, the pre-discovery HIRES data 
seem to show fading power and semi-amplitude of the signal, and 
this continues when adding the newly published HIRES data.
The descending trend of the power continues when the CARMENES data are added,
which by themselves are not consistent with a periodic signal at 11.44\,d.
}   

\label{fig1.1} 
\end{figure}

The new finding in our study is the exceptionally small eccentricity 
of $e_{\rm d}$ = 0.082$_{-0.025}^{+0.043}$ for the innermost planet GJ\,876\,d.
In \citet{Rivera2010} the estimated eccentricity for GJ\,876\,d  was $e_{\rm b}$ = 0.257 $\pm$ 0.070,
which is more than 2$\sigma$ away from our estimate.
Deviations from the literature values presented in \citet{Rivera2010}, however,
can be expected as we use more data, larger observational baseline and 
a slightly larger stellar mass \citep[$M$ = 0.35~$M_\odot$ versus 0.32~$M_\odot$ used in][]{Rivera2010}.
We believe that our estimate for $e_{\rm d}$ makes sense because the GJ\,876\,d's 
close orbital separation of only $a_{\rm d}$ = 0.021 au to its host star 
is expected to cause significant tidal circularization of the orbit.
From  Fig.~\ref{fig876}, panel d)  planet d exhibits constant, nearly circular orbital evolution
and is practically unperturbed by the resonant chain of planets GJ\,876\,b, c and e.
Therefore, if GJ\,876\,d indeed settled in a nearly circular orbit, it
will not have a strong impact on the dynamical best fit solutions obtained in our study and the one by \citet{Rivera2010}.
Likely the reason for the significant $e_{\rm d}$ deviation from \citet{Rivera2010} is that 
by using more data and a longer baseline we  better constrain the resonant chain  and, thus,
we obtain more accurately the orbit of GJ\,876\,d.
We do not intend to revisit existing detailed discussions of 
the resonant architecture of the GJ\,876 system, but we note that 
the small $e_{\rm d}$ is intriguing and may be used as an input to
more intensive dynamical and formation studies of this system using all the available data.

\subsection{The peculiar case of GJ\,15\,A}
\label{33}
  
\subsubsection{The ``fading`` GJ\,15\,Ab}
\label{331}

We obtained a total of 174 precise CARMENES RV measurements for GJ\,15\,A between January 2016 and April 2017,
but most of  them were taken during technical nights, in which the main goal 
was to observe intensively several stars and test the nightly stability of the spectrograph.
To avoid data clustering from these intensive campaigns, 
we took three Doppler measurements at random to represent each of the technical nights.
Therefore, we present a total of 92 CARMENES RVs that, when combined with the 358 HIRES newly announced 
literature velocities \citep{Butler2017} spanning the time from 1997 to 2014,
comprise twenty years of precise RV measurements, providing better constraints on the orbital 
parameters of the proposed planetary companion GJ\,15\,Ab.

Figure~\ref{fig1} shows results from our RV analysis for GJ\,15\,A based on 
the HIRES and CARMENES data separately and when combined. 
Panel a) shows a GLS periodogram of the combined data, yielding
two strong periods at 11.44\,d and 45.46\,d, similar to those found by \citet{Howard2014}, 
who attributed them to the GJ\,15\,Ab planetary signal and stellar activity, respectively.
We attribute the broad long-period peak consistent with periods exceeding the combined 
HIRES and CARMENES temporal baseline (7307.525\,d) to the negative 
RV linear trend seen in the GJ\,15\,A RVs \citep{Howard2014}.
Indeed, a linear trend fit to the HIRES data alone, and to the combined data
yields an RV trend of $-$0.39 m\,s$^{-1}$\,yr$^{-1}$ (see first column of Table~\ref{table:15a}),
which removes all significant peaks beyond the $\sim$45\,d activity period seen in panels a) and b).
Since, GJ\,15\,A forms a common-proper motion pair with the M3.5\,V star GJ\,15\,B,
\citet{Howard2014} tentatively interpreted the linear trend as a small arc of the long-period binary orbit.
We follow a similar reasoning and in our Keplerian modeling of the proposed planet GJ\,15\,Ab 
we chose to simultaneously fit an additional linear term to the data.

Figure~\ref{fig1}, panel d) shows our one-planet best-fit to the combined data, phase-folded to GJ\,15\,Ab's best-fit period.
This fit is consistent with a planetary semi-amplitude of $K_{\rm b}$ = 2.13 m\,s$^{-1}$, an orbital period 
$P_{\rm b}$~=~11.441 days, eccentricity $e_{\rm b}$~=~0.093, and a linear trend of
$-$0.346 m\,s$^{-1}$\,yr$^{-1}$, corresponding to a planetary companion 
with a minimum mass of $m_{\rm b} \sin i$~=~4.1~$M_{\oplus}$ and a semi-major axis $a_{\rm b}$ = 0.074 au.
These orbital estimates and their asymmetric bootstrap uncertainties are listed in the second column of Table~\ref{table:15a}.
Our results yield conclusions similar to those by \citet{Howard2014},
showing that the combined data are consistent with the GJ\,15\,Ab planet and a linear trend.
Interestingly, the semi-amplitude of our model is lower than that in \citet{Howard2014},
who estimated $K_{\rm b}$ = 2.93 $\pm$ 0.29\,m\,s$^{-1}$, corresponding to a 
$m_{\rm b} \sin i$ = 5.3$M_{\oplus}$ super-Earth planet.

 In Fig.~\ref{fig1}, panels b) and c) we show an independent GLS search of the HIRES and the CARMENES data, respectively,
while panels e) and f) show the best-fit results for the two data sets, 
phase-folded with GJ\,15\,Ab's best-fit period from the combined data.
The RV analysis of only the HIRES data reveals that the results from the 
combined data shown in panels a) and d) are heavily dominated by HIRES. 
The 11.44 day planetary signal, the $\sim$ 45 day activity signal, and the 
remaining long-period peaks seen in panel a) are present in the HIRES RVs.
A Keplerian fit to the HIRES data alone also yields results similar to those obtained by the combined data.
A GLS periodogram of the CARMENES data alone, however, yields no significant peaks and lacks the 11.44\,d signal.
In fact, we find that our CARMENES data are more consistent with a flat model (i.e., no planet or activity present).
This is a peculiar result for a star with a large number of CARMENES visits.
Throughout the paper we have demonstrated that the CARMENES 
data for our other stars have very good agreement with the HIRES data and even better precision (see Section~\ref{sect.4}). 
Therefore, there is no reason to assume different circumstances for the particular case of GJ\,15A,
and we should have been able to detect the 11.44\,d signal in our 92 CARMENES RVs.

The question of whether the CARMENES data are consistent or not with HIRES data is addressed in Fig.~\ref{fig1.3}.
We split the 358 HIRES RVs into four subsets having approximately the same number of data points as the CARMENES 
data set: 90, 90, 90 and 88 RVs, respectively.
For each of those subsets we searched for significant GLS signals. 
The left panels in Fig.~\ref{fig1.3} show the GLS periodograms with 
horizontal lines showing the bootstrapped FAP levels of 10\%, 1\% and 0.1\%,
while the vertical lines show the 11.44\,d and 45.46\,d peaks seen in the total HIRES data set.
In the right panels in  Fig.~\ref{fig1.3}, for illustrative purposes,
we fold each data sub-set with the best period obtained for the total HIRES data and overplot
the best-fitting Keplerian signal, also for the total HIRES data (see Fig~\ref{fig1}, panels a) and b)~).
The first data set of 90 RVs was obtained between January 1997 and December 2009 and yields several significant peaks, 
one at 11.44\,d, two near 20\,d, and one at ~40\,d. 
Between January 2010 and September 2011, we can still see a significant peak near $\sim$11.5 days, but
the strongest data peak is now near 50.4\,d, followed by 54.3\,d and 45.4\,d.  
Likely GJ\,15\,A experienced an epoch of intense quasi-periodic activity that caused additional peaks near the 50 day peak.
The 90 RVs taken between August 2011 and December 2011 show significant GLS peaks at 39.6\,d and 19.1\,d (probably a harmonic), 
also likely induced by stellar activity and followed by another peak at 11.6\,d,  potentially 
due to the planet candidate, and a 7.4\,d peak, which is likely an alias.
The last three years of HIRES  data (Jan. 2012 -- Dec. 2014) do not provide any evidence for a planetary signal at 11.44\,d,
which is consistent with what we find based on the CARMENES follow-up data alone.
We note in passing that we inspected the S- and H-indices, which also show periodogram
peaks at several of the values found in the RVs, such as 44\,d and 11.4\,d.

 \begin{figure*}[tp]
\begin{center}$
\begin{array}{cc} 
\includegraphics[width=9cm]{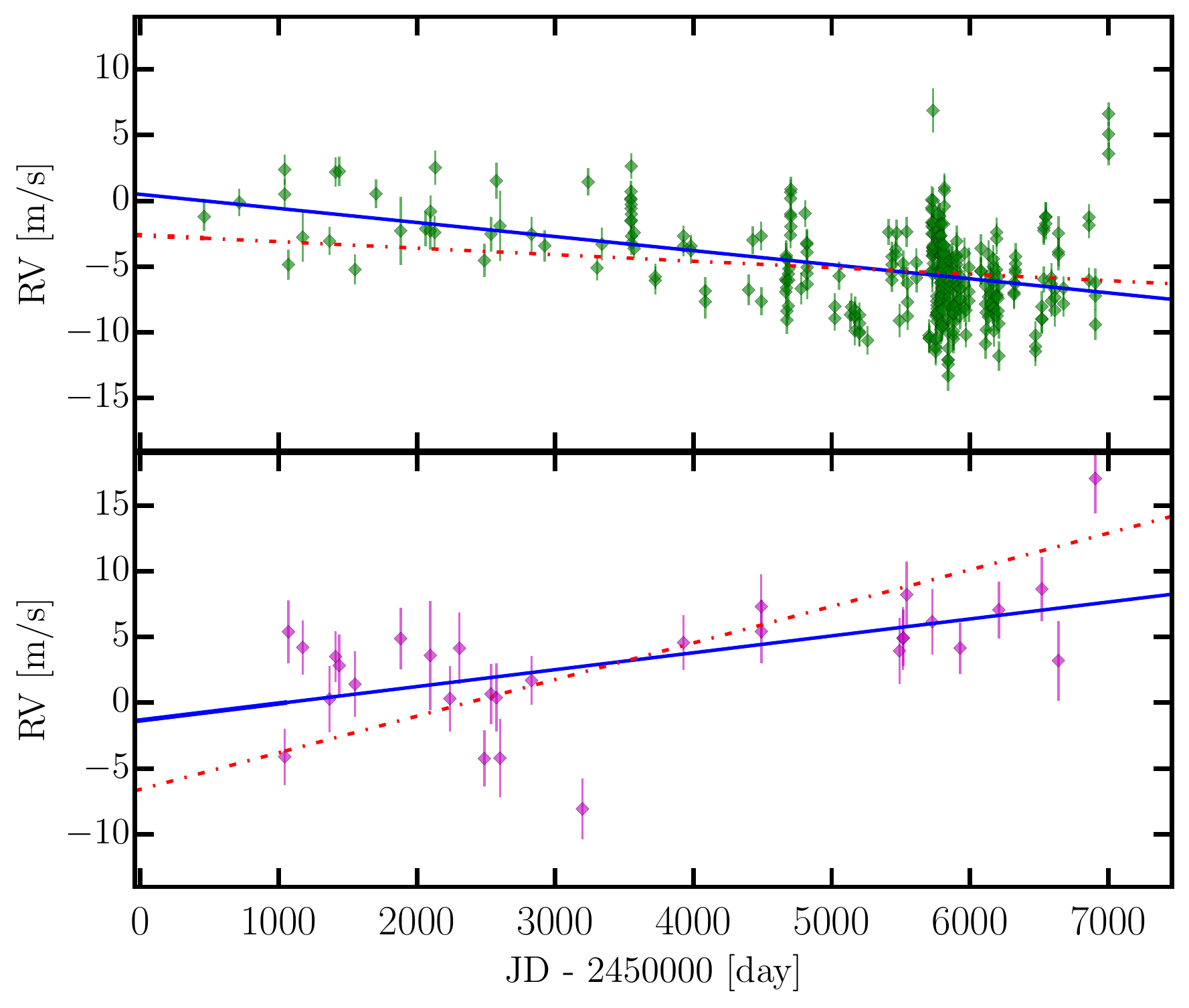}
\put(-180,195){GJ\,15\,A} \put(-220,195){a)}
\put(-180,105){GJ\,15\,B}
\includegraphics[width=9cm]{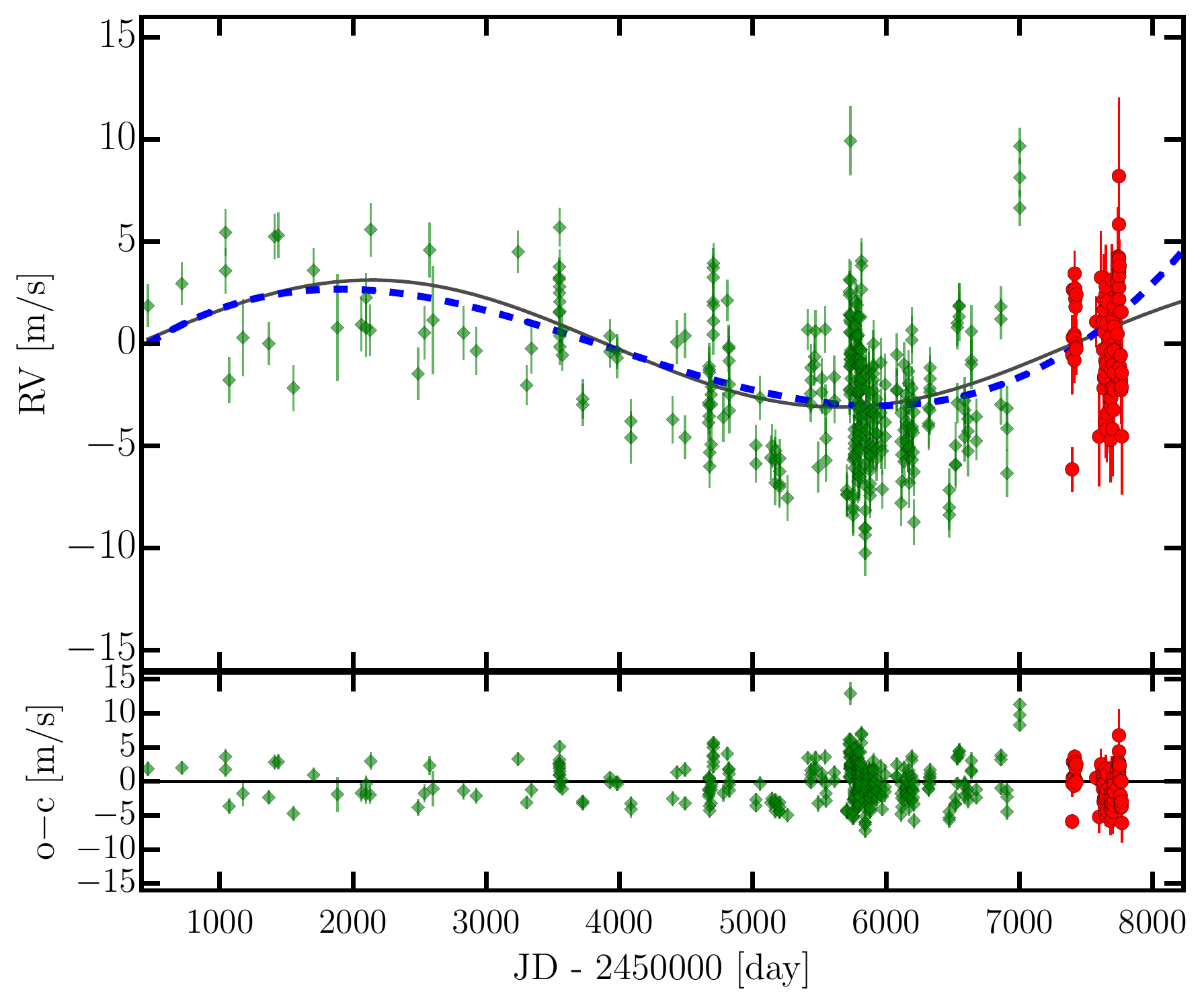}
\put(-150,195){GJ\,15\,Ac} \put(-220,195){b)}
\end{array} $
\end{center}

\caption{
Panel a) From top to bottom, HIRES RV data for GJ\,15\,A and GJ\,15\,B 
modeled with a linear trend. Blue continuous lines represent the best 
linear trend model of the GJ\,15\,A and B datasets, while the red dot-dashed lines
show the  linear trend expected from the mass ratio of the two stars under the assumption that the 
other binary companion has the correct trend estimate.
Panel b) Time series RVs for GJ\,15\,A obtained with  HIRES (green diamonds)
and CARMENES (red circles). The fit to the data (black solid line) is a suggestive 
Keplerian model consistent with a long-period Saturn mass planet on a circular orbit. 
The dashed blue line represents a third order polynomial approximation
to this orbit, which is needed for the statistical analysis (see text).
The lower panel shows the residuals to the circular fit.
A long-period planet yields a significant improvement over a simpler model fitting only a linear trend. 
}   

\label{fig11} 
\end{figure*}
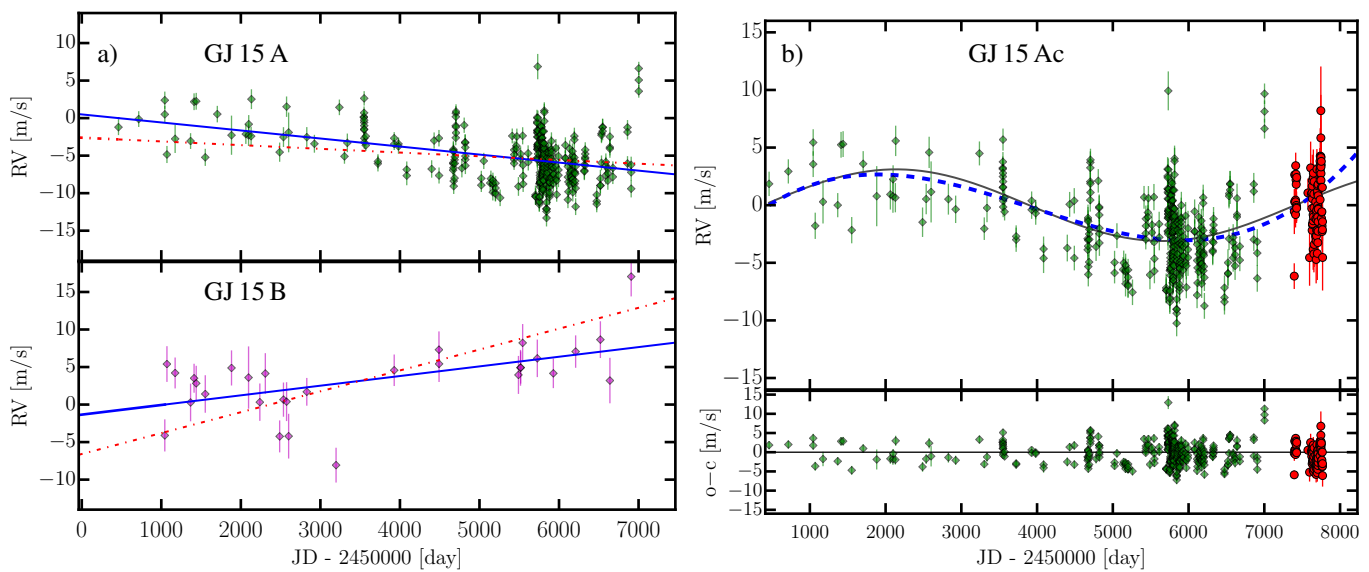

Further, we systematically analyze the GLS power spectrum for the 11.4-day period
as a function of the number of observations, the results of which are shown in Fig.~\ref{fig1.1}.
We note that instead of the raw Scargle power for this test we adopted the 
original GLS power formalism, which defines the power as 
$p(\omega)$ = ($\chi_{\rm flat}^2 - \chi_{\omega}^2$) /  $\chi_{\rm flat}^2$,
where $\chi_{\rm flat}^2$ is the $\chi^2$ of a flat model applied to the data, while
$\chi_{\omega}^2$ is the $\chi^2$ of a sinusoidal  model having frequency $\omega$.
Here p($\omega$) is the relative $\chi^2$ improvement in a comparison of the two models.
It is close to zero when the sinusoidal model does not represent an improvement
over the flat model, while is close to unity in the case of a strong improvement
\citep[for more details, see][]{Zechmeister2009}.
Thus, the GLS power is a very informative quantity to study whether a given frequency in the data 
arises due to a signal rather than from noise. 
In particular, with our $p(\omega)$ vs. N$_{\rm data}$ test we probe 
if by adding more data we increase the evidence for the GJ\,15\,Ab signal or if we just add noise.
We started with the first 90 RVs available for GJ\,15\,A
as after this subset of data  (N$_{\rm data}$ > 90) the 11.4\,d peak was the strongest in the GLS,
and we kept adding data and recomputing the power spectrum.
Interestingly, the GLS power and semi-amplitude decrease even before the announcement of the planet discovery.
The power keeps decreasing when we add the CARMENES RVs.

At the moment, we are puzzled by the strong periodicity seen in the early 
HIRES data and the absence of this signal at later HIRES epochs and in our CARMENES data.
It seems that we are able to construct a Keplerian fit for GJ\,15\,A 
thanks to the contribution of the HIRES data obtained at earlier epochs prior to the planet announcement. 
More data taken at later epochs are consistent with noise and 
the semi-amplitude $K_{\rm b}$ of the planet signal decreases monotonously.
Currently, the weighted $rms_{\rm HIRES}$ scatter of the Keck data for GJ\,15\,A after removing the planet signal is about 2.61 m\,s$^{-1}$.
This value is actually larger the $rms_{\rm CARMENES}$ = 2.40\,m\,s$^{-1}$ of the CARMENES data without removing any signal.

We conclude that the CARMENES data show no evidence for the planet's existence, nor do the post-discovery HIRES data.
However, the HIRES Doppler data prior to the planet announcement do show a significant 11.44\,d periodicity. 
Incidentally, the 11.44\,d signal seen in the HIRES data is 
almost exactly 1/4 of the second strongest GLS peak at 45.46 days (see Fig.~\ref{fig1}). 
If the latter is due to the  stellar rotation period proposed by \citet{Howard2014},
then the 11.44 day signal could be an overtone of the activity.

\begin{table*}[ht]

\centering  
  

\caption{Best fit Keplerian parameters for GJ\,15\,A constructed based 
on the combined CARMENES and HIRES RVs.}   
\label{table:15a}  

\centering      
\begin{tabular}{l r r r r r r r r}     
\hline 
\noalign{\vskip 0.5mm}
\hline
\noalign{\vskip 0.5mm}

\makebox[0.1\textwidth][l]{\hspace{49 mm}Only      \hspace{15 mm}One-planet\hspace{8.5 mm} Long-period\hspace{9.5 mm} } \\
\makebox[0.1\textwidth][l]{\hspace{45 mm}lin. trend\hspace{14 mm}fit update$^a$\hspace{10 mm}planet fit$^b$\hspace{19 mm} Two-planet fit$^b$} \\

\cline{2-2}\cline{4-4}\cline{6-6}\cline{8-8}\noalign{\vskip 0.9mm} 


Orb. param.\hspace{22.5 mm} &\hspace{8.5 mm} && GJ\,15\,Ab & & GJ\,15\,Ac  &&GJ\,15\,Ab\hspace{9.5 mm}GJ\,15\,Ac     \\     
\hline 
\noalign{\vskip 0.9mm}
%
$K$  [m\,s$^{-1}$]                &\dots & & 2.13$_{-0.10}^{+0.27}$     & & 3.11$_{-0.30}^{+0.36}$       && 2.05$_{-0.10}^{+0.25}$\hspace{10.5 mm}2.92$_{-0.33}^{+0.37}$     \\  \noalign{\vskip 0.9mm}
$P$ [d]                           &\dots & & 11.441$_{-0.002}^{+0.004}$ & & 7024.8$_{-628.6}^{+972.0}$   && 11.443$_{-0.002}^{+0.003}$\hspace{4.5 mm}7837.6$_{-920.9}^{+1401.4}$   \\  \noalign{\vskip 0.9mm}
$e$                               &\dots & & 0.093$_{-0.010}^{+0.152}$  & & 0.0 (fixed)                  && 0.137$_{-0.032}^{+0.124}$\hspace{8.3 mm}0.0 (fixed)     \\  \noalign{\vskip 0.9mm}
$\varpi$ [deg]                    &\dots &  & 106.4$_{-35.7}^{+131.5}$   & & 0.0 (fixed)                  && 114.1$_{-36.6}^{+69.4}$\hspace{8.3 mm}0.0 (fixed)      \\  \noalign{\vskip 0.9mm}
$M$ [deg]                         &\dots & & 305.9$_{-38.4}^{+112.1}$   & & 274.1$_{21.6}^{+22.1}$       && 321.6$_{-272.5}^{+15.5}$\hspace{9.0 mm}303.4$_{-29.8}^{+20.5}$     \\ \noalign{\vskip 0.9mm}

$a$ [au]                          &\dots & & 0.074$_{-0.001}^{+0.001}$  & & 5.351$_{-0.356}^{+0.445}$   && 0.074$_{-0.001}^{+0.001}$\hspace{8.0 mm}5.756$_{-0.494}^{+0.626}$   \\  \noalign{\vskip 0.9mm}
$m  \sin i$  [$M_\oplus$]         &\dots & & 4.144$_{-0.309}^{+0.428}$  & & 51.77$_{-5.76}^{+5.47}$   && 3.98$_{-0.29}^{+0.38}$\hspace{9.0 mm}50.35$_{-6.78}^{+6.88}$   \\  \noalign{\vskip 3.9mm} 

$\gamma_{\rm HIRES}$~[m\,s$^{-1}$] &5.53$_{-0.78}^{+0.67}$             & & 4.90$_{-0.89}^{+0.80}$    & & 2.47$_{-0.33}^{+0.26}$     &  \makebox[\dimexpr(\width-9em)][l]{2.42$_{-0.37}^{+0.29}$} & \\ \noalign{\vskip 0.9mm}
$\gamma_{\rm CARM.}$~[m\,s$^{-1}$] &9.60$_{-1.22}^{+1.10}$             & & 8.14$_{-1.45}^{+1.44}$    & & 1.40$_{-0.96}^{+1.29}$   & \makebox[\dimexpr(\width-9em)][l]{1.40$_{-1.10}^{+1.44}$}& \\  \noalign{\vskip 0.9mm}

Trend  [m\,s$^{-1}$\,yr$^{-1}$]   &$-$0.391$_{-0.053}^{+0.059}$ & & $-$0.346$_{-0.062}^{+0.067}$  & &  \dots  &\makebox[\dimexpr(\width-9em)][l]{\dots} &                  \\ \noalign{\vskip 3.9mm}

$rms$~[m\,s$^{-1}]$               &2.98 &     & 2.60                       & & 2.89   &\makebox[\dimexpr(\width-9em)][l]{2.53}&                     \\ \noalign{\vskip 0.9mm}
$rms_{\rm HIRES}$~[m\,s$^{-1}]$   &3.07 &     & 2.62                       & & 2.97   &\makebox[\dimexpr(\width-9em)][l]{2.52}&                     \\ \noalign{\vskip 0.9mm}
$rms_{\rm CARM.}$~[m\,s$^{-1}]$   &2.40 &     & 2.53                       & & 2.44   &\makebox[\dimexpr(\width-9em)][l]{2.59}&                     \\ \noalign{\vskip 0.9mm}

$\chi_{\nu}^2$                   &7.22 &  & 5.54                       & & 6.71   &\makebox[\dimexpr(\width-9em)][l]{5.18}&                   \\ \noalign{\vskip 0.9mm}

Valid for                &                           &&  && &&    \\  \noalign{\vskip 0.9mm}

 $T_0$ [JD-2450000]      &   \makebox[\dimexpr(\width-12em)][l]{461.771}                &&  && &&    \\  \noalign{\vskip 0.9mm}
   
\noalign{\vskip 0.5mm}
\hline

\end{tabular}

 \tablefoot{
 a - GJ\,15\,Ab planet is doubtful,  b - these two fits including a long-period circular planet 
 are only suggestive. 
 }

\end{table*}

\subsubsection{A long-period planet companion?}

As we mentioned in Section~\ref{331}, 
the long-period orbital motion of the GJ\,15\,AB binary 
is a plausible explanation of the possible linear RV trend seen in the HIRES data for GJ\,15\,A.
We measured a trend of $-$0.39 $\pm$ 0.02 m\,s$^{-1}$\,yr$^{-1}$ for GJ\,15\,A,
while given the binary mass ratio $\mu$\,=\,0.391 $\pm$ 0.042 (for GJ\,15\,B we 
derived a mass of $M$ = 0.162 $\pm$ 0.016 $M_\odot$), 
we would also expect a positive trend of 1.00 $\pm$ 0.05 m\,s$^{-1}$\,yr$^{-1}$ in the RV data of GJ\,15\,B. 
Fortunately, this can be tested as the literature HIRES data for GJ\,15\,B
consist of 30 RVs with a similar temporal base line as for GJ\,15\,A.
In Fig.~\ref{fig11} panel\,a) we show the HIRES RV time series for GJ\,15\,A and GJ\,15\,B.
With blue continuous lines we plot their best fit linear trend model, 
while with red dot-dashed lines we plot their predicted trend assuming 
the respective other binary companion has the correct trend estimate. 
In the sparse HIRES data of GJ\,15\,B we indeed measured a 
marginally significant positive trend of 0.47 $\pm$ 0.08 m\,s$^{-1}$\,yr$^{-1}$, which 
is a factor of two smaller than expected, but is $\sim$ 6$\sigma$ away from the predicted value.
Therefore, given the binary mass ratio of GJ\,15\,AB the observed trends are not both 
compatible with the binary orbit at the same time.

To our knowledge, preliminary orbits of the visual binary GJ\,15\,AB were calculated
only by \citet{Lippincott1972} and \citet{Romanenko2014}.
Due to the long period these astrometric observations covered only short arcs of the orbit. 
The binary orbital solution proposed by \citet{Lippincott1972} assumed an
eccentricity $e$ = 0 and argument of periastron $\omega$ = 0$^\circ$, and
implies a period of $P$ = 2600 yr, inclination $i$\,=\,61$^\circ$, ascending node $\Omega$ = 45$^\circ$ and 
time of periastron passage $t_{\rm 0}$ = A.D. 1745.
Using these orbital parameters, and considering that the astrometric solutions have 180$^\circ$ ambiguity in $\Omega$
we derive a trend of $\pm$0.63 m\,s$^{-1}$\,yr$^{-1}$ for the orbital phase where the GJ\,15\,A HIRES data are obtained.
This is in a reasonable agreement with the trend we determine for GJ\,15\,A, but inconsistent with the trend for GJ\,15\,B.
The more recent binary solution from \citet{Romanenko2014} yields $P$ = 1253 yr, $e$ = 0.59,
$\omega$ = 331$^\circ$, $i$ = 46$^\circ$, $\Omega$ = 234$^\circ$ and $t_{\rm 0}$ = A.D. 2327, from which 
we derived an RV trend for GJ\,15\,A of about $\pm$ 0.1\,m\,s$^{-1}$
for the time of the HIRES observations.

If we assume that \citet{Romanenko2014} provided the more realistic binary orbit and the 
expected mutual RV acceleration between GJ\,15\,A and B is small at present epochs, 
then the HIRES data for GJ\,15\,A yield the interesting possibility of 
a long-period orbital motion of a sub-stellar companion.
In Fig.~\ref{fig1}, panel b) we showed that the HIRES RV measurements 
for GJ\,15\,A are consistent with a 6997\,d significant signal,
which can be modeled well with a low-amplitude, long-period sine-like velocity curve (see Fig.~\ref{fig11}).
We note that in this case no trend is included in the model\footnote{However, 
if the measured marginally significant trend for GJ\,15\,B is 
real then a smaller trend is expected for GJ\,15\,A (see Fig.~\ref{fig11}a ),
which would lower the amplitude of the long period signal}.

To test whether one can indeed make a good case for a long-period planet, we 
fit the combined data with a long-period circular Keplerian 
term (i.e., we fixed $e$ = 0, $\varpi$ = 0$^\circ$, since at this point we cannot 
provide meaningful constrains for these parameters based on the available data).
In Fig.~\ref{fig11}, panel b) we illustrate the results from this test.
The HIRES and the CARMENES RV time series spanning over twenty years are 
well modeled (black sine curve) with a Saturn-mass planet at $a$ = 5.35 au ($P$ = 7024.8\,d).
Orbital parameters and uncertainties for this 
long-period circular planet are shown in  the third column in Table~\ref{table:15a}.
This suggestive fit has $\chi_{\nu}^2$ = 6.71 and  $rms$ = 2.89\,m\,s$^{-1}$,
while by fitting only a linear trend we obtain $\chi_{\nu}^2$ = 7.22 and $rms$ = 2.98\,m\,s$^{-1}$.
Nonetheless, the statistical comparison between a linear trend model and a sine model is complicated by the 
fact that they are not nested within each other.
Therefore, our adopted F-test approach from \citet{Bevington2003}  
is not a appropriate test to validate whether a circular Keplerian indeed leads to a significant improvement
(both models have $\gamma_{\rm HIRES}$, $\gamma_{\rm CARM.}$, but 
each model is constructed of parameters that the other model does not have, e.g., $P$, $K$, $M_0$ v.s. $a_0$$t$).
Nested models can be obtained, however, by extending a
simple slope model to a third-order polynomial, which like the sine fit has five fitting parameters 
($\gamma_{\rm HIRES}$, $\gamma_{\rm CARM.}$ + $a_0$$t$ + $a_1$$t^2$ + $a_2$$t^3$)
and for the temporal extent of the available data is a very good approximation of a circular Keplerian model.

A third order polynomial fit to the combined data has $\chi_{\nu}^2$ = 6.73 and $rms$ = 2.92\,m\,s$^{-1}$ 
and shows a very good agreement with the adopted long-period circular planetary model
 (see Fig.~\ref{fig11}, panel b), blue dashed curve). An F-test yields a FAP = 5.4$\times$10$^{-8}$
that the two additional fitting parameters, which approximate the sine fit significantly improve the linear trend model.
Thus we concluded that a circular Keplerian fit is justified, and perhaps 
indicates the existence of a long-period planet around GJ\,15\,A.

Including the 11.44\,d signal representing the putative planet GJ\,15\,Ab leads to similar results.
Fitting for the GJ\,15\,Ab planet simultaneously on the one hand with a linear trend
and on the other, with a third order polynomial
suggests that the latter is better with a FAP of 1.7$\times$10$^{-7}$. 
A suggestive two-planet Keplerian fit including the short period planet GJ\,15\,Ab
and the possible long-period circular planet GJ\,15\,Ac is shown in the last two columns of Table~\ref{table:15a}. 
Unfortunately, assuming that a long-period planet is indeed the right model 
does not improve the detectability of the putative GJ\,15\,Ab planet. 
We repeated the same test as in Section~\ref{331}, but
including the determined long-period circular model instead of the linear trend
and once again the latest HIRES and new CARMENES data did not strengthen the case for the close-in planet.

For GJ\,15\,A our CARMENES data cover only about fourteen months, or 
$\sim$5\% of the putative long-period planetary orbit and they currently cannot provide 
strong evidence for the existence of the outer planet.
In addition, as can be seen from Fig.~\ref{fig11}, the CARMENES data 
do not overlap with the HIRES data and, thus, their mutual RV offset parameter is not well constrained.
Therefore, the constraints on the long-period fit are poor and currently rely only on the HIRES data. 
We plan to continue our CARMENES monitoring of GJ\,15\,A, which will allow us to extend our temporal baseline 
and see if our CARMENES data are consistent with a positive RV trend 
as expected from the existence of a long-period planet around GJ\,15\,A. 
Until then, the question whether GJ\,15\,A has two, one,  or zero planets remains open.

\section{CARMENES vs. HARPS and HIRES}
\label{sect.4}

\begin{figure}[tbp]
\begin{center}$
\begin{array}{cc} 
\includegraphics[width=9cm]{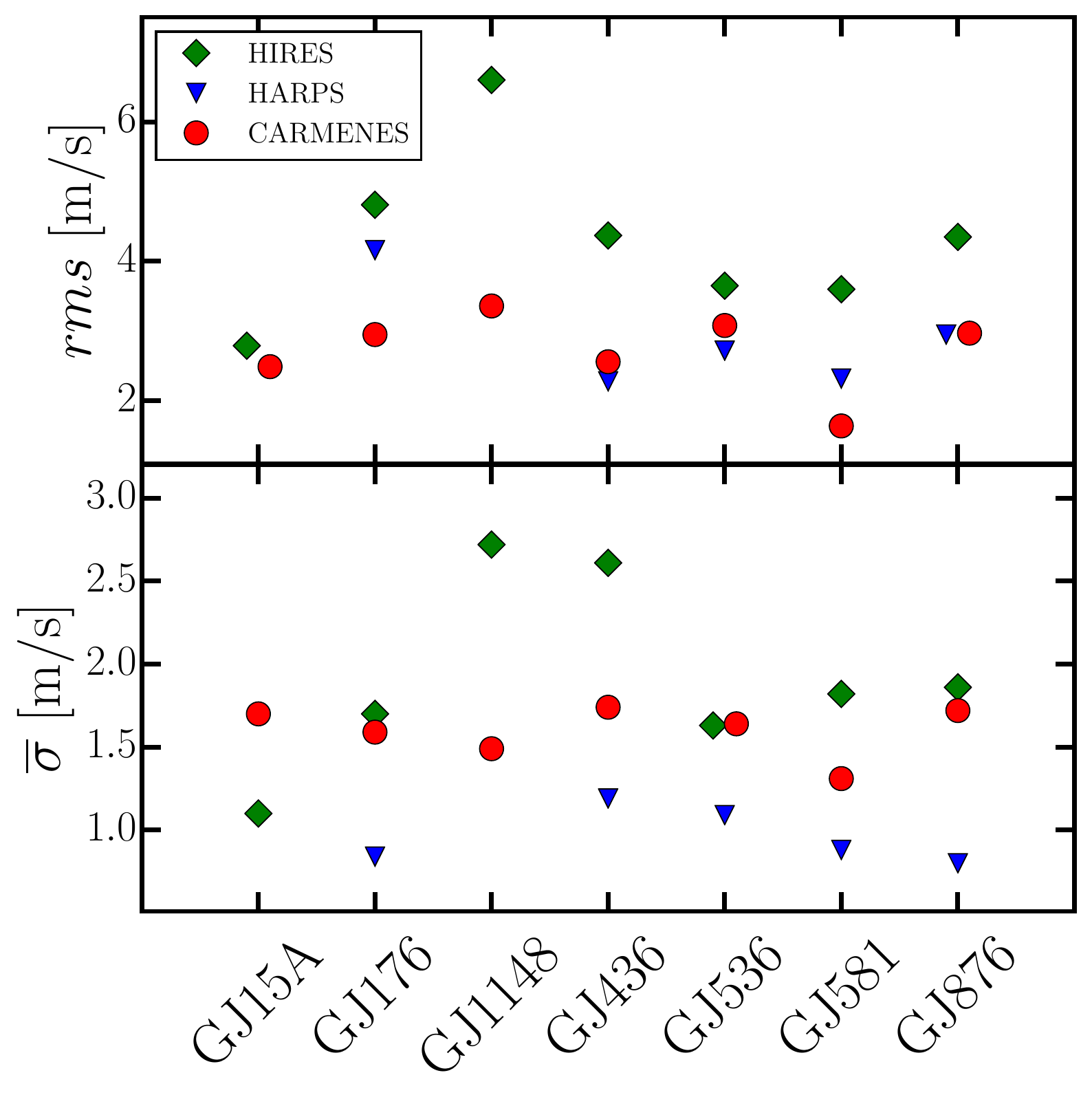}
\end{array} $
\end{center}

\caption{ 
Weighted mean $rms$ scatter around the best fits (top) and mean formal 
RV uncertainties (bottom) for HARPS, HIRES and CARMENES.
The CARMENES data are more precise and have a lower scatter when compared to HIRES.
The internal errors of HARPS are usually smaller than CARMENES, but comparable in terms of RV scatter.
The $rms$ scatter for GJ\,15\,A and GJ\,1148 is averaged from
the orbital solutions for these two targets given in 
Table~\ref{table:15a} and Table~\ref{table:1148}, respectively.
}   

\label{fig9} 
\end{figure}

We compare the $rms$ velocity dispersion around the best fit models  
for the three largest  RV data sets used in this paper, namely CARMENES, HARPS and HIRES.
Figure ~\ref{fig9} compares the external dispersion (weighted $rms$)
and typical internal uncertainties between the three data sets.
CARMENES and HIRES data are available for all seven stars, while 
GJ\,15\,A and GJ\,1148 were not observed with HARPS.

The CARMENES optical velocities have similar RV precision and 
overall scatter when compared to the RV measurements conducted with HIRES and HARPS.
In fact, our RV analysis shows that CARMENES data are more precise and have
smaller formal RV errors than the HIRES RVs. The only exception where the
HIRES data seem to have smaller RV errors than CARMENES is GJ\,15\,A. 
CARMENES however, always shows better weighted $rms$ results to the fits.
The formal RV errors of HARPS are usually lower than those from CARMENES, 
but in terms of RV scatter they seem to be comparable.
The two cases where CARMENES shows a larger scatter compared to HARPS are GJ\,436 and GJ\,536.
For GJ\,176 and GJ\,581 CARMENES is better and for GJ\,876 equally good as HARPS.

Nevertheless, we note a few important implications for a fair comparison of the instrument precision.
(1) The CARMENES formal RV errors for these stars were boosted from 
$\sim$1.0\,m\,s$^{-1}$ to $\sim$1.6\,m\,s$^{-1}$ by the NZP correction, which added 
(in quadrature) the NZP uncertainties to the RV uncertainties delivered by SERVAL.
(2) Both our CARMENES spectra and the HARPS data were reduced with our SERVAL pipeline.
Our HARPS  RV measurements resulted in a better RV precision than the one obtained with the official ESO-HARPS pipeline.
(3) The HIRES data set from \citep{Butler2017} includes a considerable 
number of RVs obtained before the HIRES CCD upgrade in 2004, which then improved the instrument performance. 
Since in our analysis we do not distinguish between pre- and post-upgrade data,
the $rms_{\rm HIRES}$ is some average taken over the two parts of the dataset.  
(4) M dwarfs can change their activity level with time.
Therefore, the data from the instruments to be compared may have been affected
by different stellar jitter levels, which complicates the analysis of the achievable RV precision for a given star. 

Overall, our results show that the visual channel of CARMENES is capable of achieving a
comparable performance to HARPS and HIRES.

\section{Summary and conclusions}
\label{Sec4}

We present precise optical radial velocity measurements for seven known M-dwarf planet hosts
obtained during the first 15 months of CARMENES operations. 
These planetary systems are the presumably single planet systems: 
GJ\,15\,A \citep{Howard2014},
GJ\,436 \citep{Lanotte2014},
GJ\,176 \citep{Forveille2009}, and 
GJ\,536 \citep{Mascareno2017}, 
and the confirmed multiple planet systems: 
GJ\,1148 \citep{Haghighipour2010},
GJ\,581 \citep{Mayor2009}, 
and 
GJ\,876 \citep{Rivera2010}.
These systems were previously intensively observed with  
high-precision optical spectrographs such as HARPS and HIRES,
yielding a large number of high-precision Doppler measurements,
which we use as an excellent benchmark for assessing the precision of the new optical CARMENES data.
We find that the large number of HIRES and HARPS data together with 
the new visible-channel CARMENES data yield improved orbital planetary parameters for these systems.

For GJ\,176 and GJ\,536 we present updated orbital solutions similar to those listed in the literature.
We have only $\sim$20--30 new RV CARMENES measurements, 
that on their own cannot independently confirm the planets around these stars. 
Our data are, however, consistent with the planetary signals for GJ\,176\,b and GJ\,536\,b, showing very 
small residual values comparable to those from HARPS and HIRES. 
For these two stars, the CARMENES data strengthen the one-planet orbital solutions.

Our 113 RVs for GJ\,436 are sufficient to independently confirm the well-studied
short-period Neptune mass ($m_{\rm b} \sin i$ = 22.2 $M_{\oplus}$) companion around this star. 
We find full consistency between the CARMENES, HARPS, and HIRES data, 
leading to a refined orbital elements and physical parameters for GJ\,436\,b.

In the case of the already intensively studied multiple planet systems 
GJ\,581 and GJ\,876, the limited number of additional CARMENES observations are 
found to be consistent with the HIRES and HARPS data. 
They follow the best-fit dynamical solutions with very low scatter levels.
The best fits for GJ\,581 and GJ\,876 successfully survived 10 Myr 
of precise dynamical simulations in agreement with our current understanding for these systems. 
We find, however, a significantly smaller eccentricity for GJ\,876\,d
than the one often cited in the literature \citep[e.g.,][]{Rivera2010}.
To our knowledge, our three- and four-planet dynamical models for GJ\,581 and GJ\,876, respectively,
are based on all available high precision RV data and provide a benchmark for 
more comprehensive dynamical and statistical analyses.

The CARMENES data shed new light on two systems. 
On the one hand, the planetary nature of the 11.44\,d signal reported in GJ\,15\,A
seems controversial since it is absent in the CARMENES data alone, but also in later RVs from HIRES. 
We speculate that the 11.44\,d signal seen in the early HIRES data could be related to stellar activity. 
On the other hand, our analysis of the GJ\,15\,A data reveals 
a possible planet with a period of $\sim$7026\,d and a minimum mass of $\sim$52~$M_\oplus$.

Based on our CARMENES data we confirm GJ\,1148\,b and we discover a new outer planet in the GJ\,1148 system.
We note, however, that \citet{Butler2017} have already mentioned
discover the second planetary signal  in their more extended HIRES data,
but they classified this signal as a ``planetary candidate'', and did not provide an orbital solution.
Based on the combined HIRES and CARMENES data 
for GJ\,1148\,c we derived a period of $P_{\rm c}$ = 533\,d, eccentricity $e_{\rm c}$ = 0.36, 
and minimum mass $m_{\rm c} \sin i$ = 68~$M_\oplus$. 
Our two-planet dynamical model is now consistent with two Saturn-mass
planets on eccentric orbits with $e_{\rm b}$ = 0.39 and $e_{\rm c}$ = 0.34
and semi-major axes $a_{\rm b}$ = 0.166 au and $a_{\rm c}$ = 0.912 au.
We find that this configuration is stable for at least 10 Myr 
and very likely dynamically stable on the Gyr time scale.

The CARMENES survey is taking radial-velocity time-series measurements of $\sim$300 nearby M-dwarf
stars in an attempt to find Earth-mass planets in their habitable zones. 
In addition, we aim to find additional multiple planetary systems like GJ\,581, GJ\,876 and GJ\,1148
and to place further constraints on planet formation and orbital evolution around low-mass stars.
As CARMENES is a new instrument, a critical point was to 
test the overall capabilities in terms of RV precision and long-term stability of the spectrograph.
Based on the results presented  in this paper, we conclude that the 
visible-light spectrograph of the CARMENES instrument has the precision needed to discover 
exoplanets resembling our Earth, which could provide a habitable 
environment suitable for sustaining life around nearby M-dwarf stars.
An analogous analysis of the performance of the near-infrared spectrograph and 
pipeline will be presented in a separate paper.

\begin{acknowledgements}
  CARMENES is an instrument for the Centro Astron\'omico Hispano-Alem\'an de
  Calar Alto (CAHA, Almer\'{\i}a, Spain). 
  CARMENES is funded by the German Max-Planck-Gesellschaft (MPG), 
  the Spanish Consejo Superior de Investigaciones Cient\'{\i}ficas (CSIC),
  the European Union through FEDER/ERF FICTS-2011-02 funds, 
  and the members of the CARMENES Consortium 
  (Max-Planck-Institut f\"ur Astronomie,
  Instituto de Astrof\'{\i}sica de Andaluc\'{\i}a,
  Landessternwarte K\"onigstuhl,
  Institut de Ci\`encies de l'Espai,
  Insitut f\"ur Astrophysik G\"ottingen,
  Universidad Complutense de Madrid,
  Th\"uringer Landessternwarte Tautenburg,
  Instituto de Astrof\'{\i}sica de Canarias,
  Hamburger Sternwarte,
  Centro de Astrobiolog\'{\i}a and
  Centro Astron\'omico Hispano-Alem\'an), 
  with additional contributions by the Spanish Ministry of Economy,
  the German Science Foundation (DFG),
  the Klaus Tschira Stiftung,
  the states of Baden-W\"urttemberg and Niedersachsen,
  the DFG Research Unit FOR2544 "Blue Planets around Red Stars,
  and by the Junta de Andaluc\'{\i}a. 
This work has made use of data from the European Space Agency (ESA)
mission {\it Gaia} (\url{https://www.cosmos.esa.int/gaia}), processed by
the {\it Gaia} Data Processing and Analysis Consortium (DPAC,
\url{https://www.cosmos.esa.int/web/gaia/dpac/consortium}). Funding
for the DPAC has been provided by national institutions, in particular
the institutions participating in the {\it Gaia} Multilateral Agreement.
This work used the {\it Systemic~Console} package \citep{Meschiari2009} for cross-checking our Keplerian and Dynamical fits
and the python package {\it astroML} \citep{VanderPlas2012} for the calculation of the GLS periodogram.
The IEEC-CSIC team acknowledges support by the Spanish Ministry of Economy and Competitiveness (MINECO) and the Fondo Europeo de Desarrollo Regional 
(FEDER) through grant ESP2016-80435-C2-1-R, as well as the support of the Generalitat de Catalunya/CERCA programme.
The IAA-CSIC team acknowledges support by the Spanish Ministry of Economy and Competitiveness (MINECO) through grants AYA2014-54348-C03-01 
and AYA2016-79425-C3-3-P as well as FEDER funds.
The UCM team acknowledges support by the Spanish Ministry of Economy and Competitiveness (MINECO) from projects
AYA2015-68012-C2-2-P and AYA2016-79425- C3-1,2,3-P and the Spanish Ministerio de 
Educaci\'on, Cultura y Deporte, programa de
Formación de Profesorado Universitario, under grant FPU15/01476.
T.T. and M.K. thank to Jan Rybizki for the very helpful discussion in the early phases of this work.
VJSB are supported by grant AYA2015-69350-C3-2-P from the Spanish Ministry of Economy
and Competiveness (MINECO)
J.C.S acknowledges funding support from Spanish public funds for research under 
project ESP2015-65712-C5-5-R (MINECO/FEDER), and under Research Fellowship program "Ram\'on y Cajal"
with reference RYC2012-09913 (MINECO/FEDER).
The contributions of M.A. were supported by DLR (Deutsches Zentrum f\"ur Luft- und Raumfahrt) 
through the grants50OW0204 and 50OO1501.
J.L.-S. acknowledges the Office of Naval Research Global (award no. N62909-15- 1-2011) for support. 
CdB acknowledges that this work has been supported by Mexican 
CONACyT research grant CB-2012-183007 and the Spanish Ministry 
of Economy and Competitivity through projects AYA2014-54348-C3-2-R.
J.I.G.H., and R.R. acknowledge financial support from the Spanish Ministry project MINECO AYA2014-56359-P. 
J.I.G.H. also acknowledges financial support from the Spanish MINECO under the 2013 Ram\`on y Cajal program MINECO RYC- 2013-14875. 
V. Wolthoff acknowledges funding from the DFG Research Unit FOR2544 
'Blue Planets around Red Stars', project no. RE 2694/4-1.
We thank the anonymous referee for the excellent comments that helped to improve the quality of this paper.

\end{acknowledgements}

\bibliographystyle{aa}

\bibliography{carm_bib}

\clearpage

\begin{appendix} 

\label{appendix}

 \setcounter{table}{0}
\renewcommand{\thetable}{A\arabic{table}}

\setcounter{figure}{0}
\renewcommand{\thefigure}{A\arabic{figure}}

\begin{figure}[ht]

 \begin{center}$
\begin{array}{cc} 
\includegraphics[width=9cm]{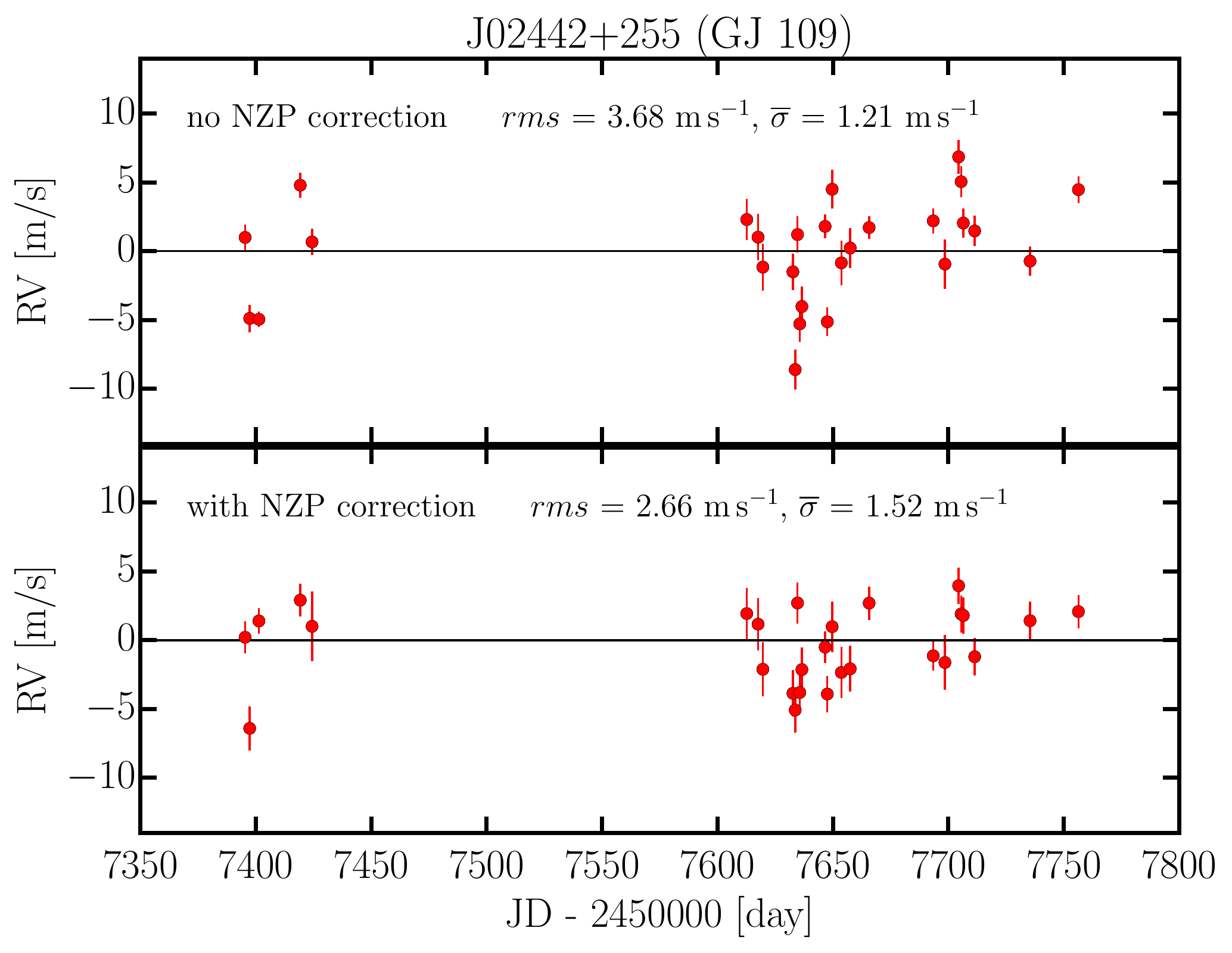} 
\\
\includegraphics[width=9cm]{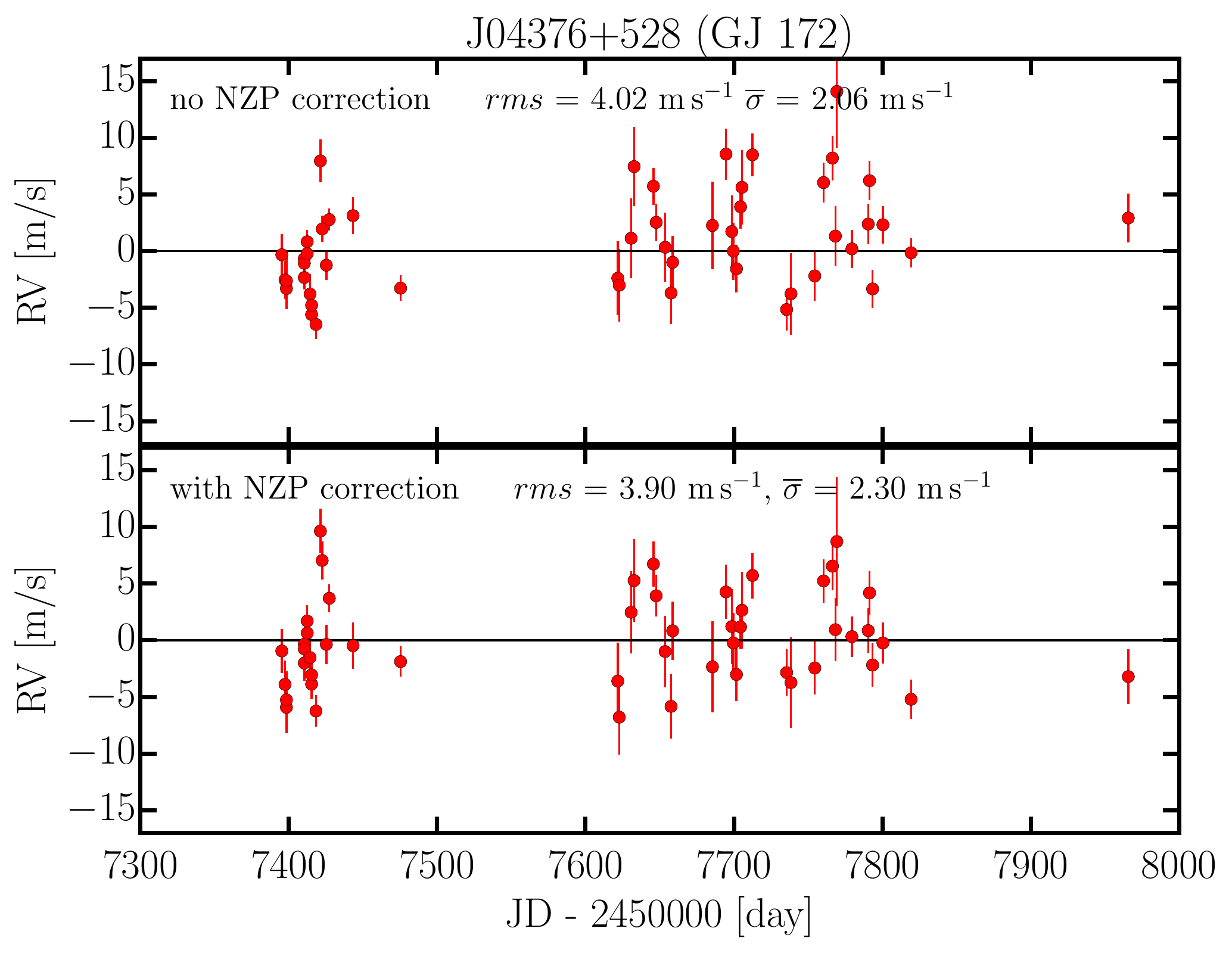}  
\\
\includegraphics[width=9cm]{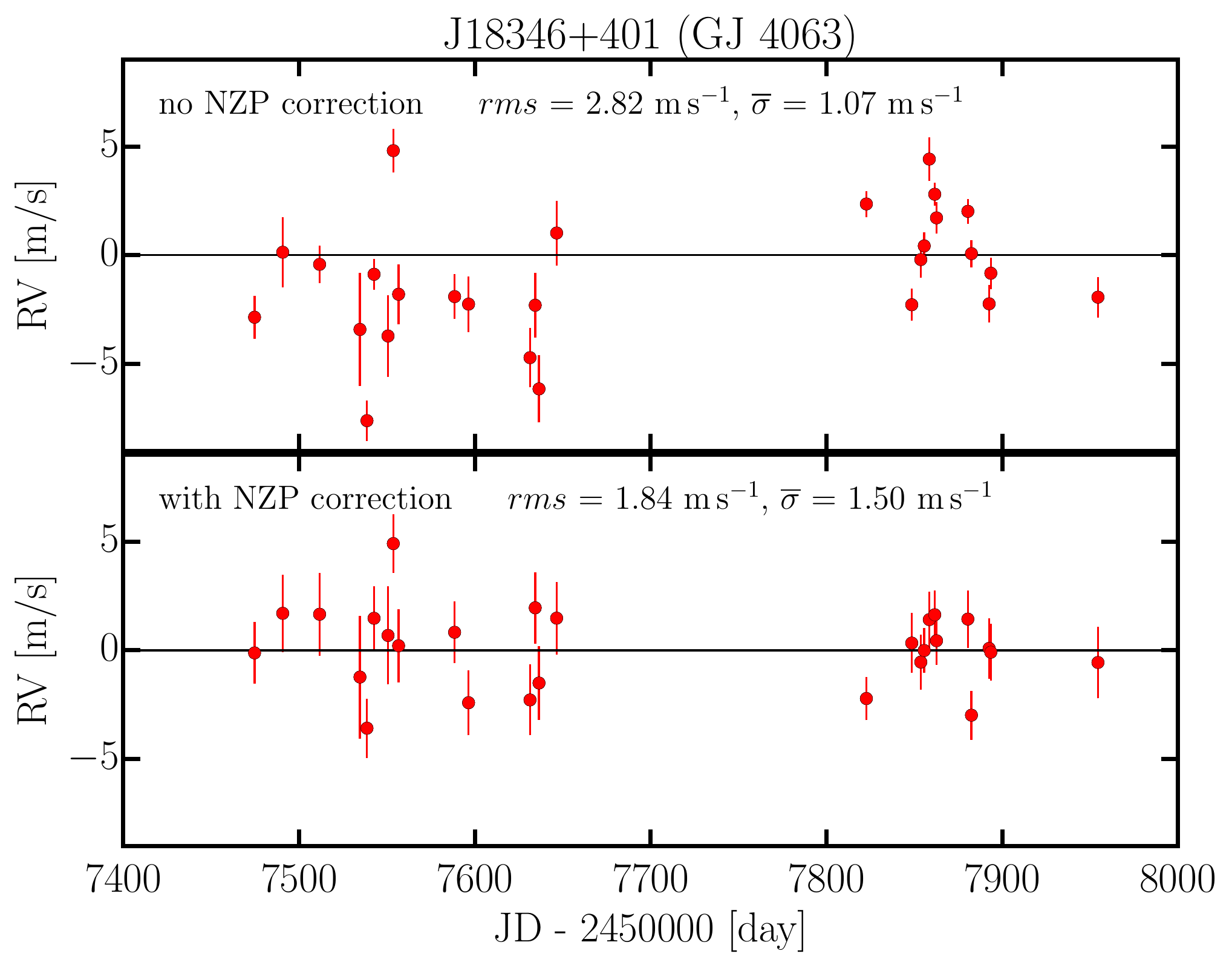}  
\end{array} $
\end{center}

\caption{ 
Pre- and post- NZP correction RV scatter for three CARMENES targets
that we consider as "RV-quiet" stars. 
The top two panels are for GJ\,109, the middle two show GJ\,172 and the bottom two GJ\,4063.
The NZP correction leads to lower $rms$ scatter for almost all RV-quiet stars
in the CARMENES sample (see Fig.~\ref{FigGam:A2}).
Note that the mean measurement error $\overline\sigma$ increases somewhat due to the propagation of the error of the NZP correction.
}
\label{FigGam:A1}
\end{figure}

\begin{figure}[ht]

 \begin{center}$
\begin{array}{cc} 
\includegraphics[width=9cm]{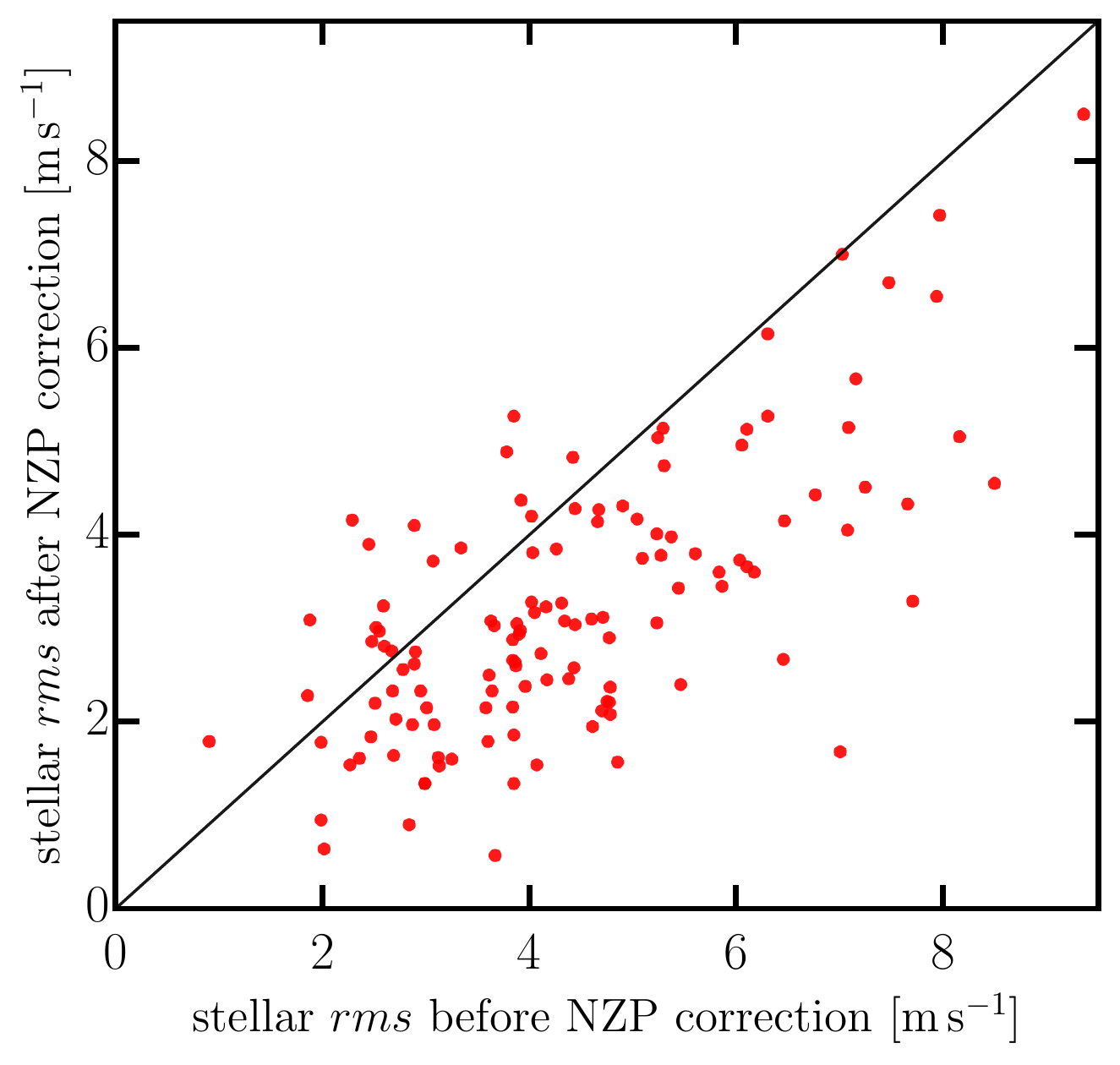}  
\end{array} $
\end{center}

\caption{
Improvement of the stellar $rms$ due to NZP correction.  
For this plot we have selected a sample of 126 CARMENES RV-quiet stars,
which have at least 10 RV measurements and an overall $rms$ scatter less than 10\,m\,s$^{-1}$.
On average the RV scatter for these stars is reduced when the NZP correction is applied.
}
\label{FigGam:A2}
\end{figure}

\begin{table*}
\label{table:A1} 
\caption{CARMENES Doppler measurements for GJ\,15\,A.} 

\centering  

\begin{tabular}{c c c c c c c c} 

\hline\hline    
\noalign{\vskip 0.5mm}

Epoch [JD] & RV [m\,s$^{-1}$] & $\sigma_{RV}$ [m\,s$^{-1}$] & H$\alpha$ & $\sigma_{\rm H\alpha}$  & SNR  & Exp. time [sec] \\  

\hline     
\noalign{\vskip 0.5mm}    

2457395.246   &   -4.8   &    1.1   &   0.9653  &   0.0007 &   183  &   600 \\ 
2457396.296   &   0.8   &    1.9   &   0.9648  &   0.0006 &   195  &   600 \\ 
2457400.265   &   1.7   &    1.1   &   0.9643  &   0.0006 &   228  &   600 \\ 
2457401.294   &   4.0   &    0.8   &   0.9655  &   0.0004 &   296  &   400 \\ 
2457412.315   &   1.8   &    1.0   &   0.9579  &   0.0003 &   347  &   601 \\ 
2457412.321   &   1.7   &    1.0   &   0.9587  &   0.0005 &   242  &   300 \\ 
2457412.325   &   0.6   &    1.1   &   0.9598  &   0.0005 &   215  &   300 \\ 
2457414.301   &   4.8   &    1.1   &   0.9589  &   0.0010 &   125  &   500 \\ 
2457415.294   &   3.9   &    1.0   &   0.9588  &   0.0006 &   209  &   600 \\ 
2457418.310   &   4.1   &    0.9   &   0.9584  &   0.0005 &   245  &   700 \\ 
2457419.349   &   3.6   &    1.5   &   0.9589  &   0.0009 &   140  &   600 \\ 
2457421.265   &   3.2   &    0.8   &   0.9593  &   0.0005 &   247  &   300 \\ 
2457423.295   &   1.4   &    1.2   &   0.9582  &   0.0006 &   213  &   500 \\ 
2457425.281   &   1.2   &    1.4   &   0.9539  &   0.0008 &   156  &   501 \\ 
2457426.307   &   1.1   &    1.1   &   0.9575  &   0.0006 &   199  &   1200 \\ 
2457427.298   &   3.8   &    1.1   &   0.9560  &   0.0006 &   206  &   500 \\ 
2457575.673   &   2.5   &    1.2   &   0.9593  &   0.0006 &   231  &   240 \\ 
2457597.563   &   -3.2   &    2.4   &   0.9646  &   0.0020 &   66  &   27 \\ 
2457611.580   &   4.6   &    2.3   &   0.9521  &   0.0020 &   66  &   30 \\ 
2457621.603   &   2.4   &    2.1   &   0.9561  &   0.0019 &   68  &   16 \\ 
2457626.529   &   1.1   &    1.8   &   0.9494  &   0.0019 &   69  &   24 \\ 
2457630.491   &   2.5   &    2.0   &   0.9560  &   0.0019 &   68  &   30 \\ 
2457631.660   &   -0.8   &    1.8   &   0.9653  &   0.0019 &   67  &   26 \\ 
2457632.535   &   3.1   &    2.0   &   0.9657  &   0.0020 &   66  &   23 \\ 
2457633.460   &   3.5   &    2.2   &   0.9546  &   0.0019 &   70  &   40 \\ 
2457634.624   &   2.0   &    2.0   &   0.9639  &   0.0018 &   68  &   21 \\ 
2457635.553   &   -0.3   &    2.3   &   0.9649  &   0.0020 &   68  &   21 \\ 
2457636.501   &   -0.2   &    1.8   &   0.9562  &   0.0020 &   66  &   25 \\ 
2457644.415   &   0.0   &    2.3   &   0.9557  &   0.0012 &   67  &   375 \\ 
2457644.428   &   0.6   &    1.6   &   0.9628  &   0.0011 &   111  &   118 \\ 
2457645.557   &   -2.4   &    1.8   &   0.9559  &   0.0012 &   116  &   294 \\ 
2457646.406   &   3.8   &    1.3   &   0.9593  &   0.0018 &   104  &   424 \\ 
2457647.387   &   4.2   &    2.1   &   0.9628  &   0.0019 &   72  &   141 \\ 
2457649.614   &   0.4   &    2.1   &   0.9601  &   0.0020 &   68  &   20 \\ 
2457653.505   &   2.4   &    2.3   &   0.9511  &   0.0015 &   65  &   70 \\ 
2457654.720   &   -2.8   &    1.6   &   0.9553  &   0.0008 &   88  &   67 \\ 
2457655.395   &   1.2   &    1.3   &   0.9579  &   0.0008 &   160  &   140 \\ 
2457657.388   &   3.4   &    1.5   &   0.9564  &   0.0015 &   173  &   237 \\ 
2457658.476   &   -2.0   &    2.0   &   0.9586  &   0.0014 &   83  &   69 \\ 
2457677.457   &   -1.3   &    1.7   &   0.9565  &   0.0013 &   93  &   50 \\ 
2457678.339   &   1.7   &    1.6   &   0.9624  &   0.0015 &   99  &   95 \\ 
2457679.296   &   0.2   &    2.0   &   0.9677  &   0.0014 &   84  &   71 \\ 
2457684.359   &   -3.3   &    2.1   &   0.9618  &   0.0020 &   95  &   113 \\ 
2457685.333   &   -1.7   &    2.1   &   0.9667  &   0.0013 &   66  &   122 \\ 
2457688.418   &   -1.3   &    1.7   &   0.9629  &   0.0014 &   95  &   71 \\ 
2457689.303   &   0.7   &    1.5   &   0.9616  &   0.0005 &   96  &   46 \\ 
2457690.392   &   0.9   &    1.1   &   0.9613  &   0.0005 &   227  &   300 \\ 
2457690.396   &   0.1   &    1.0   &   0.9621  &   0.0006 &   245  &   300 \\ 
2457690.505   &   -0.8   &    1.0   &   0.9617  &   0.0005 &   284  &   300 \\ 
2457694.315   &   4.5   &    1.7   &   0.9538  &   0.0017 &   99  &   61 \\ 
2457698.361   &   -0.8   &    1.7   &   0.9582  &   0.0023 &   76  &   152 \\ 
2457699.470   &   -2.8   &    2.3   &   0.9597  &   0.0005 &   52  &   152 \\ 
2457703.338   &   2.4   &    0.8   &   0.9580  &   0.0004 &   282  &   300 \\ 
2457703.379   &   2.6   &    0.9   &   0.9579  &   0.0004 &   302  &   300 \\ 
2457703.388   &   3.1   &    0.9   &   0.9586  &   0.0004 &   298  &   300 \\ 
2457704.392   &   1.4   &    1.6   &   0.9549  &   0.0013 &   99  &   41 \\ 
2457705.297   &   0.4   &    1.5   &   0.9642  &   0.0013 &   94  &   87 \\ 
2457706.380   &   -1.9   &    1.6   &   0.9570  &   0.0012 &   97  &   47 \\ 
2457709.439   &   1.1   &    1.5   &   0.9569  &   0.0006 &   101  &   74 \\ 
2457710.323   &   3.2   &    2.1   &   0.9569  &   0.0007 &   195  &   150 \\ 
2457710.336   &   2.2   &    2.1   &   0.9584  &   0.0007 &   175  &   150 \\ 
  
\hline           
\end{tabular}
\end{table*}

\begin{table*}
\caption{CARMENES Doppler measurements for GJ\,15\,A (continue from Table~A1).} 
\label{table:GJ15A2} 

\centering  
\begin{tabular}{c c c c c c c c} 

\hline\hline    
\noalign{\vskip 0.5mm}

Epoch [JD] & RV [m\,s$^{-1}$] & $\sigma_{RV}$ [m\,s$^{-1}$] & H$\alpha$ & $\sigma_{\rm H\alpha}$  & SNR  & Exp. time [sec] \\  

\hline     
\noalign{\vskip 0.5mm}

2457710.410   &   -0.9   &    2.2   &   0.9570  &   0.0007 &   164  &   150 \\ 
2457711.274   &   -0.0   &    1.8   &   0.9584  &   0.0013 &   102  &   61 \\ 
2457712.436   &   1.0   &    1.6   &   0.9570  &   0.0008 &   99  &   52 \\ 
2457735.255   &   4.2   &    1.3   &   0.9555  &   0.0008 &   150  &   150 \\ 
2457735.268   &   4.9   &    1.2   &   0.9563  &   0.0009 &   146  &   150 \\ 
2457735.281   &   5.2   &    1.1   &   0.9574  &   0.0007 &   156  &   150 \\ 
2457735.293   &   4.5   &    1.3   &   0.9591  &   0.0007 &   142  &   150 \\ 
2457736.284   &   4.4   &    2.0   &   0.9575  &   0.0007 &   196  &   150 \\ 
2457736.340   &   3.6   &    2.0   &   0.9595  &   0.0007 &   136  &   150 \\ 
2457736.421   &   1.9   &    1.9   &   0.9603  &   0.0008 &   185  &   150 \\ 
2457738.310   &   5.7   &    2.4   &   0.9594  &   0.0008 &   70  &   49 \\ 
2457746.361   &   9.6   &    3.8   &   0.9628  &   0.0008 &   104  &   57 \\ 
2457746.498   &   5.6   &    3.7   &   0.9572  &   0.0010 &   140  &   150 \\ 
2457746.515   &   7.2   &    3.7   &   0.9570  &   0.0007 &   130  &   152 \\ 
2457747.342   &   5.5   &    2.1   &   0.9592  &   0.0007 &   210  &   150 \\ 
2457747.381   &   4.9   &    2.1   &   0.9601  &   0.0006 &   176  &   150 \\ 
2457747.425   &   4.9   &    2.1   &   0.9609  &   0.0007 &   194  &   150 \\ 
2457748.279   &   4.1   &    1.5   &   0.9609  &   0.0008 &   142  &   150 \\ 
2457748.295   &   4.7   &    1.5   &   0.9590  &   0.0009 &   145  &   150 \\ 
2457748.311   &   3.6   &    1.4   &   0.9592  &   0.0010 &   153  &   150 \\ 
2457752.331   &   5.2   &    1.3   &   0.9604  &   0.0013 &   103  &   54 \\ 
2457753.298   &   3.0   &    1.7   &   0.9614  &   0.0012 &   98  &   43 \\ 
2457754.290   &   0.3   &    1.4   &   0.9584  &   0.0013 &   97  &   40 \\ 
2457755.351   &   0.8   &    1.5   &   0.9616  &   0.0012 &   106  &   54 \\ 
2457760.292   &   0.8   &    1.5   &   0.9618  &   0.0014 &   96  &   76 \\ 
2457762.275   &   -0.4   &    1.4   &   0.9590  &   0.0012 &   104  &   44 \\ 
2457763.271   &   -0.9   &    1.6   &   0.9607  &   0.0012 &   94  &   51 \\ 
2457766.283   &   2.9   &    1.7   &   0.9621  &   0.0013 &   101  &   39 \\ 
2457767.316   &   -0.7   &    2.1   &   0.9612  &   0.0013 &   102  &   102 \\ 
2457768.296   &   -0.0   &    1.7   &   0.9574  &   0.0008 &   97  &   277 \\ 
2457769.296   &   -3.1   &    2.9   &   0.9551  &   0.0009 &   101  &   151 \\ 
  
\hline           
\end{tabular}
\end{table*}

\begin{table*}
\caption{CARMENES Doppler measurements for GJ\,176.} 
\label{table:GJ176} 

\centering  

\begin{tabular}{c c c c c c c c} 

\hline\hline    
\noalign{\vskip 0.5mm}

Epoch [JD] & RV [m\,s$^{-1}$] & $\sigma_{RV}$ [m\,s$^{-1}$] & H$\alpha$ & $\sigma_{\rm H\alpha}$  & SNR  & Exp. time [sec] \\  

\hline     
\noalign{\vskip 0.5mm}     

2457395.392   &   -8.1   &    1.4   &   0.9476  &   0.0016 &   85  &   650 \\ 
2457397.433   &   -7.3   &    1.6   &   0.9449  &   0.0012 &   106  &   1000 \\ 
2457398.440   &   -3.6   &    2.1   &   0.9447  &   0.0015 &   91  &   1100 \\ 
2457414.333   &   -9.8   &    1.9   &   0.9977  &   0.0017 &   82  &   700 \\ 
2457415.339   &   -8.2   &    1.2   &   0.9892  &   0.0013 &   107  &   900 \\ 
2457421.372   &   -9.1   &    0.9   &   0.9568  &   0.0008 &   154  &   550 \\ 
2457423.352   &   -9.6   &    1.4   &   0.9576  &   0.0011 &   122  &   400 \\ 
2457425.319   &   -5.5   &    1.5   &   0.9556  &   0.0017 &   79  &   1200 \\ 
2457442.296   &   -4.1   &    1.0   &   0.9694  &   0.0007 &   175  &   650 \\ 
2457443.374   &   -1.4   &    1.5   &   0.9680  &   0.0012 &   111  &   1500 \\ 
2457636.593   &   0.5   &    1.8   &   0.9323  &   0.0021 &   69  &   204 \\ 
2457657.621   &   -2.6   &    4.0   &   0.9858  &   0.0053 &   29  &   637 \\ 
2457658.664   &   -9.9   &    1.4   &   0.9536  &   0.0011 &   120  &   402 \\ 
2457698.521   &   -7.0   &    1.9   &   0.9456  &   0.0021 &   66  &   602 \\ 
2457699.561   &   -5.6   &    1.5   &   0.9575  &   0.0017 &   83  &   502 \\ 
2457704.481   &   -13.5   &    1.3   &   0.9479  &   0.0015 &   92  &   216 \\ 
2457705.513   &   -11.5   &    1.4   &   0.9416  &   0.0014 &   95  &   198 \\ 
2457711.423   &   -11.6   &    1.8   &   0.9352  &   0.0014 &   99  &   201 \\ 
2457712.501   &   -8.6   &    1.4   &   0.9392  &   0.0015 &   92  &   375 \\ 
2457753.433   &   0.4   &    1.5   &   0.9412  &   0.0014 &   96  &   175 \\ 
2457754.431   &   -4.7   &    1.3   &   0.9469  &   0.0013 &   98  &   192 \\ 
2457756.424   &   -5.4   &    1.2   &   0.9531  &   0.0013 &   105  &   709 \\ 
2457760.380   &   -0.5   &    1.4   &   0.9539  &   0.0014 &   97  &   203 \\ 
  
\hline           
\end{tabular}
\end{table*}

\begin{table*}
\caption{CARMENES Doppler measurements for GJ\,436.} 
\label{table:GJ436} 

\centering  

\begin{tabular}{c c c c c c c c c} 

\hline\hline    
\noalign{\vskip 0.5mm}

Epoch [JD] & RV [m\,s$^{-1}$] & $\sigma_{RV}$ [m\,s$^{-1}$] & H$\alpha$ & $\sigma_{\rm H\alpha}$  & SNR  & Exp. time [sec] \\  

\hline     
\noalign{\vskip 0.5mm}    

2457395.714   &   -29.4   &    1.2   &   0.9685  &   0.0012 &   119  &   1000 \\ 
2457418.708   &   -4.1   &    0.9   &   0.9680  &   0.0012 &   120  &   800 \\ 
2457422.600   &   -37.6   &    1.4   &   0.9674  &   0.0010 &   142  &   1500 \\ 
2457422.619   &   -35.5   &    1.4   &   0.9692  &   0.0011 &   129  &   1500 \\ 
2457422.641   &   -37.0   &    1.4   &   0.9693  &   0.0009 &   152  &   1500 \\ 
2457424.598   &   -21.1   &    3.2   &   0.9578  &   0.0032 &   45  &   1800 \\ 
2457426.474   &   -1.3   &    1.0   &   0.9635  &   0.0011 &   126  &   800 \\ 
2457426.567   &   -1.7   &    1.0   &   0.9662  &   0.0012 &   119  &   1001 \\ 
2457426.626   &   -3.5   &    1.2   &   0.9646  &   0.0013 &   112  &   1000 \\ 
2457426.721   &   -7.0   &    1.1   &   0.9638  &   0.0014 &   106  &   1000 \\ 
2457427.484   &   -32.0   &    1.1   &   0.9685  &   0.0012 &   115  &   1100 \\ 
2457427.650   &   -32.9   &    1.2   &   0.9693  &   0.0012 &   118  &   1500 \\ 
2457427.710   &   -35.9   &    1.0   &   0.9683  &   0.0011 &   131  &   1500 \\ 
2457430.530   &   -33.2   &    1.5   &   0.9653  &   0.0018 &   81  &   1501 \\ 
2457430.659   &   -31.7   &    1.2   &   0.9634  &   0.0015 &   96  &   1800 \\ 
2457440.696   &   -33.3   &    1.2   &   0.9681  &   0.0009 &   156  &   1400 \\ 
2457441.615   &   -28.6   &    1.5   &   0.9614  &   0.0019 &   77  &   900 \\ 
2457442.578   &   -7.2   &    0.8   &   0.9622  &   0.0009 &   161  &   800 \\ 
2457444.613   &   -11.2   &    1.6   &   0.9636  &   0.0020 &   73  &   900 \\ 
2457444.625   &   -9.2   &    1.3   &   0.9629  &   0.0017 &   81  &   900 \\ 
2457449.482   &   -32.0   &    1.4   &   0.9670  &   0.0010 &   141  &   900 \\ 
2457449.558   &   -27.4   &    1.5   &   0.9689  &   0.0009 &   153  &   900 \\ 
2457466.513   &   -14.0   &    1.5   &   0.9731  &   0.0016 &   90  &   750 \\ 
2457466.524   &   -14.5   &    1.7   &   0.9681  &   0.0021 &   67  &   750 \\ 
2457467.476   &   -37.7   &    2.3   &   0.9628  &   0.0030 &   50  &   1200 \\ 
2457472.570   &   -34.0   &    1.1   &   0.9670  &   0.0010 &   146  &   800 \\ 
2457473.545   &   -17.8   &    1.4   &   0.9710  &   0.0013 &   110  &   1000 \\ 
2457474.521   &   -16.6   &    1.5   &   0.9670  &   0.0012 &   121  &   900 \\ 
2457475.620   &   -35.9   &    1.2   &   0.9683  &   0.0013 &   112  &   800 \\ 
2457476.480   &   -5.1   &    1.4   &   0.9633  &   0.0009 &   165  &   901 \\ 
2457476.496   &   -4.2   &    1.4   &   0.9663  &   0.0008 &   168  &   900 \\ 
2457477.470   &   -24.0   &    1.1   &   0.9665  &   0.0011 &   131  &   800 \\ 
2457488.468   &   -37.0   &    1.0   &   0.9707  &   0.0009 &   159  &   900 \\ 
2457489.451   &   -15.6   &    1.9   &   0.9654  &   0.0024 &   61  &   800 \\ 
2457490.475   &   -20.2   &    1.3   &   0.9674  &   0.0015 &   94  &   800 \\ 
2457492.444   &   -4.9   &    1.5   &   0.9654  &   0.0015 &   94  &   1200 \\ 
2457493.529   &   -35.0   &    1.2   &   0.9657  &   0.0011 &   123  &   901 \\ 
2457494.495   &   -29.4   &    1.4   &   0.9627  &   0.0012 &   120  &   800 \\ 
2457499.415   &   -36.2   &    1.7   &   0.9723  &   0.0022 &   67  &   800 \\ 
2457503.383   &   -8.7   &    1.4   &   0.9636  &   0.0013 &   114  &   800 \\ 
2457504.395   &   -32.8   &    1.5   &   0.9627  &   0.0016 &   90  &   800 \\ 
2457505.399   &   -11.0   &    1.5   &   0.9671  &   0.0016 &   93  &   800 \\ 
2457509.408   &   -34.9   &    1.2   &   0.9705  &   0.0013 &   110  &   900 \\ 
2457510.455   &   -22.8   &    1.1   &   0.9684  &   0.0015 &   99  &   800 \\ 
2457511.340   &   -10.6   &    1.5   &   0.9702  &   0.0009 &   157  &   750 \\ 
2457511.351   &   -12.5   &    1.5   &   0.9696  &   0.0009 &   161  &   750 \\ 
2457511.402   &   -13.2   &    1.5   &   0.9709  &   0.0009 &   150  &   750 \\ 
2457511.440   &   -14.9   &    1.6   &   0.9712  &   0.0010 &   137  &   750 \\ 
2457511.476   &   -15.8   &    1.6   &   0.9684  &   0.0011 &   127  &   750 \\ 
2457511.524   &   -17.7   &    1.5   &   0.9724  &   0.0010 &   135  &   750 \\ 
2457511.567   &   -20.1   &    1.6   &   0.9726  &   0.0014 &   103  &   750 \\ 
2457511.606   &   -20.8   &    1.7   &   0.9701  &   0.0015 &   96  &   750 \\ 
2457511.617   &   -22.0   &    1.8   &   0.9705  &   0.0017 &   89  &   750 \\ 
2457512.478   &   -38.5   &    1.1   &   0.9650  &   0.0010 &   134  &   900 \\ 
2457525.421   &   -34.7   &    2.5   &   0.9590  &   0.0020 &   68  &   1100 \\ 
2457527.346   &   -9.5   &    1.7   &   0.9630  &   0.0013 &   112  &   800 \\ 
2457529.352   &   -2.3   &    1.3   &   0.9618  &   0.0012 &   120  &   800 \\ 
2457530.343   &   -24.4   &    1.5   &   0.9637  &   0.0016 &   89  &   800 \\ 
2457531.343   &   -28.8   &    2.4   &   0.9610  &   0.0026 &   56  &   800 \\ 
2457532.390   &   1.2   &    1.7   &   0.9630  &   0.0018 &   78  &   900 \\

\hline           
\end{tabular}
\end{table*}

\begin{table*}
\caption{CARMENES Doppler measurements for GJ\,436 (continue from Table~\ref{table:GJ436}).} 
\label{table:GJ436_2} 

\centering  

\begin{tabular}{c c c c c c c c c} 

\hline\hline    
\noalign{\vskip 0.5mm}

Epoch [JD] & RV [m\,s$^{-1}$] & $\sigma_{RV}$ [m\,s$^{-1}$] & H$\alpha$ & $\sigma_{\rm H\alpha}$  & SNR  & Exp. time [sec] \\  

\hline     
\noalign{\vskip 0.5mm}    

2457533.402   &   -33.5   &    2.4   &   0.9614  &   0.0026 &   55  &   713 \\ 
2457533.415   &   -34.4   &    2.1   &   0.9581  &   0.0021 &   67  &   900 \\ 
2457534.364   &   -13.9   &    1.4   &   0.9609  &   0.0014 &   100  &   800 \\ 
2457535.351   &   -16.4   &    1.5   &   0.9633  &   0.0016 &   78  &   800 \\ 
2457536.424   &   -35.8   &    4.9   &   0.9750  &   0.0068 &   23  &   1000 \\ 
2457539.379   &   -27.9   &    1.3   &   0.9619  &   0.0011 &   115  &   900 \\ 
2457540.376   &   -6.0   &    1.3   &   0.9646  &   0.0011 &   120  &   900 \\ 
2457541.409   &   -35.8   &    1.5   &   0.9959  &   0.0010 &   132  &   1000 \\ 
2457543.407   &   -22.7   &    1.6   &   0.9637  &   0.0019 &   74  &   1000 \\ 
2457544.380   &   -35.7   &    1.3   &   0.9620  &   0.0011 &   115  &   1001 \\ 
2457545.363   &   -1.4   &    1.8   &   0.9644  &   0.0014 &   97  &   800 \\ 
2457550.364   &   -1.7   &    3.0   &   0.9599  &   0.0039 &   37  &   900 \\ 
2457552.374   &   -36.3   &    1.6   &   0.9628  &   0.0011 &   115  &   900 \\ 
2457553.377   &   -2.7   &    1.3   &   0.9658  &   0.0016 &   86  &   900 \\ 
2457554.389   &   -35.0   &    1.3   &   0.9603  &   0.0015 &   92  &   1000 \\ 
2457556.400   &   -12.2   &    1.4   &   0.9594  &   0.0017 &   83  &   1000 \\ 
2457557.401   &   -38.2   &    1.7   &   0.9638  &   0.0014 &   98  &   1000 \\ 
2457558.371   &   -5.7   &    1.2   &   0.9673  &   0.0011 &   117  &   1000 \\ 
2457559.379   &   -24.1   &    1.1   &   0.9627  &   0.0011 &   122  &   1000 \\ 
2457563.388   &   -21.2   &    1.6   &   0.9653  &   0.0015 &   88  &   900 \\ 
2457564.363   &   -12.3   &    1.2   &   0.9610  &   0.0013 &   101  &   900 \\ 
2457565.358   &   -37.6   &    1.4   &   0.9768  &   0.0016 &   81  &   900 \\ 
2457567.363   &   -22.9   &    5.0   &   0.9628  &   0.0027 &   49  &   900 \\ 
2457570.354   &   -30.9   &    1.8   &   0.9601  &   0.0011 &   114  &   800 \\ 
2457571.349   &   -22.0   &    2.1   &   0.9570  &   0.0014 &   98  &   800 \\ 
2457572.358   &   -12.4   &    1.6   &   0.9626  &   0.0013 &   97  &   800 \\ 
2457573.363   &   -33.4   &    2.5   &   0.9628  &   0.0023 &   61  &   800 \\ 
2457574.360   &   1.0   &    1.3   &   0.9595  &   0.0012 &   112  &   1000 \\ 
2457575.361   &   -24.1   &    1.4   &   0.9609  &   0.0010 &   129  &   1000 \\ 
2457584.374   &   -24.6   &    1.6   &   0.9603  &   0.0013 &   104  &   800 \\ 
2457586.358   &   -29.4   &    1.6   &   0.9610  &   0.0012 &   106  &   800 \\ 
2457587.369   &   -5.5   &    3.0   &   0.9582  &   0.0036 &   40  &   800 \\ 
2457591.351   &   -30.2   &    1.6   &   0.9634  &   0.0015 &   103  &   900 \\ 
2457593.349   &   -8.6   &    1.9   &   0.9587  &   0.0017 &   93  &   700 \\ 
2457594.352   &   -30.8   &    2.2   &   0.9580  &   0.0023 &   68  &   321 \\ 
2457595.332   &   -8.4   &    2.6   &   0.9228  &   0.0021 &   71  &   312 \\ 
2457596.329   &   -11.7   &    7.1   &   0.8381  &   0.0032 &   41  &   319 \\ 
2457597.334   &   -31.1   &    2.0   &   0.9433  &   0.0022 &   70  &   337 \\ 
2457688.721   &   -19.6   &    1.9   &   0.9621  &   0.0027 &   54  &   320 \\ 
2457691.718   &   -24.4   &    3.6   &   0.9686  &   0.0063 &   27  &   27 \\ 
2457691.721   &   -21.6   &    3.6   &   0.9611  &   0.0061 &   26  &   27 \\ 
2457691.728   &   -17.6   &    3.7   &   0.9728  &   0.0079 &   21  &   18 \\ 
2457691.732   &   -15.7   &    5.5   &   0.9803  &   0.0090 &   19  &   17 \\ 
2457692.740   &   -29.0   &    1.1   &   0.9689  &   0.0016 &   92  &   227 \\ 
2457693.731   &   -10.3   &    1.1   &   0.9664  &   0.0014 &   98  &   310 \\ 
2457695.728   &   -10.3   &    1.5   &   0.9685  &   0.0024 &   60  &   601 \\ 
2457699.675   &   -31.7   &    1.7   &   0.9693  &   0.0025 &   61  &   701 \\ 
2457703.662   &   -11.0   &    1.3   &   0.9627  &   0.0014 &   104  &   408 \\ 
2457704.669   &   -18.5   &    1.4   &   0.9675  &   0.0022 &   66  &   687 \\ 
2457705.704   &   -36.0   &    1.3   &   0.9663  &   0.0016 &   92  &   881 \\ 
2457706.677   &   0.2   &    1.3   &   0.9582  &   0.0014 &   98  &   662 \\ 
2457706.734   &   -3.0   &    1.4   &   0.9593  &   0.0016 &   92  &   474 \\ 
2457712.732   &   -22.4   &    1.8   &   0.9624  &   0.0022 &   67  &   686 \\ 
  
\hline           
\end{tabular}
\end{table*}

\begin{table*}
\caption{CARMENES Doppler measurements for GJ\,536.} 
\label{table:GJ536} 

\centering  

\begin{tabular}{c c c c c c c c c} 

\hline\hline    
\noalign{\vskip 0.5mm}

Epoch [JD] & RV [m\,s$^{-1}$] & $\sigma_{RV}$ [m\,s$^{-1}$] & H$\alpha$ & $\sigma_{\rm H\alpha}$  & SNR  & Exp. time [sec] \\  

\hline     
\noalign{\vskip 0.5mm}    

2457397.764   &   -0.8   &    2.1   &   0.9333  &   0.0015 &   84  &   800 \\ 
2457414.718   &   -14.7   &    1.1   &   0.9275  &   0.0009 &   138  &   900 \\ 
2457415.733   &   -12.2   &    1.1   &   0.9258  &   0.0009 &   146  &   701 \\ 
2457419.767   &   -8.2   &    1.1   &   0.9283  &   0.0008 &   156  &   700 \\ 
2457477.599   &   -11.4   &    1.4   &   0.9386  &   0.0012 &   102  &   600 \\ 
2457490.566   &   -14.5   &    2.2   &   0.9291  &   0.0018 &   70  &   801 \\ 
2457536.442   &   -11.8   &    2.0   &   0.9321  &   0.0019 &   69  &   900 \\ 
2457539.448   &   -4.4   &    1.9   &   0.9282  &   0.0013 &   102  &   1001 \\ 
2457542.474   &   -9.6   &    1.9   &   0.9288  &   0.0012 &   104  &   901 \\ 
2457760.755   &   -6.9   &    1.4   &   0.9349  &   0.0014 &   94  &   275 \\ 
2457761.718   &   -9.8   &    1.2   &   0.9437  &   0.0013 &   100  &   247 \\ 
2457768.763   &   -14.3   &    1.7   &   0.9311  &   0.0015 &   90  &   1002 \\ 
2457771.760   &   -14.5   &    2.1   &   0.9243  &   0.0018 &   71  &   1201 \\ 
2457779.687   &   -13.3   &    1.4   &   0.9366  &   0.0017 &   75  &   1502 \\ 
2457788.673   &   -15.3   &    2.4   &   0.9295  &   0.0018 &   72  &   1202 \\ 
2457790.700   &   -12.7   &    2.0   &   0.9274  &   0.0021 &   63  &   1001 \\ 
2457791.656   &   -10.3   &    2.4   &   0.9334  &   0.0025 &   53  &   482 \\ 
2457793.638   &   -6.9   &    1.7   &   0.9347  &   0.0015 &   88  &   482 \\ 
2457798.625   &   -13.0   &    1.2   &   0.9417  &   0.0011 &   114  &   320 \\ 
2457799.673   &   -8.7   &    1.5   &   0.9259  &   0.0014 &   91  &   399 \\ 
2457806.656   &   -15.2   &    1.9   &   0.9322  &   0.0016 &   80  &   795 \\ 
2457817.634   &   -5.3   &    1.7   &   0.9308  &   0.0013 &   100  &   795 \\ 
2457819.648   &   -5.4   &    1.3   &   0.9283  &   0.0008 &   169  &   674 \\ 
2457824.566   &   -8.7   &    1.1   &   0.9291  &   0.0008 &   167  &   590 \\ 
2457830.657   &   -13.3   &    1.3   &   0.9257  &   0.0009 &   138  &   796 \\ 
2457857.499   &   -10.1   &    1.2   &   0.9311  &   0.0007 &   168  &   796 \\ 
2457877.474   &   -12.4   &    1.5   &   0.9247  &   0.0011 &   112  &   795 \\ 
2457879.491   &   -11.0   &    2.3   &   0.9276  &   0.0020 &   64  &   1601 \\ 
  
\hline           
\end{tabular}
\end{table*}

\begin{table*}
\caption{CARMENES Doppler measurements for GJ\,581.} 
\label{table:GJ581} 

\centering  

\begin{tabular}{c c c c c c c c c} 

\hline\hline    
\noalign{\vskip 0.5mm}

Epoch [JD] & RV [m\,s$^{-1}$] & $\sigma_{RV}$ [m\,s$^{-1}$] & H$\alpha$ & $\sigma_{\rm H\alpha}$  & SNR  & Exp. time [sec] \\  

\hline     
\noalign{\vskip 0.5mm}    

2457415.763   &   -15.3   &    1.0   &   0.9995  &   0.0011 &   131  &   900 \\ 
2457418.761   &   9.0   &    1.0   &   1.0017  &   0.0015 &   103  &   750 \\ 
2457422.737   &   -9.6   &    1.5   &   0.9958  &   0.0011 &   131  &   900 \\ 
2457466.712   &   10.7   &    1.3   &   0.9915  &   0.0012 &   119  &   800 \\ 
2457476.635   &   -3.4   &    1.4   &   0.9923  &   0.0011 &   142  &   800 \\ 
2457490.597   &   -14.4   &    1.3   &   0.9931  &   0.0018 &   85  &   1200 \\ 
2457493.600   &   11.5   &    1.5   &   0.9964  &   0.0017 &   92  &   801 \\ 
2457503.558   &   3.2   &    1.4   &   0.9968  &   0.0011 &   137  &   800 \\ 
2457505.545   &   -4.3   &    1.5   &   0.9989  &   0.0018 &   87  &   800 \\ 
2457510.570   &   -2.9   &    1.1   &   1.0001  &   0.0011 &   136  &   800 \\ 
2457543.465   &   -6.8   &    1.6   &   0.9982  &   0.0019 &   85  &   900 \\ 
2457606.332   &   6.4   &    1.6   &   0.9831  &   0.0013 &   117  &   645 \\ 
2457793.709   &   3.3   &    1.6   &   0.9928  &   0.0019 &   81  &   575 \\ 
2457798.689   &   -3.7   &    1.3   &   0.9925  &   0.0014 &   108  &   368 \\ 
2457799.712   &   5.5   &    1.2   &   0.9940  &   0.0017 &   89  &   473 \\ 
2457800.737   &   -4.0   &    1.2   &   0.9902  &   0.0014 &   110  &   326 \\ 
2457806.710   &   -15.9   &    1.3   &   0.9888  &   0.0017 &   92  &   948 \\ 
2457817.671   &   -13.4   &    1.6   &   1.0060  &   0.0019 &   82  &   947 \\ 
2457824.675   &   -13.4   &    1.1   &   0.9937  &   0.0011 &   141  &   948 \\ 
2457875.545   &   -2.5   &    1.0   &   1.0041  &   0.0010 &   156  &   670 \\ 
  
\hline           
\end{tabular}
\end{table*}

\begin{table*}
\caption{CARMENES Doppler measurements for GJ\,876.} 
\label{table:GJ876}

\centering  

\begin{tabular}{c c c c c c c c c} 

\hline\hline    
\noalign{\vskip 0.5mm}

Epoch [JD] & RV [m\,s$^{-1}$] & $\sigma_{RV}$ [m\,s$^{-1}$] & H$\alpha$ & $\sigma_{\rm H\alpha}$  & SNR  & Exp. time [sec] \\  

\hline     
\noalign{\vskip 0.5mm}    

2457554.666   &   20.4   &    1.7   &   1.0144  &   0.0027 &   70  &   501 \\ 
2457556.659   &   26.7   &    1.4   &   1.0230  &   0.0019 &   101  &   800 \\ 
2457558.666   &   -8.6   &    1.3   &   1.0428  &   0.0018 &   112  &   360 \\ 
2457574.670   &   -404.7   &    1.4   &   1.0223  &   0.0017 &   111  &   400 \\ 
2457596.613   &   -464.7   &    1.7   &   1.0253  &   0.0031 &   64  &   132 \\ 
2457608.617   &   -197.8   &    1.6   &   1.0102  &   0.0021 &   86  &   372 \\ 
2457611.603   &   -87.4   &    2.2   &   0.9974  &   0.0042 &   46  &   131 \\ 
2457625.599   &   -159.6   &    1.3   &   0.9891  &   0.0025 &   75  &   101 \\ 
2457634.430   &   -374.5   &    1.4   &   0.9831  &   0.0025 &   73  &   131 \\ 
2457635.436   &   -375.0   &    4.3   &   1.0033  &   0.0066 &   29  &   201 \\ 
2457636.443   &   -404.8   &    1.9   &   1.0015  &   0.0031 &   59  &   131 \\ 
2457643.520   &   -383.5   &    1.5   &   1.0091  &   0.0021 &   79  &   164 \\ 
2457644.508   &   -376.5   &    1.7   &   1.0396  &   0.0022 &   81  &   269 \\ 
2457645.524   &   -376.3   &    1.8   &   0.9851  &   0.0029 &   69  &   1200 \\ 
2457650.486   &   -438.0   &    1.6   &   1.0233  &   0.0030 &   62  &   131 \\ 
2457654.518   &   -472.2   &    1.9   &   1.0045  &   0.0019 &   62  &   130 \\ 
2457657.434   &   -473.2   &    1.4   &   0.9998  &   0.0024 &   92  &   255 \\ 
2457673.386   &   -46.9   &    1.4   &   1.0038  &   0.0022 &   70  &   131 \\ 
2457678.359   &   14.8   &    1.4   &   0.9972  &   0.0030 &   77  &   301 \\ 
2457679.315   &   -5.3   &    2.3   &   1.0084  &   0.0033 &   57  &   131 \\ 
2457684.369   &   -121.3   &    2.4   &   1.0090  &   0.0027 &   52  &   262 \\ 
2457685.339   &   -143.9   &    1.7   &   1.0028  &   0.0023 &   65  &   261 \\ 
2457689.310   &   -240.7   &    1.4   &   0.9986  &   0.0023 &   73  &   132 \\ 
2457694.318   &   -366.0   &    1.4   &   0.9970  &   0.0019 &   68  &   132 \\ 
2457694.321   &   -364.5   &    1.1   &   0.9999  &   0.0031 &   92  &   181 \\ 
2457698.376   &   -423.5   &    1.8   &   1.0016  &   0.0019 &   56  &   301 \\ 
2457705.299   &   -367.7   &    1.5   &   0.9968  &   0.0019 &   89  &   190 \\ 
2457711.277   &   -449.2   &    1.7   &   1.0038  &   0.0015 &   89  &   283 \\ 
  
\hline           
\end{tabular}
\end{table*}

\begin{table*}
\caption{CARMENES Doppler measurements for GJ\,1148.} 
\label{table:GJ1148} 

\centering  

\begin{tabular}{c c c c c c c c c} 

\hline\hline    
\noalign{\vskip 0.5mm}

Epoch [JD] & RV [m\,s$^{-1}$] & $\sigma_{RV}$ [m\,s$^{-1}$] & H$\alpha$ & $\sigma_{\rm H\alpha}$  & SNR  & Exp. time [sec] \\  

\hline     
\noalign{\vskip 0.5mm}    

2457414.659   &   -70.2   &    1.3   &   0.9902  &   0.0023 &   75  &   1500 \\ 
2457419.694   &   -74.6   &    1.2   &   0.9841  &   0.0017 &   98  &   1100 \\ 
2457476.533   &   -18.0   &    1.5   &   0.9955  &   0.0016 &   101  &   1200 \\ 
2457510.439   &   -9.4   &    1.3   &   0.9862  &   0.0020 &   80  &   1200 \\ 
2457529.388   &   -51.9   &    1.4   &   0.9940  &   0.0015 &   97  &   1200 \\ 
2457754.710   &   -21.3   &    1.5   &   0.9812  &   0.0020 &   83  &   710 \\ 
2457755.760   &   -3.0   &    4.5   &   0.9850  &   0.0095 &   20  &   1317 \\ 
2457761.604   &   3.3   &    1.0   &   0.9817  &   0.0020 &   85  &   759 \\ 
2457802.575   &   6.1   &    1.2   &   0.9868  &   0.0016 &   102  &   973 \\ 
2457806.388   &   11.1   &    3.8   &   0.9686  &   0.0083 &   24  &   1802 \\ 
2457808.566   &   -5.2   &    2.4   &   0.9801  &   0.0026 &   65  &   1802 \\ 
2457814.639   &   -20.2   &    1.3   &   0.9767  &   0.0011 &   142  &   1471 \\ 
2457815.709   &   -22.9   &    1.3   &   0.9814  &   0.0014 &   121  &   1802 \\ 
2457817.534   &   -25.4   &    1.5   &   0.9824  &   0.0023 &   74  &   1802 \\ 
2457818.542   &   -31.3   &    1.7   &   0.9800  &   0.0018 &   94  &   1802 \\ 
2457819.513   &   -35.9   &    1.3   &   0.9780  &   0.0011 &   144  &   1740 \\ 
2457821.507   &   -39.6   &    1.1   &   0.9753  &   0.0011 &   142  &   1299 \\ 
2457822.540   &   -41.6   &    1.0   &   0.9737  &   0.0011 &   138  &   1208 \\ 
2457823.547   &   -47.0   &    1.3   &   0.9735  &   0.0011 &   141  &   1297 \\ 
2457824.549   &   -50.4   &    1.0   &   0.9710  &   0.0011 &   142  &   1802 \\ 
2457828.529   &   -62.4   &    1.4   &   0.9757  &   0.0024 &   69  &   1802 \\ 
2457829.517   &   -63.8   &    1.5   &   0.9727  &   0.0028 &   60  &   1802 \\ 
2457830.522   &   -67.3   &    1.2   &   0.9823  &   0.0014 &   120  &   1802 \\ 
2457833.528   &   -62.5   &    1.0   &   0.9782  &   0.0011 &   142  &   1719 \\ 
2457834.668   &   -56.6   &    1.7   &   0.9853  &   0.0027 &   63  &   1802 \\ 
2457848.448   &   -1.0   &    1.3   &   0.9744  &   0.0012 &   137  &   1801 \\ 
2457852.628   &   -10.9   &    1.0   &   0.9789  &   0.0012 &   130  &   1802 \\ 
2457853.450   &   -14.3   &    1.1   &   0.9783  &   0.0012 &   136  &   1646 \\ 
2457855.513   &   -20.7   &    1.0   &   0.9816  &   0.0011 &   139  &   1452 \\ 
2457856.461   &   -22.4   &    1.5   &   0.9797  &   0.0012 &   134  &   1668 \\ 
2457857.449   &   -24.9   &    0.9   &   0.9918  &   0.0011 &   142  &   1802 \\ 
2457858.456   &   -28.5   &    0.9   &   0.9736  &   0.0012 &   138  &   1556 \\ 
2457859.473   &   -32.6   &    1.1   &   0.9950  &   0.0014 &   112  &   1801 \\ 
2457860.448   &   -31.6   &    4.1   &   0.9411  &   0.0081 &   22  &   899 \\ 
2457861.461   &   -37.3   &    1.0   &   0.9774  &   0.0012 &   136  &   1460 \\ 
2457862.495   &   -40.1   &    1.0   &   0.9839  &   0.0012 &   136  &   1802 \\ 
2457863.449   &   -40.6   &    2.8   &   0.9748  &   0.0053 &   34  &   1802 \\ 
2457864.454   &   -45.5   &    1.6   &   0.9820  &   0.0024 &   70  &   1801 \\ 
2457866.443   &   -53.2   &    1.4   &   0.9813  &   0.0011 &   140  &   1802 \\ 
2457875.494   &   -69.7   &    1.1   &   0.9743  &   0.0013 &   122  &   1801 \\ 
2457876.447   &   -59.7   &    1.3   &   0.9830  &   0.0011 &   146  &   1374 \\ 
2457877.399   &   -48.2   &    1.1   &   0.9818  &   0.0013 &   124  &   1802 \\ 
2457880.399   &   -9.6   &    1.2   &   0.9840  &   0.0011 &   141  &   1407 \\ 
2457881.389   &   -5.8   &    1.6   &   0.9824  &   0.0011 &   145  &   1715 \\ 
2457882.435   &   0.4   &    1.2   &   0.9805  &   0.0011 &   139  &   1634 \\ 
2457883.425   &   0.3   &    1.3   &   0.9761  &   0.0020 &   83  &   1801 \\ 
2457886.501   &   -1.4   &    1.7   &   0.9808  &   0.0018 &   93  &   1801 \\ 
2457887.493   &   4.2   &    1.5   &   0.9772  &   0.0020 &   81  &   1800 \\ 
2457888.436   &   2.7   &    1.4   &   0.9790  &   0.0011 &   145  &   1488 \\ 
2457889.401   &   -1.6   &    1.1   &   0.9839  &   0.0012 &   135  &   1801 \\ 
2457890.474   &   -3.2   &    1.1   &   0.9791  &   0.0011 &   153  &   1720 \\ 
2457891.470   &   -8.2   &    1.9   &   0.9764  &   0.0011 &   143  &   1623 \\ 
  
\hline           
\end{tabular}
\end{table*}

\end{appendix}

\end{document}